\documentclass[a4paper,11pt]{article}
\usepackage{jcappub} % for details on the use of the package, please see the JINST-author-manual
\usepackage{xspace}
\usepackage{txfonts}
\usepackage[normalem]{ulem}
\usepackage{fontawesome5}
\usepackage{hyperref}

\usepackage[table]{xcolor}
\usepackage[dvipsnames]{xcolor}
\definecolor{Teal}{HTML}{F02DC6}
\definecolor{gray}{gray}{0.9}

\newcommand{\rd}{\mathrm{d}}

\title{\fontsize{24}{35}\selectfont{The Super-Sample Covariance of Line-Intensity Mapping Power Spectrum}}

\author[a]{Sefa Pamuk,}
\author[a]{José Luis Bernal,}
\author[b]{and Azadeh Moradinezhad Dizgah}
\affiliation[a]{Instituto de F\'isica de Cantabria, Edificio Juan Jord\'a, Avenida de los Castros, 39005 Santander, Spain}
\affiliation[b]{Laboratoire d’Annecy de Physique Theorique (LAPTh), CNRS/USMB, 99 Chemin de Bellevue BP110 - Annecy - F-74941 - ANNECY CEDEX - FRANCE}

\emailAdd{pamuk@ifca.es, jlbernal@ifca.es, azadeh.moradinezhad@lapth.cnrs.fr}

\abstract{
In this work, we provide the first derivation of the line-intensity mapping (LIM) power spectrum super-sample covariance (SSC) from first principles, and also derive as a by-product the non-Gaussian in-box contributions to the covariance for the first time. Previous studies have typically modelled the LIM power spectrum covariance using either the Gaussian approximation or estimates obtained from mocks or the data itself, neglecting uncertainties related to whether the limited volume surveyed sits in a cosmological overdensity. This contribution, known as the SSC
or, depending on the context, the field-to-field variance, cannot be estimated from the data, but it is crucial for a correct inference of \textit{global} quantities, i.e., for ensemble-averaged parameters rather than the actual values just \textit{within} the patch of the Universe observed. For our derivation, we employ a combination of the halo model and standard perturbation theory that allows us to capture the nonlinearity and non-Gaussianity of the covariance.  After a successful validation of our predictions against painted N-body simulations, we explore different scenarios related to current and future LIM experiments, quantifying the relative importance of the non-Gaussian in-box and SSC. We find that the newly derived contributions to the LIM power spectrum covariance are crucial at intermediate and small scales, especially for cases in which the covariance is not dominated by instrumental noise. We find that the relative relevance of the SSC with respect to the other covariance contributions is roughly independent of the survey volume, but does depend on the specific response of the power spectrum to large-scale modes for each line and redshift. 
Therefore, the impact of the SSC will be increasingly significant for parameter inference from the current and the next generation high signal-to-noise LIM surveys. 
}

\begin{document}
\maketitle
\flushbottom

\section{Introduction}
\label{sec:intro}
Line intensity mapping (LIM)~\cite{Bernal_2022, Chang:2026ake} has emerged as a novel technique to probe the large-scale structure of the Universe. This methodology uses observations with low-aperture telescopes and spectrometers to measure the integrated flux from a particular line of sight without requiring high-significance, resolved detections. By targeting well-identifiable spectral lines and making use of the good spectral resolution, LIM reconstructs the redshift of the measured emission and builds three-dimensional maps of line intensity fluctuations, a tracer of the underlying matter field. Despite its low angular resolution and not resolving individual sources, LIM has the potential to survey distant volumes into cosmic history, providing an excellent independent and complementary probe to other observables of cosmic large-scale structure (LSS) such as galaxy clustering and weak lensing. 
For instance, the $21\,\mathrm{cm}$ neutral hydrogen line has been proposed to probe the post-reionisation large-scale structure~\cite{Cunnington:2025sdr, CHIME:2023til, SKA:2018ckk} and give a unique window into the era of reionisation, which is normally inaccessible to conventional wide-field galaxy surveys \cite{Lidz2008ApJ...680..962L, Pierre_Christian_2013}. Other currently explored lines include CO rotational lines~\cite{Cleary_2022, Breysse_2022, Karkare_2022}, [CII] fine structure line~\cite{concerto_overview, 10.1117/1.JATIS.7.4.044004} and Lyman-$\alpha$~\cite{niemeyer2025lyalphaintensitymappinghetdex}, among others. Different spectral lines probe halos with different properties, allowing LIM to trace the underlying dark matter field and reconstruct the LSS also in an inherently multi-tracer way. 
The most widely used method to obtain cosmological information from three-dimensional line-intensity maps is to measure the power spectrum. Cross-correlations with galaxy surveys have enabled high-significance measurements (see e.g., Refs.~\cite{Cunnington_2022, 10.1093/mnras/staf195, CHIME:2023til, niemeyer2025lyalphaintensitymappinghetdex}). Meanwhile, recent measurements of the auto-power spectra~\cite{Paul:2023yrr, CHIME:2025cee} at small, nonlinear scales, and upcoming detections at cosmological scales on the horizon~\cite{Cunnington:2025sdr} are demonstrating the feasibility of LIM. As observations become sensitive enough to extract astrophysical and cosmological information from them, an accurate assessment of the uncertainties is crucial for a correct interpretation of the results and their implications.

There are different approaches to estimate the covariance for LSS. For instance, galaxy clustering studies have usually relied on numerical estimates from simulated mocks to account for all contributions and observational effects from modes inside the volume probed \cite{Kitaura:2015uqa,Blot:2018oxk,Colavincenzo:2018cgf,eBOSS:2020wwo,Ereza:2023zmz,Forero-Sanchez:2024bjh,Maus:2026wsb}. As precision has increased, accurate covariance estimation requires a large number of mocks, which poses a significant computational challenge. Hybrid approaches introducing theoretical understanding of the covariance have been adopted in order to circumvent the high computational cost \cite{Wadekar:2019rdu,Taruya:2020qoy,Zhao:2024xit,Hadzhiyska:2026wts,Farina:2026kji}. While current volumes and precision do not impose hard computational requirements for LIM observations,  additional challenges in the covariance estimation are present: after foreground cleaning and systematics treatment, low levels of unknown residuals may still affect the variance of the observations significantly. This is why, in order to estimate the covariance, on top of approaches using simulations  (see e.g., Refs.~\cite{COMAP:2018kem, niemeyer2025lyalphaintensitymappinghetdex}), 
data-based approaches using e.g., jackknife estimates are also applied to account for all contributions from the observed volume (see e.g., Ref.~\cite{Carucci:2024qpm}).
 
Various contributions to the covariance of LSS tracers and their physical origin are theoretically well-understood. The first contribution, which dominates on large scales, is the cosmic variance. It arises naturally assuming that each Fourier mode is an independent Gaussian random variable. Thus the power spectrum estimates for different Fourier modes are independent, but only a limited number of modes are available to their estimation. We refer to this contribution as the `Gaussian covariance' \cite{Scoccimarro:1999kp}. There are two additional non-Gaussian contributions to the covariance. First, the nonlinear evolution of structure introduces non-Gaussianities and mode coupling in the matter distribution and its biased tracers \textit{within} the probed volume, (see Refs. e.g.,~\cite{Scoccimarro:1999kp, Harnois-Deraps:2011ixh,Harnois-Deraps:2012kbb, Repp:2015jja, Bertolini:2015fya, Mohammed:2016sre, Wadekar:2019rdu, Kobayashi:2023vpu} for derivations in the context of galaxy clustering and cosmic shear). We will refer to this contribution as the `in-box non-Gaussian covariance'. The second one is related to the uncertainty of whether the observed volume corresponds to an overall overdense or underdense region of the Universe, hence related to perturbations \textit{beyond} the size of the survey. This contribution is known as the `super-sample covariance'~\cite{Hamilton:2005dx,dePutter:2011ah,Takada_2013,Li:2014sga,Li:2014jra, Chan:2017fiv,Schreiner:2024grf} (SSC). In other words, long-wavelength modes outside the survey volume modulate the amplitude of matter fluctuations (and thus any other observable) relative to the cosmic mean. At the level of perturbation theory, these unconstrained large-scale modes that lie outside the survey volume allow for correlation between pairs of small-scale modes. After averaging over the super-survey fluctuations, this gives rise to a connected covariance between otherwise independent Fourier modes \cite{Hamilton:2005dx}. The SSC has been well studied in the literature also for the case of galaxy clustering and cosmic shear~\cite{Li:2017qgh,euclid_ssc, kids_cov, Barreira:2017fjz}.

LIM surveys have only accounted so far for the in-box covariance, neglecting the contribution from SSC. This approach is appropriate for determining the detection significance of a given measurement or for inferring a parameter \textit{within the observed volume} (e.g., the hydrogen density in the volume probed). Nonetheless, the inference of global parameters (e.g., how much hydrogen density there is, what the cosmic star-formation rate is or what the cosmological parameters are in the whole Universe) from a limited sample requires accounting for the uncertainty of whether 
the volume probed lies in an underdensity or an overdensity, and for the mode coupling due to modes larger than the survey itself. In order to fill this gap, this work focuses on the derivation and validation of the estimation of the LIM power spectrum SSC.

Previous studies focused on other LSS observables have found that the SSC is relevant at scales for which the variance $\sigma^2_V$ of 
the super-survey mode (i.e., the background mode) 
is larger than the cosmic variance: roughly speaking, scales for which $\sigma^2_W\equiv\sigma^2_V/V \sim N_{\rm modes}^{-1}$, where $N_{\rm modes}$ is the number of modes measured. Therefore, for small surveys, the SSC is relevant at all scales, but, contrary to intuition, the SSC is still relevant for large surveys at small scales, where the cosmic variance is very small.

In the context of LIM, we expect a sizable SSC contribution to the power spectrum covariance. Long-wavelength modes will result in a larger or smaller number of collapsed objects, impacting the number of galaxies that emit the spectral line of interest. Therefore, the measured intensity fluctuations depend on whether the volume probed sits on a large-scale overdensity or underdensity.\footnote{21-cm LIM from cosmic dawn and the epoch of reionisation is similarly affected by the SSC. In this case, the long-wavelength modes determine the number of collapsed objects which host the sources that heat and ionise the gas.} It is worth noting that, in galaxy clustering, the two contributions to the SSC, commonly referred to as the `beat-coupling' and `local average' terms, partially cancel each other~\cite{dePutter:2011ah}, to a degree that depends on the estimator and is weaker for the FKP estimator used in practice~\cite{Wadekar:2019rdu}. In LIM, on the other hand, the observables are absolute intensities and are not normalised to an in-survey estimated mean, so no such cancellation occurs.
Despite all this, there has just been an empirical study of the SSC based on simulations~\cite{G20}. This work estimated the SSC (referred to as field-to-field variance) directly from the power spectrum measurements in different subvolumes of a lightcone simulation. In addition, following the intuition exposed above, the authors found that the number count of bright sources within small volumes varies strongly, with the corresponding impact on the power spectrum measurements in the shot-noise regime.

In this work, we derive an analytical expression, for the first time, of the total non-Gaussian covariance of the LIM power spectrum, including the in-box non-Gaussian and SSC contributions. Perturbation theory is a very common approach to predict galaxy clustering summary statistics with high accuracy; however, the LIM theoretical framework usually relates the fluctuations to a line-luminosity halo relation, given that LIM measurements are also sensitive to astrophysics. This 
makes the halo model a very suitable approach for its modelling. We acknowledge the benefits of both techniques; hence, we merge both approaches, extending the halo model \cite{COORAY_2002, Asgari_2023} by combining it with standard perturbation theory (SPT) of large-scale structure~\cite{Bernardeau_2002}, similar to Ref.~\cite{MoradinezhadDizgah:2021dei}, which combines the halo model and the effective field theory of the large-scale structure. 

We derive the full real-space power-spectrum covariance matrix in this framework, isolate the SSC and validate our predictions using [CII] LIM simulations constructed by painting halos in N-body simulations with [CII] line luminosities. 
The relative importance of each contribution to the covariance depends significantly on the line and the corresponding specifics of the halo-line connection; in particular, the relevant properties are the mass range of the halos that dominate the line flux, and the survey geometry and redshift. Assuming simple survey geometries, we quantify the relevance of the SSC with respect to the in-box covariance for different lines and experiments. This demonstrates that the SSC must be taken into account for accurate global parameter inference from LIM power-spectrum measurements. 
We accompany this work with the public release of a newly developed code called \texttt{SSLimPy}.\footnote{\href{https://github.com/sefa76/SSLimPy}{https://github.com/sefa76/SSLimPy} \faGithub}

This work is structured as follows. Section~\ref{sec:theo_cov} includes a general derivation of all contributions to the LIM power spectrum covariance in real space. We specify our model in section~\ref{sec:limhalos}, discussing the proposed extension of the halo model and showing the explicit expressions for the covariance contributions. We validate our theoretical prediction of the SSC with simulations in section~\ref{sec:sim}, and show the relevance of the SSC with respect to the in-box covariance for different cases in section~\ref{sec:forecast}. Finally, we present our conclusions in section~\ref{sec:conc}. The finer details of the SPT and halo model considered in this work are discussed in appendix~\ref{sec:appendixA}, while we show our implementation of the trispectrum in appendix~\ref{sec:trispectrum_terms}. An approximate treatment of the effects of finite survey resolution on the power spectrum covariance is presented in appendix~\ref{sec:survey}. 

\section{The full real-space covariance}
\label{sec:theo_cov}
In this section, we derive the full real-space covariance for the LIM power spectrum. We will distinguish between three contributions: the Gaussian covariance, the `in-box' non-Gaussian covariance (i.e., the non-Gaussian contributions sourced by the gravitational evolution of the perturbations \textit{within} the volume probed) and the SSC. We neglect redshift-space distortions in this derivation. We refer the interested reader to Ref.~\cite{Wadekar:2019rdu} for a derivation in the presence of redshift-space distortions in the context of galaxy clustering,\footnote{A finite volume is also subject to a coherent super-survey tidal field, of amplitude comparable to that of long-wavelength density ~\cite{Akitsu:2016leq}. Being traceless, it does not contribute to the spherically averaged power spectrum at leading order, but will need to be included in the axisymmetric extension.} and leave their inclusion, along with anisotropic resolution limits, and other effects like primordial non-Gaussianity \cite{Castorina:2020blr} and local source density fluctuations~\cite{Wadekar:2019rdu} for future work.

We start by introducing the definitions that we will use throughout the paper. Our Fourier convention is: 
\begin{equation}
    f(\boldsymbol{k}) = \int_V\:f(\boldsymbol{x})\,e^{-i\,\boldsymbol{k}\,\boldsymbol{x}}\,\mathrm{d}^3\boldsymbol{x}\:,\\ \qquad\qquad\qquad 
    f(\boldsymbol{x}) = \frac{1}{V} \sum_{\boldsymbol{k}}\:f(\boldsymbol{k})\,e^{i\,\boldsymbol{k}\,\boldsymbol{x}}\:,
\end{equation}
where $\boldsymbol{x}$ and $\boldsymbol{k}$ denote the three-dimensional position in configuration space and its Fourier conjugate wave vector, respectively, and $V$ refers to the observed volume. In this work, we will use the discrete Fourier space since our observables are always defined to be measured on finite volumes. For a density fluctuation $\delta_\mathrm{L}$ in the linear regime, the power spectrum is defined as  
\begin{equation}
    \langle \delta_\mathrm{L}(\boldsymbol{k})\,\delta_\mathrm{L}(\boldsymbol{k}')\rangle = V\,P(k)\,\delta_{\boldsymbol{k}+\boldsymbol{k}',0}\:,
\end{equation}
and similarly for higher-order correlation functions. The symbol $\delta_{\boldsymbol{a},\boldsymbol{b}}$ is the Kronecker delta, the discrete limit of the three-dimensional Dirac delta. 

We define the survey window function $W$, which relates the underlying observable $\mathcal{O}$ to the observable as measured by the survey $\mathcal{O}_W$:
\begin{equation}
    \mathcal{O}_W(\boldsymbol{x}) = W(\boldsymbol{x})\,\mathcal{O}(\boldsymbol{x})\:,\qquad\qquad\qquad
    \mathcal{O}_W(\boldsymbol{k}) = \frac{1}{V}\,\sum_{\boldsymbol{k}'}\:W(\boldsymbol{k}-\boldsymbol{k}')\,\mathcal{O}(\boldsymbol{k}')\:.
\end{equation}
We will encounter the Fourier transformation of integer powers of $W$ as well as their real-space integral: \begin{align}
    W_n(\boldsymbol{k}) &= \frac{1}{V^{n-1}}\:\sum_{\boldsymbol{k}_1\dots\,\boldsymbol{k}_n}\:\delta_{\boldsymbol{k}_1+\dots+\boldsymbol{k}_{n},\boldsymbol{k}}\,\prod_{i=1}^n\:W(\boldsymbol{k}_i)\:,\\
    V_n &= \int_V\:W^n(\boldsymbol{x})\,d^3\boldsymbol{x} = W_n(0)\:.
    \label{eq:Vn}
\end{align} 
The simplest survey window function is a top-hat window that is unity inside the survey window and zero outside it. For this function, $W_n=W$ and $V_n=V$ for all $n$. We derive all results for an arbitrary survey window function. However, we assume binary window functions for both the numerical validation of the covariance model against simulations and the estimation of the covariance contributions and their relative importance, considering a cubic and a cylindrical geometry, respectively.

The power spectrum of the observed temperature fluctuations $\delta_\mathrm{T}^\mathrm{obs}(\boldsymbol{x}) = W(\boldsymbol{x})\,\left[T(\boldsymbol{x})-\langle T\rangle_\mathrm{survey}\right]$ is given by a convolution with the observational mask $W$:
\begin{align}
     \left\langle\delta_\mathrm{T}^\mathrm{obs}(\boldsymbol{k})\,\delta_\mathrm{T}^\mathrm{obs}(\boldsymbol{k}')\right\rangle &=\frac{1}{V}\sum_{\boldsymbol{q}_1}W(\boldsymbol{k}-\boldsymbol{q}_1)\,W(\boldsymbol{k}'+\boldsymbol{q}_1)\,P_\mathrm{TT}(q_1)\:,
\end{align}
where $P_{\rm TT}$ is the ensemble average temperature power spectrum. Using ergodicity, we can replace the stochastic average with a volume average and recover an unbiased estimator of the power spectrum after normalising the survey function:
\begin{equation}
\widehat{P}_\mathrm{TT}(k_i)=\frac{1}{V_2\,N_i}\,\sum_{\boldsymbol{k}\in\boldsymbol{V}_{k_i}}\:\delta_\mathrm{T}^\mathrm{obs}(\boldsymbol{k})\,\delta_\mathrm{T}^\mathrm{obs}(-\boldsymbol{k})\:,
\end{equation}
where $k_i$ denotes the central wave number of a bin $\boldsymbol{V}_{k_i}$, and $N_i\approx V\, V_{k_i}\,(2\,\pi)^{-3}$ 
is the number of discrete wave vectors inside the bin, where $V_{k_i}$ is the Fourier-space volume of bin $k_i$. 

We can now compute the covariance of this estimator, for which we find a correlator of four $\delta_\mathrm{T}$. We split this correlator into the disconnected part, which we compute with Wick's probability theorem, and the connected part, which will become the trispectrum and consists of the contributions $T_0$, from the nonlinear evolution of modes within the survey, and the SSC. 
The covariance is given by
\begin{equation}
\begin{split}
    \mathrm{Cov}(k_i,k_j) &=\left\langle\widehat{P}_\mathrm{TT}(k_i)\,\widehat{P}_\mathrm{TT}(k_j)\right\rangle - \left\langle\widehat{P}_\mathrm{TT}(k_i)\right\rangle\,\left\langle\widehat{P}_\mathrm{TT}(k_j)\right\rangle\\
    &=\frac{1}{V_2^2\,N_i\,N_j}\,\sum_{\boldsymbol{k}_1\in\boldsymbol{V}_{\boldsymbol{k}_i}}\sum_{\boldsymbol{k}_2\in\boldsymbol{V}_{\boldsymbol{k}_j}}\left\lbrace\left\langle\delta_\mathrm{T}^\mathrm{obs}(\boldsymbol{k}_1)\,\delta_\mathrm{T}^\mathrm{obs}(-\boldsymbol{k}_1)\,\delta_\mathrm{T}^\mathrm{obs}(\boldsymbol{k}_2)\,\delta_\mathrm{T}^\mathrm{obs}(-\boldsymbol{k}_2)\right\rangle\right.\\
    &\left.\phantom{\frac{1}{V_W^2}}-\left\langle\delta_\mathrm{T}^\mathrm{obs}(\boldsymbol{k}_1)\,\delta_\mathrm{T}^\mathrm{obs}(-\boldsymbol{k}_1)\right\rangle\,\left\langle\delta_\mathrm{T}^\mathrm{obs}(\boldsymbol{k}_2)\,\delta_\mathrm{T}^\mathrm{obs}(-\boldsymbol{k}_2)\right\rangle  \right\rbrace \\ &= {\rm Cov}^{\rm G}(k_i,k_j) + {\rm Cov}^{\rm NG}(k_i,k_j)  \\
    &=\frac{1}{V_2^2\,V^2\,N_i\,N_j}\,\sum_{\boldsymbol{k}_1\in\boldsymbol{V}_{\boldsymbol{k}_i}}\sum_{\boldsymbol{k}_2\in\boldsymbol{V}_{\boldsymbol{k}_j}}\left\lbrace\sum_{\boldsymbol{q}_1,\boldsymbol{q}_2}\:P_\mathrm{TT}(q_1)\,P_\mathrm{TT}(q_2)\right.\\
    &\phantom{\frac{1}{V_W^2}}\times\left[W(\boldsymbol{k}_1-\boldsymbol{q}_1)\,W(\boldsymbol{k}_2+\boldsymbol{q}_1)\,W(-\boldsymbol{k}_1-\boldsymbol{q}_2)\,W(-\boldsymbol{k}_2+\boldsymbol{q}_2)\right.\\
    &\phantom{\frac{1}{V_W^2}}\quad+\left.W(\boldsymbol{k}_1-\boldsymbol{q}_1)\,W(-\boldsymbol{k}_2+\boldsymbol{q}_1)\,W(-\boldsymbol{k}_1-\boldsymbol{q}_2)\,W(\boldsymbol{k}_2+\boldsymbol{q}_2)\right] \\
    &\phantom{\frac{1}{V_W^2}}+\frac{1}{V}\sum_{\boldsymbol{q}_1, \boldsymbol{q}_2, \boldsymbol{q}_3, \boldsymbol{q}_4}\:W(\boldsymbol{q}_1)\,W(\boldsymbol{q}_2)\,W(\boldsymbol{q}_3)\,W(\boldsymbol{q}_4)\,\delta_{\boldsymbol{q}_1+\boldsymbol{q}_2+\boldsymbol{q}_3+\boldsymbol{q}_4,0}\\
    &\left.\phantom{\frac{1}{V_W^2}\frac{1}{V}}\times T_\mathrm{T}(\boldsymbol{k}_1-\boldsymbol{q}_1,-\boldsymbol{k}_1-\boldsymbol{q}_2,\boldsymbol{k}_2-\boldsymbol{q}_3,-\boldsymbol{k}_2-\boldsymbol{q}_4)\right\rbrace\:,
\end{split}
\label{eq:cov_conn_and_disconn}
\end{equation}
where $T_{\rm T}$ denotes the temperature trispectrum.

We will start our discussion of the individual terms with the disconnected part in \eqref{eq:cov_conn_and_disconn}: the first term in the last equality. The main structure of this term consists of two power spectra convolved in a complex manner with a combination of window functions. 
Let us consider first the large-volume limit. In this limit, the window functions are sharply peaked, essentially confining the argument to a region of size $q_\mathrm{smooth}$. The first and second window functions, thus confine $\boldsymbol{k}_i\sim \boldsymbol{q}_i$, for modes $k_i\gg q_\mathrm{smooth}$:
\begin{equation}
    \frac{1}{V}\,\sum_{\boldsymbol{q}_1}\:P_\mathrm{TT}(q_1)\,W(\boldsymbol{k}_1-\boldsymbol{q}_1)\,W(\boldsymbol{k}_2+\boldsymbol{q}_1)
    \approx P_\mathrm{TT}(k_1)\,W_2(\boldsymbol{k}_1+\boldsymbol{k}_2)\:. 
 \end{equation}
What we are left with is the Fourier transform of the square of the window function. Following our logic, $W_{2}$ can also be assumed to be highly peaked. Inside the sums, we can write\,
\begin{equation}
    \sum_{\boldsymbol{k}_1\in\boldsymbol{V}_{\boldsymbol{k}_i}}\sum_{\boldsymbol{k}_2\in\boldsymbol{V}_{\boldsymbol{k}_j}}\dots \left[ \left|W_{2}(\boldsymbol{k}_1+\boldsymbol{k}_2)\right|^2+\left|W_{2}(\boldsymbol{k}_1-\boldsymbol{k}_2)\right|^2 \right] \approx \sum_{\boldsymbol{k}_1\in\boldsymbol{V}_{\boldsymbol{k}_i}}\sum_{\boldsymbol{k}_2\in\boldsymbol{V}_{\boldsymbol{k}_j}}\dots V_4\,V\,\left[\, \delta_{\boldsymbol{k}_1+\boldsymbol{k}_2,0} + \delta_{\boldsymbol{k}_1-\boldsymbol{k}_2,0}\right] \label{eq:binoverlap_Gcov}
\end{equation}
Here, we have used the fact that the window function itself is sharply peaked, confining $\boldsymbol{q}_1\sim\boldsymbol{k}_1$ to pull the power spectrum outside of the sum.  Inserting this into the first term of \eqref{eq:cov_conn_and_disconn}, we find
\begin{equation}
    \mathrm{Cov}^\mathrm{G}(k_i,k_j) \approx \,\frac{V_4\,V}{V_2^2}\times\frac{2\,\delta_{i,j}}{N_i}\times\frac{1}{N_i}\sum_{\boldsymbol{k}_1\in\boldsymbol{V}_{\boldsymbol{k}_i}}P^2_\mathrm{TT}(k_1)\:,
    \label{eq:Gaussiancov}
\end{equation}
where we have used that $\boldsymbol{k}_1$ and $\boldsymbol{k}_2$ can only coincide if they are in the same bin. Here we can identify the effective volume $V_{\rm eff}\equiv V_2^2\,V_4^{-1}$, the standard expression of the cosmic variance in the second factor, and the bin average of the square of the power spectrum in the last factor. 

Foregoing the approximations in equation \eqref{eq:binoverlap_Gcov} introduces a correlation between two wave numbers that are separated by less than $q_\mathrm{smooth}$. The correlation between bins is enhanced by the finite width of the wave number bins~\cite{Wadekar:2019rdu}. 

Depending on experimental specifications, the effective experimental noise power spectrum should also be added to equation~\eqref{eq:Gaussiancov}. The thermal noise of the experiment, assuming single-dish observations, is usually assumed to be Gaussian-distributed with zero mean and variance given by the radiometer equation:
\begin{equation}
    \sigma^2_\mathrm{T} = \frac{(T_{\rm sys} / \eta)^2}{t_{\rm pix}\,\delta\nu}\:,
\end{equation}
where $T_{\rm sys} / \eta$ is the effective system temperature, $\delta\nu$ is the channel frequency width, and $t_{\rm pix}$ is the total effective integration time per pixel, including contributions from all detectors, polarisations, and repeated observations. The noise power spectrum assuming constant $\sigma^2_{\rm T}$ for the whole survey is then given as:
\begin{equation}
    P_\mathrm{TN} = \sigma_\mathrm{T}^2\,V_\mathrm{vox}\:,
\end{equation}
where $V_\mathrm{vox}$ is the comoving voxel volume, given by the product of the pixel area and the radial depth corresponding to $\delta \nu$.

Since the thermal noise is assumed to be Gaussian and uncorrelated with the signal, it modifies the Gaussian covariance through the replacement of the signal power spectrum with the total power $P_\mathrm{TT} + P_\mathrm{TN}$. The non-Gaussian contributions, such as the trispectrum contribution and the SSC, remain unaffected at leading order. Within our large volume approximation, the Gaussian covariance is given by:
\begin{equation}
    \mathrm{Cov}^\mathrm{G}(k_i,k_j) \approx \,\frac{V_4\,V}{V_2^2}\times\frac{2\,\delta_{i,j}}{N_i}\times\frac{1}{N_i}\sum_{\boldsymbol{k}_1\in\boldsymbol{V}_{\boldsymbol{k}_i}}\left[P_\mathrm{TT}(k_1)+P_\mathrm{TN}\right]^2\:,
    \label{eq:Gaussian_Noise_cov}
\end{equation}
A survey is usually defined by its total area and observation time. Therefore, we compute the quantities $t_\mathrm{pix}$ and $V_\mathrm{vox}$ from the angular resolution, the bandwidth, and the number of independent detectors. We assume that the pixel size is given by the full-width half-maximum of the beam, and the voxel frequency width is given by the channel width. Additionally, we will assume an even scanning of the survey area.\footnote{Realistic observations introduce further effects such as those from a more complex survey geometry, masking, map-making choices, and other observational effects in the power spectrum measurements. These can lead to mode-mixing and can generate correlations between different $k$ bins. Nevertheless, typically, the noise covariance remains close to diagonal.}
This gives us good reason to use these idealised expressions for a preliminary study. The Gaussian covariance of the LIM power spectrum Legendre multipoles with all observational terms can be found in Ref.~\cite{Bernal_2019}.

Let us now focus on the second term of the covariance, the connected part. The Fourier modes entering the arguments of the window functions in the connected trispectrum in the expression of the covariance are confined to survey scales. It is therefore convenient to characterise them with a wavevector often referred to as the ``beat mode''.
For the discussion of the connected, non-Gaussian part, we define the beat mode as $\boldsymbol{\varepsilon}=\boldsymbol{q}_1+\boldsymbol{q}_2$~\cite{Hamilton:2005dx}, which leads to a more physically intuitive expression,

\begin{equation}
\begin{split}
    \mathrm{Cov}^\mathrm{NG}(k_i,k_j)&=\frac{1}{V_2^2\,V^3\,N_i\,N_j}\sum_{\boldsymbol{k}_1\in\boldsymbol{V}_{\boldsymbol{k}_i}}\sum_{\boldsymbol{k}_2\in\boldsymbol{V}_{\boldsymbol{k}_j}}\sum_{\boldsymbol{q}_1,\boldsymbol{q}_2,\boldsymbol{q}_3,\boldsymbol{q}_4}\,W(\boldsymbol{q}_1)\,W(\boldsymbol{q}_2)\,W(\boldsymbol{q}_3)\,W(\boldsymbol{q}_4)\\ 
    &\phantom{\frac{1}{V_W^2\,V^3\,N_i\,N_j}}T_\mathrm{T}(\boldsymbol{k}_1+\boldsymbol{q}_1,-\boldsymbol{k}_1+\boldsymbol{q}_2,\boldsymbol{k}_2+\boldsymbol{q}_3,-\boldsymbol{k}_2+\boldsymbol{q}_4)\,\delta_{\boldsymbol{q}_1+\boldsymbol{q}_2+\boldsymbol{q}_3+\boldsymbol{q}_4,0}\\
    &=\frac{1}{V_2^2\,V^3\,N_i\,N_j}\sum_{\boldsymbol{k}_1\in\boldsymbol{V}_{\boldsymbol{k}_i}}\sum_{\boldsymbol{k}_2\in\boldsymbol{V}_{\boldsymbol{k}_j}}\sum_{\boldsymbol{q}_1,\boldsymbol{q}_2,\boldsymbol{\varepsilon}}T_\mathrm{T}(\boldsymbol{q}_1,-\boldsymbol{q}_1+\boldsymbol{\varepsilon},\boldsymbol{q}_3,-\boldsymbol{q}_3-\boldsymbol{\varepsilon})\\
    &\phantom{\frac{1}{V_W^2\,V^3\,N_i\,N_j}}W(\boldsymbol{q}_1-\boldsymbol{k}_1)\,W(\boldsymbol{\varepsilon}-\boldsymbol{q}_1+\boldsymbol{k}_1)\,W(\boldsymbol{q}_3-\boldsymbol{k}_2)\,W(-\boldsymbol{\varepsilon}-\boldsymbol{q}_3+\boldsymbol{k}_2)\:,
\end{split}
\end{equation}
where the last equality includes a redefinition of $\boldsymbol{q}_1+\boldsymbol{k}_1\longrightarrow\boldsymbol{q}_1$ and $\boldsymbol{q}_3+\boldsymbol{k}_2\longrightarrow\boldsymbol{q}_3$ to be able to interpret the inner wave vectors. In the large-volume approximation, the first and third survey window functions confine $\boldsymbol{q}_1 \sim \boldsymbol{k}_1$ and $\boldsymbol{q}_3 \sim \boldsymbol{k}_2$. The remaining windows confine $\boldsymbol{\varepsilon} \sim 0$, thus we can identify it as a long-wavelength mode.

Studying the limit of this squeezed configuration, we can find that the two external wave vectors $\boldsymbol{k}_1$ and $\boldsymbol{k}_2$ decorrelate~\cite{Takada_2013}. We can thus schematically write the terms within the sums above as:
\begin{equation}
    \lim_{\varepsilon\rightarrow0} T_\mathrm{T}(\boldsymbol{k}_1+\boldsymbol{\varepsilon}, -\boldsymbol{k}_1+\boldsymbol{\varepsilon}, \boldsymbol{k}_2-\boldsymbol{\varepsilon},-\boldsymbol{k}_2-\boldsymbol{\varepsilon}) = T_\mathrm{T}(\boldsymbol{k}_1,-\boldsymbol{k}_1,\boldsymbol{k}_2,-\boldsymbol{k}_2)+\frac{\mathrm{d}P_\mathrm{TT}}{\mathrm{d}\delta_\mathrm{b}}(\boldsymbol{k}_1)\,\frac{\mathrm{d}P_\mathrm{TT}}{\mathrm{d}\delta_\mathrm{b}}(\boldsymbol{k}_2)\,P(\varepsilon)\:, \label{eq:large-scale-expansion}
\end{equation}
where $\mathrm{d} P_\mathrm{TT} /\mathrm{d}\delta_\mathrm{b}$ is the response function of the power spectrum to background modes $\delta_{\rm b}$ smoothed over the sizes of the survey volume.

The first term in equation \eqref{eq:large-scale-expansion} is the collapsed trispectrum contribution, which is the non-Gaussian in-box covariance, given by
\begin{align}
    \mathrm{Cov}_{T_0}^\mathrm{NG}(k_i,k_j)&=\frac{V_4}{V_2^2}\,\frac{1}{N_i\,N_j}\,\sum_{\boldsymbol{k}_1\in\boldsymbol{V}_{\boldsymbol{k}_i}}\sum_{\boldsymbol{k}_2\in\boldsymbol{V}_{\boldsymbol{k}_j}}T_\mathrm{T}(\boldsymbol{k}_1,-\boldsymbol{k}_1,\boldsymbol{k}_2,-\boldsymbol{k}_2)\:. \label{eq:cov_ng}
\end{align}
For the simple survey case, the prefactor is again just $V^{-1}$. We can identify the non-Gaussian in-box covariance with the angular average of the squeezed trispectrum over $V_\mathrm{eff}$. In turn, the second term is what is known as the SSC, given by
\begin{equation}
    \mathrm{Cov}^\mathrm{NG}_\mathrm{SSC}(k_i,k_j) = \frac{1}{N_i \,N_j}\sum_{\boldsymbol{k}_1\in\boldsymbol{V}_{\boldsymbol{k}_i}}\sum_{\boldsymbol{k}_2\in\boldsymbol{V}_{\boldsymbol{k}_j}}\frac{\mathrm{d}P_\mathrm{TT}}{\mathrm{d}\delta_\mathrm{b}}(\boldsymbol{k}_1)\,\frac{\mathrm{d}P_\mathrm{TT}}{\mathrm{d}\delta_\mathrm{b}}(\boldsymbol{k}_2)\\
    \frac{1}{V\,V_2^2}\sum_{\boldsymbol{\varepsilon}}\left|W_2(\boldsymbol{\varepsilon})\right|^2\,P(\varepsilon)\:. \label{eq:cov_ssc}
\end{equation}

Physically, the response function represents the systematic shifts of the measured power spectrum inside a finite volume due to a smooth background mode $\delta_{\rm b}$. 
The response functions can be computed directly from a given theoretical framework or using a separate-Universe approach \cite{Takada_2013,Barreira:2017kxd,Barreira:2017fjz}. In the latter method, the main assumption is that the large-scale overdensity, in essence, leads the local volume to evolve like a separate Universe with a higher matter density. The large-scale overdensity enhances the local growth of structure and varies the local expansion history of the Universe, slightly altering the comoving scales with respect to the global average. Additionally, the number of halos depends on these large-scale overdensities, an argument used in the past to explain the halo bias within the peak--background split formalism~\cite{1974ApJ...187..425P}. Refs.~\cite{Takada_2013, Li:2014sga} showed that both methods lead to very similar results for the matter power spectrum SSC. In this work, we will use the halo-model method, for which we find an equivalent result to that of galaxy clustering~\cite{Wadekar:2019rdu}.

The SSC has the form of a product of two shell-averaged response functions. The third factor is related to the variance of the matter field background modes within the survey, $\sigma^2_V$. In this case, while the factor $V^{-1}$ in front normalises the sum over background modes $\boldsymbol{\varepsilon}$, the dependence on the volume is non-trivially encoded in the normalised functions $V_2^{-1}\, W_2$. For smaller volumes, these functions become less peaked, allowing more modes to contribute to $\sigma^2_V$. The resulting scaling is therefore not exactly $\propto V^{-1}$, but depends on the shape of the linear power spectrum over the modes admitted by the window. 

However, $\sigma_V^2$ still roughly scales like $V^{-1}$, despite its non-trivial dependence on the survey window. Thus, when comparing the relative importance of the SSC to the Gaussian and in-box non-Gaussian covariance, the volume dependence of $\sigma_V^2$ provides an approximately common scaling to the other contributions. The amplitude of the SSC additionally depends on the specific response function. This, in turn, depends on the redshift, the halo--line luminosity relation and the scale. The response therefore determines how important the SSC is relative to the other covariance contributions, but the relative importance is roughly independent of the survey volume. This is further explored in section~\ref{subsec:relative}

\section{The LIM halo model with Standard Perturbation Theory}

\label{sec:limhalos}
In this section, we review the formalism for the LIM correlation functions in Fourier space. We combine the halo model~\cite{Asgari_2023, COORAY_2002} with SPT~\cite{Bernardeau_2002}, in a similar spirit to Ref.~\cite{MoradinezhadDizgah:2021dei}, to capture the nonlinear LIM fluctuations and their higher-order statistics. In what follows, we will assume that all emission comes from within halos. This assumption neglects radiation transfer beyond the extent of the halo, which is accurate for all cases except Lyman-$\alpha$ intensity mapping (see, e.g., Refs.~\cite{LujanNiemeyer:2022rby, LujanNiemeyer:2022cte}), for which the effect is relevant even at relatively large scales~\cite{LujanNiemeyer:2024dyv}. 
Furthermore, as a first approach to these calculations in the context of LIM, we limit our study to real space, neglecting the effect of redshift-space distortions. Additional effects that depend on the observational angle between the line of sight and the wave vector modes are left for future work. Among other things, this choice foregoes the treatment of the effects of resolution limits (an effective description can be found in appendix~\ref{sec:survey}) and line broadening due to peculiar velocities of sources within a halo~\cite{COMAP:2021rny, li2024modelingnonlinearpowerspectrum}. These choices leave the LIM fluctuations isotropic. 

The first central assumption of the halo model is that all matter in the Universe is enclosed in halos, with a mean density given by the halo mass function $\langle\rd n/\rd M\rangle$, assumed to depend only on the halo mass $M$. Halos are a biased tracer of the matter distribution. As the baryonic matter clustering inside the halo emits the spectral line of interest, line-intensity fluctuations become a biased tracer of the large-scale structure itself.  Additionally, we assume that halos show a universal, spherically symmetrical shape depending only on their mass. We review below the bias expansion and adapt it for the LIM context.

\subsection{The perturbative bias expansion} 
Halos are formed from small-scale perturbations on long-wavelength modes. For example, a background mode $\delta_{\rm b}$ makes it easier or harder for the small-scale perturbation to collapse. Additionally, it was found that the presence of large-scale tidal fields influences the formation of halos. Finally, since the gravitational potential $\Phi$ and the potential $\Theta$ of velocity perturbations deviate when evolving non-linearly, we have to separately track their contributions. The effective field theory bias expansion provides us with a complete set of operators while imposing Galilean and gauge invariance. 

The functional form of this dependence, known as the bias expansion, is
\begin{equation}
\begin{split}
    \delta_\mathrm{h}(\boldsymbol{x})& = b_1\,\delta(\boldsymbol{x}) \\
    & + \frac{b_2}{2}\,\delta^2(\boldsymbol{x})+b_{\mathcal{G}_2}\,\mathcal{G}_2\left[\Phi\right](\boldsymbol{x}) \\
    & +\frac{b_3}{6}\,\delta^3(\boldsymbol{x}) + b_{\delta\mathcal{G}_2}\,\delta(\boldsymbol{x})\,\mathcal{G}_2\left[\Phi\right](\boldsymbol{x}) + b_{\mathcal{G}_3}\, \mathcal{G}_3\left[\Phi\right](\boldsymbol{x}) + b_\Gamma\,\left[\mathcal{G}_2\left[\Phi\right](\boldsymbol{x}) - \mathcal{G}_2\left[\Theta\right](\boldsymbol{x})\right] \\
    &+\dots\:,
\end{split}
 \label{eq:simple_bias}
\end{equation}
where each line corresponds to an order in the expansion, all bias coefficients depend only on $M$, and the functionals $\mathcal{G}_2$ and $\mathcal{G}_3$ are called the second- and third-order Galileon.\footnote{The expressions for the Galilean operators and the SPT kernels that will appear below are shown in the appendix~\ref{sec:appendixA}. Several references use the operator $S_2$ instead of $\mathcal{G}_2$ to describe the impact of tidal fields in the bias expansion. Both approaches are equivalent and imply different benefits and drawbacks. Note, however, that the definition of the associated biases and, in particular, the spherically averaged second-order bias, changes for each case~\cite{2018PhR...733....1D}.} This formalism neglects the formation history of halos, which can influence the particular stochastic properties and clustering of halos and their containing galaxies. In the literature, this is discussed as the halo assembly bias~\cite{assb_10.1111/j.1365-2966.2006.11230.x, assb_PhysRevLett.116.041301}.

Additionally to the operators defined here, there are stochastic operators that contribute to Eq.~\eqref{eq:simple_bias}. Within the EFT framework, these operators encode small-scale, non-deterministic contributions that are not captured by the deterministic bias expansion. This gives rise to shot-noise-like contributions to the $N$-point correlation functions. Moreover, assuming perfect Poisson sampling and working at leading order, these additional terms reduce to the familiar Poisson shot-noise terms. In this model, the correspondence is exact for point-like tracers or, equivalently, in the large-scale limit $k\to0$. With this in mind, we treat the shot-noise terms in the next section using the Poisson point process interpretation.

The large-scale mode $\delta$ undergoes nonlinear gravitational evolution and deviates away from the linear cosmological perturbations $\delta_\mathrm{L}$. Solving the gravitational equations perturbatively, we can relate $\delta$ to the linear fields of matter perturbations, following the SPT approach. The $n$-th order correction of $\delta$ and $\theta$ is given by
\begin{align}
\delta^{(n)}(\boldsymbol{k}) &= \frac{1}{V^{n-1}}\,\sum_{\boldsymbol{q}_1,\dots,\boldsymbol{q}_n}\:\delta_{\boldsymbol{q}_1+\dots+\boldsymbol{q}_n,\boldsymbol{k}}\,F_{n}(\boldsymbol{q}_1,\dots,\boldsymbol{q}_n)\,\prod_{i=1}^n\:\delta_\mathrm{L}(\boldsymbol{q}_i)\:, \label{eq:matter_spt}\\
\theta^{(n)}(\boldsymbol{k}) &= \frac{1}{V^{n-1}}\,\sum_{\boldsymbol{q}_1,\dots,\boldsymbol{q}_n}\:\delta_{\boldsymbol{q}_1+\dots+\boldsymbol{q}_n,\boldsymbol{k}}\,G_{n}(\boldsymbol{q}_1,\dots,\boldsymbol{q}_n)\,\prod_{i=1}^n\:\delta_\mathrm{L}(\boldsymbol{q}_i)\:. \label{eq:velocity_spt}
\end{align}
Combining equations~\eqref{eq:simple_bias} and the expansions above, the mildly nonlinear halo-number count perturbations can be expressed as
\begin{align}
    \delta_\mathrm{h}(\boldsymbol{k}, M) &= \sum_n \delta_\mathrm{h}^{(n)}(\boldsymbol{k}, M)\:,\\
    \delta_\mathrm{h}^{(n)}(\boldsymbol{k}, M)&=\frac{1}{V^{n-1}}\,\sum_{\boldsymbol{q}_1,\dots,\boldsymbol{q}_n}\:\delta_{\boldsymbol{q}_1+\dots+\boldsymbol{q}_n,\boldsymbol{k}}\,K_{n}(\boldsymbol{q}_1,\dots,\boldsymbol{q}_n,M)\,\prod_{i=1}^n\:\delta_\mathrm{L}(\boldsymbol{q}_i)\:,
\end{align}
where $F_n$ and $G_n$ are the SPT mode coupling kernels and $K_n$ results from their combination with the bias expansion, as shown in appendix~\ref{sec:appendixA}. Therefore, the $n$-th order correction to the halo overdensity involves $n$ powers of the linear overdensity, which we know to be small. For a reasonable bias model, we can thus require that the higher-order corrections to the halo overdensities decrease in magnitude with each order, so that the corrections after truncating the expansion are expected to be small. Within this work, we will stay at the first non-vanishing order (tree level), which does not change the conclusions of our study.\footnote{Note that SPT loop corrections may not be perturbative in some cases \cite{spt_breakdown2014JCAP...01..010B}, requiring different expansions like Lagrangian perturbation theory \cite{Matsubara_LPT_PhysRevD.77.063530} or effective field theories \cite{Baumann_EFT_2012}.}

\subsection{LIM two- and four-point statistics}
\label{sec:linemodels}
Following our assumption that all the spectral line emission comes from within halos and the choice of using the halo model, the line model is determined by the halo mass-luminosity relation. We assume that any dependence beyond the halo mass is purely stochastic and leads to a scatter of the individual halo's luminosity from the mean. We model this scatter as a mass-independent mean-preserving\footnote{For section \ref{sec:sim} only, we will use a non-mean-preserving scatter to match the model used in the painting of the $N$-body simulation from Ref.~\cite{MoradinezhadDizgah:2021dei}.} log-normal scatter $\sigma_L$. Under these assumptions, the mean and higher moments of the conditional luminosity of halos of a given mass $M$ are given by
\begin{equation}
   \langle L_X^\alpha\rangle(M) = L^\alpha(M) \, e^{\frac{1}{2}\,\alpha\,(\alpha-1)\,\sigma_L^2}\:, \label{eq:hl_moments}
\end{equation}
where $\sigma_L$, usually given in dex base, is already transformed to a natural base.

$N$-point correlations within the halo model are composed of $N$ terms that involve correlations of contributions from one to $N$ different halos. For instance, for two-point statistics, we have the one-halo and the two-halo terms. This is true in general for any Poisson point process and in particular also for our LIM model. The nonlinear LIM power spectrum is given by:
\begin{align}
    P_\mathrm{\rm TT}(k) &= \left[\mathcal{I}_1^1(k)\right]^2\,P(k)+\mathcal{I}_0^2(k,k)\:,\label{eq:lim_treelevel_pk}
\end{align}
where we have defined
\begin{equation}
    \mathcal{I}_\beta^\alpha(k_1,\dots,k_\alpha)=C_\mathrm{LT}^\alpha\,e^{\frac{1}{2}\,\alpha\,(\alpha-1)\,\sigma_L^2}\int_0^\infty\:\left\langle\frac{\mathrm{d}n}{\mathrm{d}M}\right\rangle(M)\,b_\beta(M)\prod_{j=1}^\alpha\,L(M)\,\mathcal{U}(k_j,M)\,\mathrm{d}M\:.
    \label{eq:lim_integrals}
\end{equation}
Therefore, $\mathcal{I}_0^2$ corresponds to the one-halo term, which is equivalent to the standard shot noise power spectrum. In the expression above, $\mathcal{U}$ is the Fourier transform of the intensity profile of the halo (here assumed to follow the halo density profile), the index $\beta$ runs over all the bias coefficients of equation~\eqref{eq:simple_bias}, $dn/dM$ is the halo mass function, and the geometric conversion factor $C_{\rm LT}$ transforms luminosity per volume to observed line temperature, 
\begin{equation}
    C_{\rm LT} = \frac{c^3\,(1+z)^2}{8\,\pi\,k_\mathrm{B}\,\nu^3\,H(z)}\:,
\end{equation}
where $c$ and $k_\mathrm{B}$ are the speed of light and the Boltzmann constant, respectively, $H(z)$ is the Hubble parameter, and $\nu$ is the rest-frame frequency of the line of interest.\footnote{Note that the terms $\mathcal{I}_\beta^\alpha$ have units of temperature$^\alpha$ times volume$^{\alpha-1}$.}
Although line-dependent, the weighing of each halo with its mean luminosity typically picks out less massive halos with respect to the matter power spectrum, changing the overall shape of the small-scale power spectrum.

In addition to the power spectrum, we need to compute the LIM trispectrum. Following the same logic as for the power spectrum, the connected 4-point function is given by
\begin{align}
    \langle \delta_\mathrm{T}(\boldsymbol{k}_1)\delta_\mathrm{T}(\boldsymbol{k}_2)\delta_\mathrm{T}(\boldsymbol{k}_3)\delta_\mathrm{T}(\boldsymbol{k}_4)\rangle_c &= V\, T_{\rm T}(\boldsymbol{k}_1,\boldsymbol{k}_2,\boldsymbol{k}_3,\boldsymbol{k}_4)\,\delta_{\boldsymbol{k}_1+\boldsymbol{k}_2+\boldsymbol{k}_3+\boldsymbol{k}_4,0}\,, \nonumber \\
    \text{where} \quad T_{\rm T} &= T_{\rm T}^{\rm 1h}+T_{\rm T}^{\rm 2h}+T_{\rm T}^{\rm 3h}+T_{\rm T}^{\rm 4h}\:.
\end{align}
The four-halo term, in which all four points lie in distinct halos, is the only purely clustering contribution. The remaining terms contain shot-noise-like contributions due to the underlying Poisson point process. We can compute each of the contributions from the halo trispectrum by computing the mean over the halo population. The four-halo term of the halo trispectrum reads as
\begin{equation}
\begin{split}
    T^\mathrm{4h}_\mathrm{h}&(\boldsymbol{k}_1,\boldsymbol{k}_2,\boldsymbol{k}_3,\boldsymbol{k}_4,M_1,M_2,M_3,M_4) = \\
    &6\,K_1(M_1)\,K_1(M_2)\,K_1(M_3)\,K_3(-\boldsymbol{k}_1,-\boldsymbol{k}_2,-\boldsymbol{k}_3,M_4)\,P(k_1)\,P(k_2)\,P(k_3) + \text{perms.} \\
    &+ 4\,K_1(M_1)\,K_1(M_2)\,K_2(-\boldsymbol{k}_1,\boldsymbol{k}_1+\boldsymbol{k}_3,M_3)\,K_2(-\boldsymbol{k}_2,-\boldsymbol{k}_1-\boldsymbol{k}_3,M_4) \\
    &\qquad \times P(k_1)\,P(k_2)\,P(\|\boldsymbol{k}_1+\boldsymbol{k}_3\|) + \text{perms.} \\
    &+ 4\,K_1(M_1)\,K_1(M_2)\,K_2(-\boldsymbol{k}_2,\boldsymbol{k}_2+\boldsymbol{k}_3,M_3)\,K_2(-\boldsymbol{k}_1,-\boldsymbol{k}_2-\boldsymbol{k}_3,M_4) \\
    &\qquad \times P(k_1)\,P(k_2)\,P(\|\boldsymbol{k}_2+\boldsymbol{k}_3\|) + \text{perms.}\,,
\end{split}
\label{eq:trispectrum4h}
\end{equation}
where `perms' refers to permutations of the indices for the terms in each corresponding line. 

The three-halo term of the LIM trispectrum, is related to the bispectrum of halos. In this case, it is necessary to keep track of all distinct permutations that enter the final expression i.e., the indices for which there are two points of the trispectrum which belong to the same halo.
\begin{equation}
\begin{split}
    T_\mathrm{h}^\mathrm{3h}(\boldsymbol{k}_1,\boldsymbol{k}_2,&\boldsymbol{k}_3,\boldsymbol{k}_4,M_1,M_2,M_3,M_4) =\\
    &\left[2\,K_1(M_1)\,K_1(M_2)\,K_2(-\boldsymbol{k}_1,-\boldsymbol{k}_2, M_3)\,P(k_1)\,P(k_2) \:+\text{perms.}\right]\:+\text{perms.}\:,
\end{split}
\end{equation}
where the internal permutations are among the three wave vectors $\boldsymbol{k}_1$,$\boldsymbol{k}_2$, and $\boldsymbol{k}_3+\boldsymbol{k}_4$, and the external permutations are among pairings of two of the four vectors for the sum of the third wave-vector argument. For the four-halo and three-halo terms after averaging over the halo populations, the biases in the mode coupling kernels $K_n$ do not factor out like in the power spectrum case. Thus the resulting functions are sums over different $\mathcal{I}_\beta^\alpha$. 

Finally, the one- and two- halo terms are given by.
\begin{align}
    T_\mathrm{T}^\mathrm{1h}(\boldsymbol{k}_1,\boldsymbol{k}_2,\boldsymbol{k}_3,\boldsymbol{k}_4) =& \mathcal{I}^4_0(k_1,k_2,k_3,k_4)\:,\\
    T_\mathrm{T}^\mathrm{2h}(\boldsymbol{k}_1,\boldsymbol{k}_2,\boldsymbol{k}_3,\boldsymbol{k}_4) =& \mathcal{I}_1^1(k_1)\,\mathcal{I}_1^3(k_2,k_3,k_4)\,P(k_1)\:+\text{perms.}\nonumber\\
    +&\mathcal{I}_1^2(k_1,k_2)\,\mathcal{I}_1^2(k_3,k_4)\,P(\|\boldsymbol{k}_1+\boldsymbol{k}_2\|)\:+\text{perms.}\:.
\end{align}
To compute the non-Gaussian in-box covariance, we need the shell-averaged collapsed trispectrum. Then, many of the terms in the expressions above cancel out, leaving us with simpler expressions.
This is further discussed in appendix \ref{sec:trispectrum_terms}.

Similarly, taking the limit of equation~\eqref{eq:large-scale-expansion}, we obtain an expression for the response function given by
\begin{equation}
   \frac{\mathrm{d}P_\mathrm{TT}}{\mathrm{d}\delta_\mathrm{b}}(k)= \left[\frac{68}{21}\,-\frac{1}{3}\frac{\mathrm{dlog}k^3\,P(k)}{\mathrm{dlog}k}\right]\,\left[\mathcal{I}_1^1(k)\right]^2\,P(k)
    +2\,\mathcal{I}_1^1(k)\,\left[\mathcal{I}^1_2(k)-\frac{4}{3}\,\mathcal{I}^1_{\mathcal{G}_2}(k)\right]\,P(k)
    +\mathcal{I}_1^2(k,k)\:.\label{eq:halomodelresponse}
\end{equation}
The first bracket in the equation encapsulates the linear response of the matter overdensities to the large-scale mode. The first term is the so-called growth-only response, i.e. the response of the matter growth rate of the small-scale overdensities, while the second term of the bracket can be understood as the response of the local expansion rate to the background mode. The second term arises because we compute the response of the clustering of a biased tracer of the matter field.\footnote{The factor $b_2-4b_{\mathcal{G}_2}/3$ is the local quadratic bias parameter in the spherical approximation~\cite{Eggemeier:2018qae}. The $b_{\mathcal{G}_2}$ term does not appear when $S_2$ is used instead of $\mathcal{G}_2$ for the tidal-field operator, since in that case its contribution to the quadratic local bias is proportional to the second-order Legendre polynomial.} Finally, the last term of equation \eqref{eq:halomodelresponse} is the response of the local tracer number density to a background mode, or equivalently, the shot noise response. The appearance of nonlinear biases is natural if we recall that, in the peak-background-split formalism, the local-in-density biases themselves are the response of the tracer number density to a background mode \cite{nlbiases_2016JCAP...02..018L},
\begin{equation}
    \left\langle\frac{\mathrm{d}n}{\mathrm{d}M}\right\rangle\,b_i\,=\frac{\mathrm{d}}{\mathrm{d}\delta_\mathrm{b}}\left[\left\langle\frac{\mathrm{d}n}{\mathrm{d}M}\right\rangle b_{i-1}\right]\:. \label{eq:biasderivs}
\end{equation}

It is important to note that we include only the response of the LIM power spectrum to the background density mode through the halo power spectrum. We assume that any response of the $L(M)$ relation to the background mode, for example, due to environmental effects, is negligible and effectively absorbed into the stochastic scatter. For a discussion about disentangling the effects of cosmic variance and galaxy formation from power spectrum uncertainties in the context of galaxy clustering, see Ref.~\cite{Sinigaglia:2026pqe}.

\subsection{Validity of the model}
\label{sec:limitations}
Let us discuss the theoretical limitations of this model. The model can be considered valid as long as the perturbative expansion converges, i.e., higher-order terms are smaller than the lower-order terms. This expansion is equivalent to equation~\eqref{eq:simple_bias} but for temperature perturbations, i.e., using the luminosity-weighted biases, defined as the large-scale limit of the $\mathcal{I}$ (i.e., where $\mathcal{U}(k)=1$), normalised to the mean brightness temperature, as
\begin{equation}
    b_\beta=\frac{\mathcal{I}^1_\beta(0)}{\mathcal{I}^1_0(0)}\:.
\end{equation}

To allow for a quantitative discussion, we need to assume a line and model: we consider the [CII] line, with a luminosity parametrised by the fit from Ref.~\cite{silva2015ApJ...806..209S}
\begin{equation}
    \log \frac{L_{[\mathrm{CII}]}}{L_\odot} = a\,\log \frac{\mathrm{SFR}}{M_\odot\,\mathrm{yr}^{-1}}+b\:,
    \label{eq:cii}
\end{equation}
where $a$ and $b$ are taken from Ref.~\cite{MoradinezhadDizgah:2021dei} and the star-formation rate is obtained from  Refs.~\cite{sfr2013ApJ...762L..31B, sfr2013ApJ...770...57B}. We use the Sheth-Mo-Tormen model \cite{2001MNRAS.323....1S} for the halo mass function and linear bias. Other local biases $b_2(M)$ and $b_3(M)$ can be computed through higher derivatives of \eqref{eq:simple_bias} with respect to the background mode (see equation \eqref{eq:biasderivs}). Instead, we take higher-order local biases from fitting formulas from simulations, relating them to $b_1$~\cite{nlbiases_2016JCAP...02..018L}. To compute the biases related to the tidal terms, we assume halo co-evolution relations and take the expressions presented in Ref.~\cite{coevohalo2018JCAP...07..029A}. For the halo shape, we assume the Navarro--Frenk--White profile \cite{1996ApJ...462..563N}, and the mass--concentration relation from Ref.~\cite{conc2019ApJ...871..168D}.
\begin{figure}
    \centering
    \includegraphics[width=\linewidth]{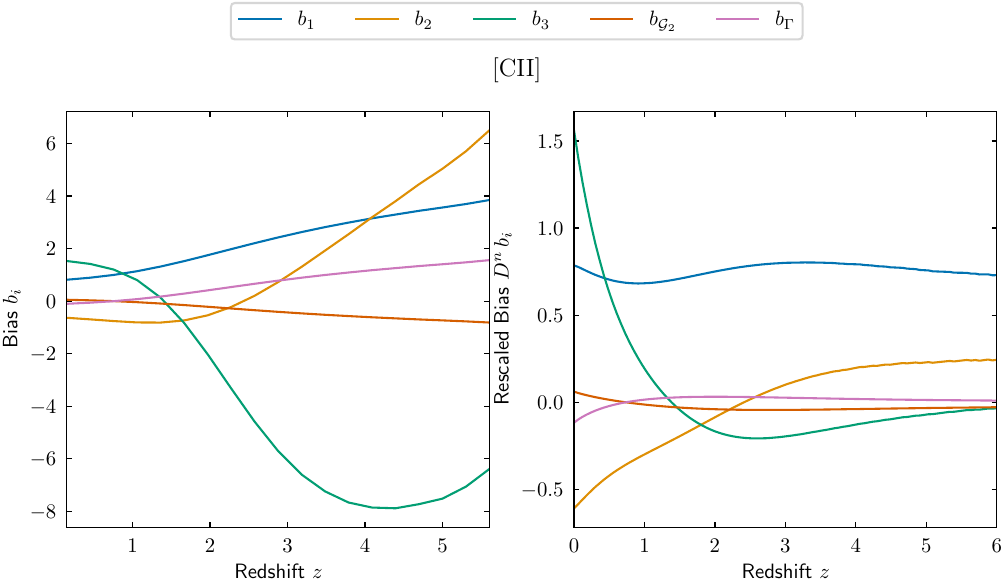}
    \caption{\emph{Left}: The redshift evolution of different line biases. We look at the first three Eulerian local matter biases, the tidal field bias and the nonlinear evolution of the tidal field. \emph{Right}: The same as the figure on the left, but we rescale an $n$-th order bias with $n$ powers of the growth factor.}
    \label{fig:bias_evolution}
\end{figure}

Contrary to the matter perturbations, the bias factors are known to grow with increasing redshift. Hence, we study in figure~\ref{fig:bias_evolution} the redshift evolution of the mean line biases to evaluate the convergence of our perturbative expansion. We see that the local bias $b_1$ grows slowly with redshift, while $b_2$ and $b_3$, after an initial slow period, grow much faster. In any case, what enters the perturbative expansion is the halo bias times a perturbation elevated to a power of the same order. Then, we also show the bias factors times the growth factor elevated to the corresponding power to show the amplitude of the actual term in the expansion. In this case, we see that the growth of the linear and quadratic bias $b_1$, $b_2$ is roughly compensated by the linear growth function $D$. This means that the resulting tree-level power spectrum stays roughly constant in amplitude. For the cubic bias $b_3$, $D^3$ decays much faster, such that the combination tends quickly to zero. The two tidal biases are comparatively small and show a relative growth roughly at the speed of $b_1$. This is expected from the halo--coevolution relations and does not change much from the matter-only case. After rescaling with $D^2$ and $D^3$, respectively, they are also quickly tending to zero.

Figure 9 in Ref.~\cite{MoradinezhadDizgah:2021dei} shows that the relevant sources of nonlinearities depend on the redshift and physical scale. At low redshifts, nonlinear matter clustering dominates the corrections to the power spectrum; conversely, at higher redshift, the physical scale at which matter becomes nonlinear is shifted to smaller scales due to the milder growth of structure, so that nonlinear matter clustering affects a more limited range of scales. In this regime, however, the tracer field is highly non-Gaussian, and nonlinear bias contributions become essential for accurate predictions of the tracer power spectrum.
In LIM, however, these trends are additionally modulated by the specifics of the line model. Thus, quantitative assessments depend significantly on the line-dependent halo-luminosity connection and its underlying halo-mass distribution.

\section{Validation with simulations}
\label{sec:sim}
In this section, we validate our theoretical prediction of the LIM power spectrum SSC from the section above using a numerical estimation from painted simulations.
Rather than using simulated lightcones (see e.g., Refs.~\cite{Sato-Polito:2022wiq, Bethermin:2022lmd}), we use the {\tt Mithra LIMSims}~\cite{MoradinezhadDizgah:2021dei}, built from the halo catalogs of the \texttt{Hidden Valley}\footnote{\hyperlink{https://cyril.astro.berkeley.edu/HiddenValley}{https://cyril.astro.berkeley.edu/HiddenValley}}~\cite{Modi:2019ewx} simulation suite. The \texttt{Hidden Valley} simulations are ($1024\, h^{-1}$ Mpc)$^3$-size dark-matter-only FastPM\footnote{\hyperlink{https://github.com/fastpm/fastpm}{https://github.com/fastpm/fastpm}}~\cite{Feng:2016yqz} simulations with a mass resolution of $8.57\times 10^7\,h^{-1}M_\odot$, which allows to resolve low-mass halos, as those usually bright in the line emission of interest. The \texttt{Mithra LIMSims} painted different CO rotational lines and the [CII] fine structure line on several redshift snapshots, using cloud-in-cell mass assignment to a regular grid of 1 $(h^{-1}{\rm Mpc})^3$ cubic cells. Finally, these simulations do not include any experimental resolution limit or noise. For this work, we focus on the real-space (i.e, without redshift-space distortions) [CII] simulation at $z=1$. 

To properly model the details of the painting and resampling routine employed by the simulations, we have to slightly tweak our halo model. Firstly, while the {\tt Mithra LIMSims} did use the same star-formation rate fitting formula from Ref.~\cite{sfr2013ApJ...762L..31B, sfr2013ApJ...770...57B} and the same mass--luminosity relation from Ref.~\cite{silva2015ApJ...806..209S}, they modelled the scatter with a non-mean-preserving single-parameter log-normal scatter. This slightly tweaks our Equation \ref{eq:hl_moments} to become:
\begin{equation}
    \langle L_X^\alpha\rangle^\mathrm{sims} = L^\alpha(M)\,e^{\frac{1}{2}\alpha^2\,\sigma^2}\:.
\end{equation}
Additionally, this painting technique does not resolve the shape of the dark matter halos. Consequently, the Fourier-transformed profile $\mathcal{U}=1$, which propagates into the definitions of the $\mathcal{I}^\alpha_\beta$, which become simply $\mathcal{I}_\beta^\alpha\to\langle T^\alpha\,b_\beta\rangle$. The cloud-in-cell resampling additionally introduces a suppression of the power spectrum at the smallest scales. To account for this, we multiply the power spectrum by an angle-averaged sinc function~\cite{Jing_resampling_2005ApJ...620..559J}, i.e., the window function for cubic voxels.

In order to estimate the SSC, we need to embed our survey volume as a subvolume of the \texttt{Mithra LIMSims}. Specifically, we can estimate the supersample covariance empirically using the `field-to-field variance', as it is referred to in Ref.~\cite{G20}. In practice, we subdivide the simulation snapshot into $8^3=512$ equally sized subvolumes. Within each subvolume, we measure the spherically averaged power spectrum in evenly spaced log $k$ bins and the mean line brightness temperature, which will be used as a proxy to calculate the background matter density. This is because we do not have access to the matter distribution on the simulation box. Therefore, to extract the background mode, we first compute the difference between the subvolume mean brightness temperature $\left\langle T \right\rangle^\mathrm{sub}$ and the whole-simulation brightness temperature $\langle T\rangle$. By using Equation \eqref{eq:biasderivs}, this quantity is related to the background mode through:
\begin{align}
    \delta T_\mathrm{b} = \left\langle T \right\rangle^\mathrm{sub} - \left\langle T \right\rangle
    =C_\mathrm{LT}\,\,e^{\frac{1}{2}\,\sigma^2}\,\int\,L(M)\,b_1(M)\,\left\langle\frac{\mathrm{d}n}{\mathrm{d}M}\right\rangle\,\mathrm{d}M\,\delta_\mathrm{b} = \left\langle T\,b_1\right\rangle \,\delta_\mathrm{b}\:.
    \label{eq:deltas_relation}
\end{align}
We use this expression to assign a given $\delta_{\rm b}$ to each subvolume, depending on the measured $\delta T_{\rm b}$, and perform a linear regression of $P_{\rm TT}$ as a function of $\delta_{\rm b}$ for each wave number bin. From this, we can empirically obtain the response function.

\begin{figure}
    \includegraphics[width=\linewidth]{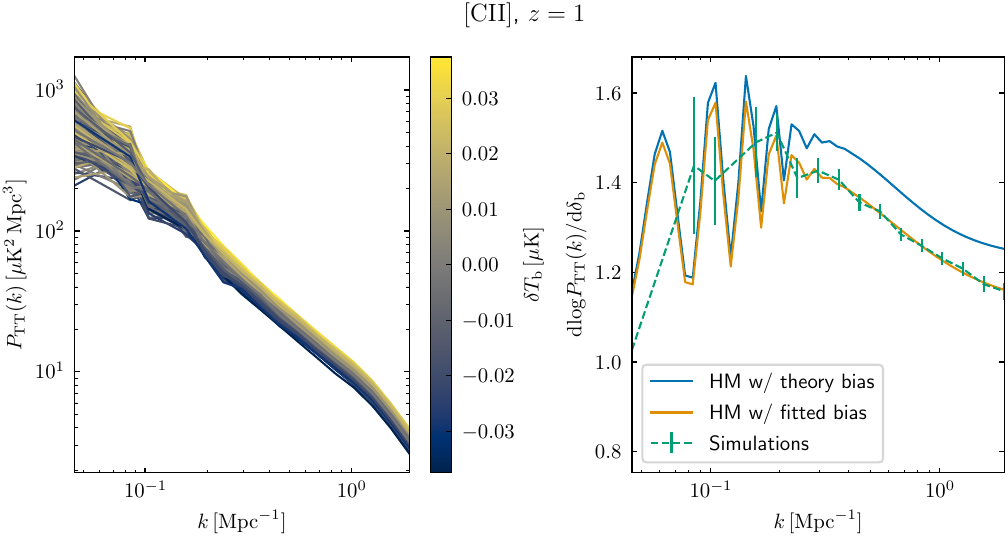}
    \caption{Measurements of the power spectrum response from simulations for the [CII] case at redshift 1. \emph{Left}: The power spectrum for each subvolume. Each curve is colour-coded according to the amplitude of the corresponding background mode temperature. \emph{Right}: The theory predictions for the power spectrum response function. We show the response function computed with biases obtained directly from our halo model in blue, and the response where biases are fitted to the simulations in yellow. The measurements of the response function from individual subboxes are presented in dashed green, with the errors estimated directly from the regression.}
    \label{fig:response_measurement}
\end{figure}

We show the [CII] power spectrum for each subvolume, colour coded according to the mean [CII] brightness temperature, and the comparison between the predicted and the measured response functions in figure~\ref{fig:response_measurement}. The power spectrum shows a remarkable correlation with the background mode. We can already distinguish by eye two regimes in the dependence on $\delta T_{\rm b}$. At smaller scales, we find a coherent trend of lower amplitudes corresponding to lower values of $\delta T_{\rm b}$. This is consistent with expectations, given the factor $(\mathcal{I}_1^1)^2$ in the clustering term of the power spectrum. At larger scales, in turn, the dependence is less defined, and we find a random scatter due to noisier measurement values caused by cosmic variance. At very small scales, where the shot-noise (or one-halo term) contribution should dominate, we find the suppression of the power spectrum due to the cloud-in-cell resampling. 

On the right-hand plot of Fig.~\ref{fig:response_measurement}, we present the response function. Here we distinguish between three different cases: the measurement from the simulations, the full theoretical prediction exposed above, and an intermediate case which consists of the theoretical prediction but fitting the bias parameters involved in the response function to match the response measured from the simulation. We see an approximate agreement between our full theoretical prediction and the measurement, especially on large scales. At small scales, the theory-only prediction overestimates the response but matches the overall shape.

Our model has freedom to vary the bias parameters that appear in our expression \eqref{eq:halomodelresponse}, which we determine by fitting to the response function and power spectrum. In practice this means that $\left\langle T\,b_1\right\rangle$ and $\left\langle T^2 \right\rangle$ are fitted to the power spectrum, and $\langle T\,(b_2-\frac{4}{3}\,b_{\mathcal{G}_2})\rangle$ and $\langle T^2\,b_1\rangle$ to the response function. Note that our model for the power spectrum is linear, for which the halo model is known to underpredict the result on intermediate scales. However, it was shown in Ref.~\cite{MoradinezhadDizgah:2021dei} that by adding loop correction terms to the power spectrum, the intermediate scales can be brought into concordance. We then follow modern halo model codes~\cite{2021MNRAS.502.1401M} and introduce a shape parameter $\eta$ modifying the transition region between the two- and the one-halo terms, leaving
\begin{equation}
    P_{\rm TT}^{\rm fitted} = \left[\left(\left\langle T\,b_1\right\rangle^2\,P\right)^\eta+ \left\langle T^2\right\rangle^\eta\right]^{1 / \eta}\:.
\end{equation}
Assuming that $\eta$ does not depend on the background modes, we can find the power spectrum response function after application of the chain rule. We present the fitted values and their respective theoretical prediction in Table \ref{tab:fit} 
They are all 5\%--15\% smaller than their respective theoretically predicted value. Comparing the fitted linear bias found in \cite{MoradinezhadDizgah:2021dei}, we find agreement within 2\%. The small mismatch in the power spectrum is expected, as the bias model for the power spectrum used in Ref.~\cite{MoradinezhadDizgah:2021dei} considers non-linear biases through the addition of loop terms. Furthermore, the mismatch between the theoretical prediction and the fitted values, also found in Ref.~\cite{MoradinezhadDizgah:2021dei}, is due to a mismatch in the predicted and measured halo mass function, as pointed out in that work.

\begin{table}[t]
    \centering
    \begin{tabular}{|c c|c c|}
    \hline
        \multicolumn{2}{|c|}{Parameter} & Prediction &  Best-fit value\\
    \hline
     $\langle T\,b_1\rangle$ & $[\mu\mathrm{K}]$ & $\phantom{-}0.27$& $\phantom{-}0.23$\\
     $\langle T^2\rangle $ & $[\mu\mathrm{K}^2\,\mathrm{Mpc}^3]$ & $\phantom{-}11.1$ & $\phantom{-}11.0$ \\
     $\left\langle T \left(b_2 -\frac{4}{3}b_{\mathcal{G}_2}\right)\right\rangle$ & $[\mu\mathrm{K}]$& $-0.18$& $-0.16$\\
     $\langle T^2 b_1\rangle$ & $[\mu\mathrm{K}^2\,\mathrm{Mpc}^3]$ & $\phantom{-}13.7$ & $\phantom{-}12.3$\\
     $\eta$& & --- & 0.76\\
    \hline
    \end{tabular}
    \caption{The theoretical predictions of parameter values describing the power spectrum and the response function of [CII] fluctuations at $z=1$ compared with the best-fit values to the measurements from \texttt{Mithra LIMSims}.}
    \label{tab:fit}
\end{table}%
By fitting the free bias parameters directly to the simulation, we can achieve excellent agreement. Adjusting the biases, we find a $\chi^2 / n_\mathrm{dof} = 0.86$ (without accounting for correlations between the different $k$ bins), which suggests a good estimate for the errors from simulations. The agreement of the shape and amplitude of the response function showcases that the extended halo model approach is well-suited to predict the response. Varying the values of the bias parameters should not be seen as fine-tuning in this case, as the bias parameters are a priori unknown nuisance parameters. Studies exploring simulation-based priors for H$_{\rm I}$ bias parameters can be found in e.g., Ref.~\cite{Sarkar:2026rfv}.

Furthermore, the intrinsic scatter of the sub-volume power spectra, stemming from the underlying cosmic variance and the in-box non-Gaussian covariance, propagates into uncertainties for our response function measurement. As a sanity check regarding the precision of our validation, we compare our theory prediction for $\langle Tb_1\rangle$ with the reported result in Ref.~\cite{MoradinezhadDizgah:2021dei}. We find our prediction returns a value 13\% larger than the simulation measurement, which we consider to be in agreement within a reasonable range given the slight differences in assumptions and the limitations of our theory discussed at the beginning of this section, and in section~\ref{sec:limitations}. 

After validating our estimation of the response function, we need to validate the estimation of the SSC. By computing the sample covariance of the different subvolume power spectra, we can estimate the full covariance from the simulation, including the super-sample contributions. We estimate the covariance as
\begin{equation}
    {\mathrm{Cov}}^\mathrm{subbox} = \frac{1}{N_\mathrm{sub}-1}\,\sum_{\mathrm{subbox}\:i}^{N_\mathrm{sub}}\left(P_\mathrm{TT}^i-\langle P_\mathrm{TT}\rangle\right)\otimes\left(P_\mathrm{TT}^i-\langle P_\mathrm{TT}\rangle\right)\:,
\end{equation}
where $N_{\rm sub}$ is the number of subdivisions and $\otimes$ denotes the outer product. Since we want to validate our estimation of the SSC, we need to isolate this contribution, i.e., remove the in-box covariance from the total estimation. We can do this by computing the covariance of each small subvolume. The ``small box" covariance can be estimated from different methods. For instance, Ref.~\cite{Li:2014sga} used multiple smaller simulations to estimate this covariance empirically. This is beyond our reach for this work, as the \texttt{Mithra LIMSims} only include one realisation.\\

In turn, we will use a linear bootstrap approximation, called the jackknife method, to estimate the in-box covariance \cite{Escoffier:2016qnf, Euclid:2025fby, Norberg:2008tg, Mohammad:2021aqc}. In this method, the entire survey volume is partitioned into equally-sized non-overlapping regions. Then, inside the individual regions, the survey functions are set to zero one by one for each different jackknife realisation, and the power spectrum is computed on this perturbed, or leave-one-out, survey. Finally, the covariance can be computed from the sample covariance of these power spectra as
\begin{equation}
    \boldsymbol{\mathrm{Cov}}^\mathrm{smallbox} = \frac{N_\mathrm{JK}-1}{N_\mathrm{JK}}\,\sum_{\mathrm{partition}\:i}^{N_\mathrm{JK}}\left(P_\mathrm{TT}^i-\langle P_\mathrm{TT}\rangle_\mathrm{JK}\right)\otimes\left(P_\mathrm{TT}^i-\langle P_\mathrm{TT}\rangle_\mathrm{JK}\right)\:,
\end{equation}
where the means are computed over all $N_{\rm JK}$ jackknife realisations of a single sub-volume.

There are two different ways to apply the expression above to compute the covariance of a small box. On the one hand, we need to estimate the in-box covariance for a volume sitting at $\delta_{\rm b}=0$. Noting also that the covariance estimate for a single small-box covariance is too noisy given the small volume, it is possible to compute the jackknife covariance from the full simulation volume and obtain the small-box covariance by rescaling it according to the expected $V^{-1} $scaling. However, this approach does not accurately reproduce the large-scale Gaussian covariance, since the window function of the full volume differs from that of the subvolume. Moreover, the number of Fourier modes on large scales only approximately follows the expected $V$ scaling due to discretisation effects. Instead, we generate 64 jackknife realisations for each subvolume of the simulation to compute the `small-box' covariance for each of them and combine them later to reduce the statistical noise of our estimate. Note that the mean of all subvolumes is the cosmic mean; combining the jackknife estimates from all subvolumes is roughly equivalent to estimating the in-box covariance for a survey of the size requested but with $\delta_{\rm b}=0$, i.e., sitting on the cosmic mean.

However, we find that the leave-one-out estimator introduces an additional SSC-like correlation. Physically, this effect is closely related to SSC: removing small subvolumes from the survey effectively changes the mean of the realisation with respect to the mean of the survey, which reintroduces their contribution as an unphysical super-survey mode. Before computing the mean of the individual covariance estimates, we correct for this effect:
\begin{equation}
    \boldsymbol{\mathrm{Cov}}^\mathrm{smallbox\:i}_\mathrm{corr} = \boldsymbol{\mathrm{Cov}}^\mathrm{smallbox}_i - (N_\mathrm{JK}-1)\,\mathrm{Var}(\delta_\mathrm{b}^{ij}-\delta{}_\mathrm{b}^{i})\,\frac{\mathrm{d}P_\mathrm{TT}}{\mathrm{d}\delta_\mathrm{b}}\otimes\frac{\mathrm{d}P_\mathrm{TT}}{\mathrm{d}\delta_\mathrm{b}}\:,
\end{equation}
where $\delta_\mathrm{b}^{i}$ is the background mode in the subvolume $i$, and $j$ denotes the jackknife realisation of said subbox. The variance is performed over the individual jackknife realisations of the same subvolume. To get a noise-reduced jackknife estimate, we then take the mean of $\boldsymbol{\mathrm{Cov}}^\mathrm{smallbox\:i}_\mathrm{corr}$ as motivated above. As this correction is not directly related to the problem at hand, we defer a systematic study of this bias in the standard jackknife covariance estimation for future work.
\begin{figure}
    \centering
    \includegraphics[width=\linewidth]{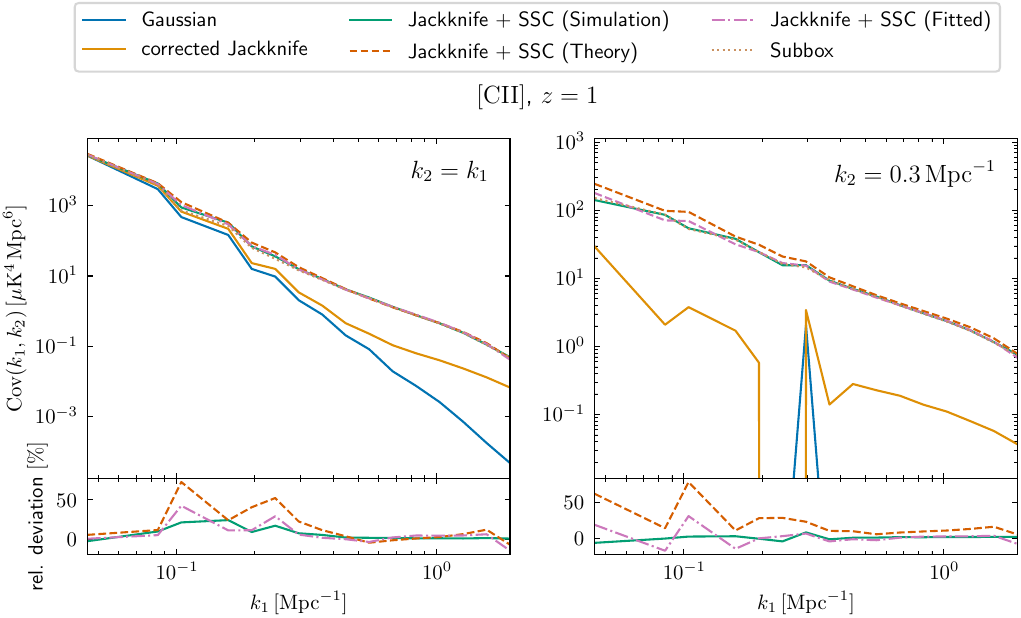}
    \caption{A comparison of the different covariance contributions. We show the Gaussian and the corrected jackknife covariance (Gaussian + non-Gaussian in-box), which should have no influence from super survey modes. Additionally, we show the across-subbox covariance (all contributions) and compare it to the corrected jackknife plus the SSC. For the SSC contribution, we show cases with either the response function measured from simulations or computed within our theoretical framework (using either the predicted or fitted bias parameters). The upper panels show the covariance values for different wavenumbers, while the lower panels show the per-cent relative difference between the covariances and the subbox covariance. \emph{Left}: We show the diagonal of the covariances. \emph{Right}: We show the covariances for a slice with a fixed wavenumber $k_2=0.3\,\mathrm{Mpc}^{-1}$}
    \label{fig:covarinace_sims}
\end{figure}

The comparisons between the different covariance contributions are presented in figure~\ref{fig:covarinace_sims}. We compare the cross subbox covariance, as an empirical estimation of the full covariance including SSC, the in-box jackknife-estimated covariance, and the sum of the in-box jackknife estimation and our prediction for the SSC. The in-box jackknife estimate serves as an estimation of the Gaussian covariance $+$ the in-box non-Gaussian covariance. The SSC prediction is computed using either the expression for the response function theoretically derived in the previous sections with the respective theoretically predicted bias values, or using the measured power spectrum response from the simulations, which is basically identical to the theoretical prediction with fitted bias parameters.

Let us focus first on the left plot, which shows the diagonal of the covariance. Here we find that all estimates of the covariances match at large scales. This is because the Gaussian contribution dominates the large-scale regime. At smaller scales, $k\gtrsim0.1\,{\rm Mpc}^{-1}$, the jackknife estimate deviates from the empirical subbox covariance, hinting at a sizable contribution from the SSC. For even smaller scales $\gtrsim 0.4\,\mathrm{Mpc}^{-1}$, the Gaussian covariance and the corrected jackknife covariance deviate.

We can identify this scale as the scale where the non-Gaussian covariance starts to contribute significantly. We can also see that by adding the SSC to the corrected in-box jackknife estimate, we are able to recover the subbox covariance well. In the right panel, we show a slice from the covariance where we have fixed one wavenumber to  $k_2=0.3\,\mathrm{Mpc}^{-1}$ and vary $k_1$ freely. As the Gaussian covariance is diagonal, its contribution only appears for $k_1=k_2$. For this case, the jackknife estimate severely underestimates the covariance, as we see that for this specific case, the SSC clearly dominates the off-diagonal contributions. However, this statement depends a lot on the survey specification, and we explore different scenarios analytically in the next section.

In general, we find that our prediction for the SSC is accurate. We can reproduce the diagonal of the covariance to a $\lesssim$30\% (50\%) level when using the fitted (fixed to theoretical values) bias parameters. When we use the response function and power spectrum measured from the simulations, the agreement is within $\lesssim$ 25\%. Note that the precision is better than these numbers for most scales, and only in 2 $k$ bins at intermediate scales do we reach the numbers cited above; instead, at smaller scales we find per cent-level precision. We find similar trends for the off-diagonal covariance. These results match the intuition from Fig.~\ref{fig:response_measurement}.

The origin of the intermediate-scale discrepancy is likely the limitations of the halo model, on which our theoretical model is based. In particular, the halo model is known to underestimate clustering on these scales (see e.g., Ref.~\cite{MoradinezhadDizgah:2021dei} for an example for LIM). Accurate predictions of these scales need proper modelling of intra-filament point correlations that are not caught by the standard Poisson point process description \cite{Ibanez:2026kuk}. This deficiency propagates directly into our SSC estimate. We expect that loop corrections to the power spectrum in the theoretical derivation of the response function can correct this small deviation. Nevertheless, the overall level of agreement validates our general derivation of the LIM SSC. 

\section{Covariance estimation for different lines and surveys}
\label{sec:forecast}

\noindent Having validated our covariance model against painted [CII] simulations, we now apply the framework more broadly. Both the response function and the relative importance of each covariance contribution depend strongly on the line and on the survey, so we consider both [CII] and the 21 cm line of neutral hydrogen. We first compare our predicted scaling relations with the field-to-field variance measured in the SIDES-Uchuu simulations in Ref. \cite{G20}, then examine how the contributions scale with survey volume and geometry, and finally quantify them for configurations representative of MeerKLASS~\cite{Cunnington:2025sdr} and CCAT-PRIME~\cite{CCAT-p_specs}.

\subsection{Response functions for [CII] and 21cm intensity mapping}
\begin{figure}
    \centering
    \includegraphics[width=0.9\linewidth]{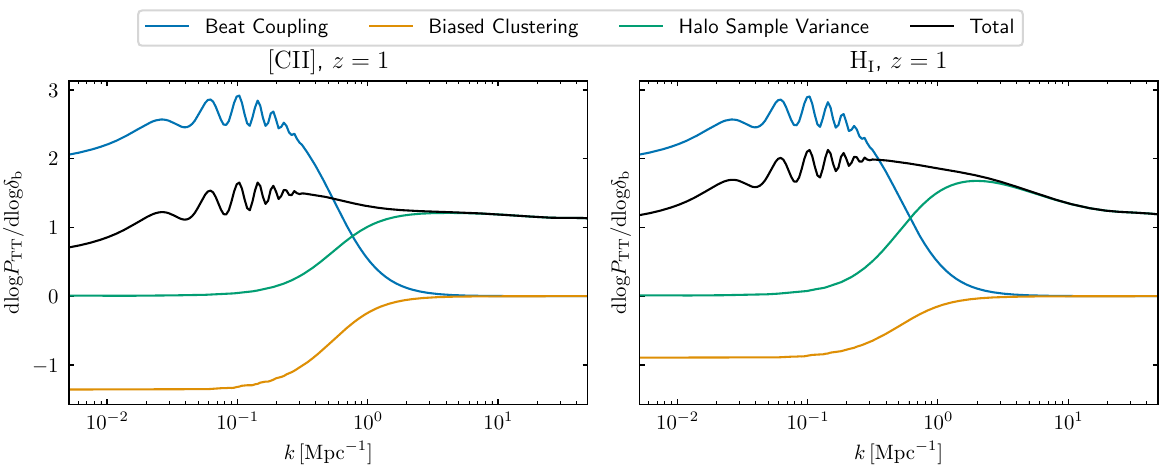}
    \caption{Comparison of logarithmic response function for [CII] on the left and H$_{\rm I}$ on the right. The different colours correspond to different contributions to the total response function.}
    \label{fig:comparasion_response_lines}
\end{figure}

\noindent We model the [CII] luminosity-halo mass relation with equation~\eqref{eq:cii}.
A commonly used parametrisation for H$_{\rm I}$ emission is in the form of a single power law with an exponential cut-off~\cite{HI_powerlaw...05..004O}:
\begin{equation}
    L_\mathrm{H_I}(M) = L_0\,\left(\frac{M}{M_{\rm min}}\right)^{\alpha_\mathrm{H_I}}\,\exp\left(-\frac{M_{\rm min}}{M}\right)\:.
\end{equation}
The parameters $L_0$, $M_{\rm min}$ and ${\alpha_\mathrm{H_I}}$ are free modelling parameters. For these, we choose $L_0=30.6\, L_\odot$, $M_{\rm min} = 1.86\times10^{11}\,h^{-1}\,M_\odot$, and ${\alpha_\mathrm{H_I}}=0.44$ taken from Ref.~\cite{Obuljen:2018kdy}. Additionally, we add a mean-preserving log-normal scatter of $\sigma_L=0.2$ dex. The resulting luminosity function weights the halo distribution differently from the [CII] case, putting more weight on higher mass halos. The comparison is presented in figure~\ref{fig:comparasion_response_lines}.  

The different mass--luminosity relations lead to different weighing of the contributions through the different biases, as well as their shape through the halo profile and relative amplitude of the one-halo and two-halo terms.
Comparing the biased clustering contributions, the ratio of second-order bias and the linear bias is larger for H$_{\rm I}$ than for $\mathrm{[CII]}$, leading to an amplification of the response function on large scales. On the other hand, the linear bias is also amplified for H$_{\rm I}$, leading to an amplification of the one-halo term response as well as shifting the scale at which this term becomes important to larger scales. Conversely, the beat coupling term is universal for all tracers of the matter power spectrum and does not change between [CII] and H$_{\rm I}$. This shows that even though the underlying halo field is the same, the different weighing of halos through the mass-luminosity relation can potentially change the response by a factor of two.
\begin{figure}
    \centering
    \includegraphics[width=0.9\linewidth]{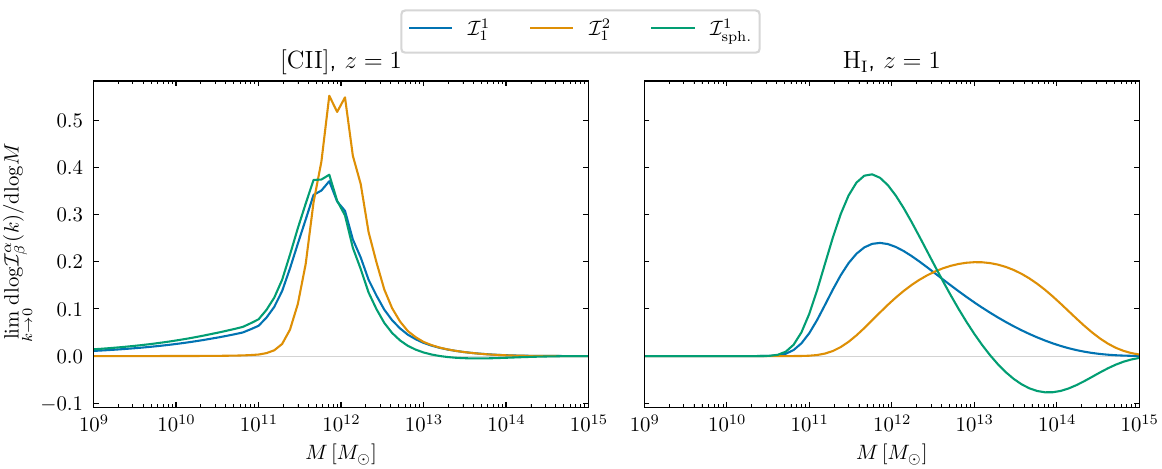}
    \caption{The differential large-scale limit of the LIM integrals featuring in the response function calculation for [CII] (left) and H$_{\rm I}$ (right).}
    \label{fig:difhalomassvsresponse}
\end{figure}
Figure \ref{fig:difhalomassvsresponse} shows which halo mass range contributes the most to each of the terms of the response function for different lines. The quantity $\mathcal{I}_1^1$ enters the beat coupling and linear biasing term, while $\mathcal{I}_\mathrm{sph.}^1$ is a shorthand notation for $\mathcal{I}_2^1-\frac{4}{3}\,\mathcal{I}_{\mathcal{G}_2}^1$, which enters the linear biasing term; finally $\mathcal{I}_1^2$ is the halo sample variance. We see that within our model, both lines get their main bulk of signal from halos of mass around $5\times10^{11}\,M_\odot$, but the distribution goes to much higher masses for ${\rm H_I}$, gaining sizeable contributions even for halos of mass $10^{14}\,M_\odot$. On the other hand, the tail of the contributions for ${\rm [CII]}$ extends to lower halo masses. All in all, the main contributing masses are more peaked for ${\rm [CII]}$. This showcases how the two different lines trace halos with different masses and therefore respond slightly differently to variations in the background density perturbation. \\

\subsection{Field-to-field variance and the SSC: scaling relations} 
\noindent Reference \cite{G20} included a study of the variations in the measured angular Fourier power spectra in small patches of their simulations, reporting a large so-called {\it field-to-field variance}. The authors provide a fit of the uncertainty related to this variation relative to the mean of the measurements as a function of the observed frequency $\nu_\mathrm{obs}$, the frequency range that fluctuations are projected over $\Delta \nu^{\rm proj}$, and the area of the survey $\Omega_\mathrm{field}$, as
\begin{equation}
    \left(\frac{\sigma}{\mu}\right) = c\,\left(\frac{\nu_\mathrm{obs}}{\nu_0}\right)^\alpha\,\left(\frac{\Delta\nu^{\rm proj}}{\Delta\nu_0^{\rm proj}}\right)^\beta\left(\frac{\Omega_\mathrm{field}}{\Omega_\mathrm{0}}\right)^\gamma\:.
    \label{eq:fitscale}
\end{equation}
The parameters $\alpha$,\footnote{This is not to be confused with the power of the mass--luminosity function, which we also denoted with $\alpha$.} $\beta$, $\gamma$, and $c$ are free parameters that were fit separately for the shot noise and the clustering regimes, which we can identify with our one- and two-halo terms, respectively. For a source of covariance, which would scale with volume alone, we would expect $\beta=-0.5$ and $\gamma=-0.5$. 
The $\nu_\mathrm{obs}$ dependence is more complicated, as it changes the redshift of the signal and mixes scaling of the volume with potential changes in the bias parameters, growth of structure, mass--luminosity relations or halo mass function. However, for the case of [CII] and on shot noise-dominated scales, they find  $\alpha=-2.1$, $\beta=-0.62$, and $\gamma=-0.45$.\footnote{Reference~\cite{G20} reports similar values of $\beta$ and $\gamma$ for the clustering
component of [CII]. The same similarity between the two regimes was found for CO, although the values themselves differ from those for [CII].}  
The dependence on the field size is therefore shallower than pure volume
scaling.

Reference~\cite{G20} notes that this shallow scaling is expected for the clustering component, but is less intuitive for the shot-noise one, for which they appeal to the fact that the measured variance of the luminosity functions exceeds the Poisson term. If the observed field-to-field variation is induced by SSC, the behaviour in both regimes is accounted for within a single mechanism. The response function of equation~\eqref{eq:halomodelresponse} contains a shot-noise response, the term $\propto I^2_1$, alongside the clustering response, so that a common background mode $\delta_{\rm b}$ modulates the one- and two-halo amplitudes coherently, consistent with the similar exponents found in the two regimes and therefore explaining the correlation between the two components in Ref.~\cite{G20}. 

As Ref.~\cite{G20} focuses on the angular power spectrum while our work studies the spherically averaged 3D power spectrum, a direct comparison with our results is not trivial. However, we expect qualitatively similar trends as described in the appendix of Ref.~\cite{Takada_2013}. 

To measure the field-to-field variance, the methodology of Ref.~\cite{G20} includes a subdivision of the simulation into smaller boxes. Then, to isolate the shot-noise regime, the position of individual sources in each smaller box is reshuffled. This methodology erases all clustering information within each sub-box, which removes all $N$-point correlation functions to vanish.

We stress that the variance measured in Ref.~\cite{G20} is the total scatter of the power spectra across subfields, and therefore also contains the Gaussian and in-box non-Gaussian contributions associated with the shot-noise power spectrum since all other clustering has been removed through the reshuffling. Hence, the estimation of the covariance using all the smaller boxes includes only the shot-noise terms of the individual covariance contributions. For small scales,  the expression for the relative variance is given by:
\begin{equation}
    \left(\frac{\sigma}{\mu}\right) = \sigma_V \, \frac{\mathcal{I}_1^2}{\mathcal{I}_0^2} + \frac{\sqrt{\mathcal{I}_0^4\,V^{-1}}}{\mathcal{I}_0^2} + \frac{2}{N_{\rm modes}}\:.\label{eq:sigmaomu}
\end{equation}
The functions $\mathcal{I}_\beta^\alpha$ in this expression are implicitly functions of $k$, but, as the sources in the subvolumes in Ref.~\cite{G20} are assumed to be point sources, we can use the $k\to0$ limit of this expression. The last term in the equation above is the Gaussian contribution, and is subdominant for $k>0.5\,\mathrm{Mpc}^{-1}$. Any significant Gaussian contribution would pull the power law indices $\beta$ and $\gamma$ to $-0.5$, but would at the same time break the assumption that the ratio $\sigma/\mu$ is independent of scale.

\begin{figure}[t]
    \centering
    \includegraphics[width=0.99\linewidth]{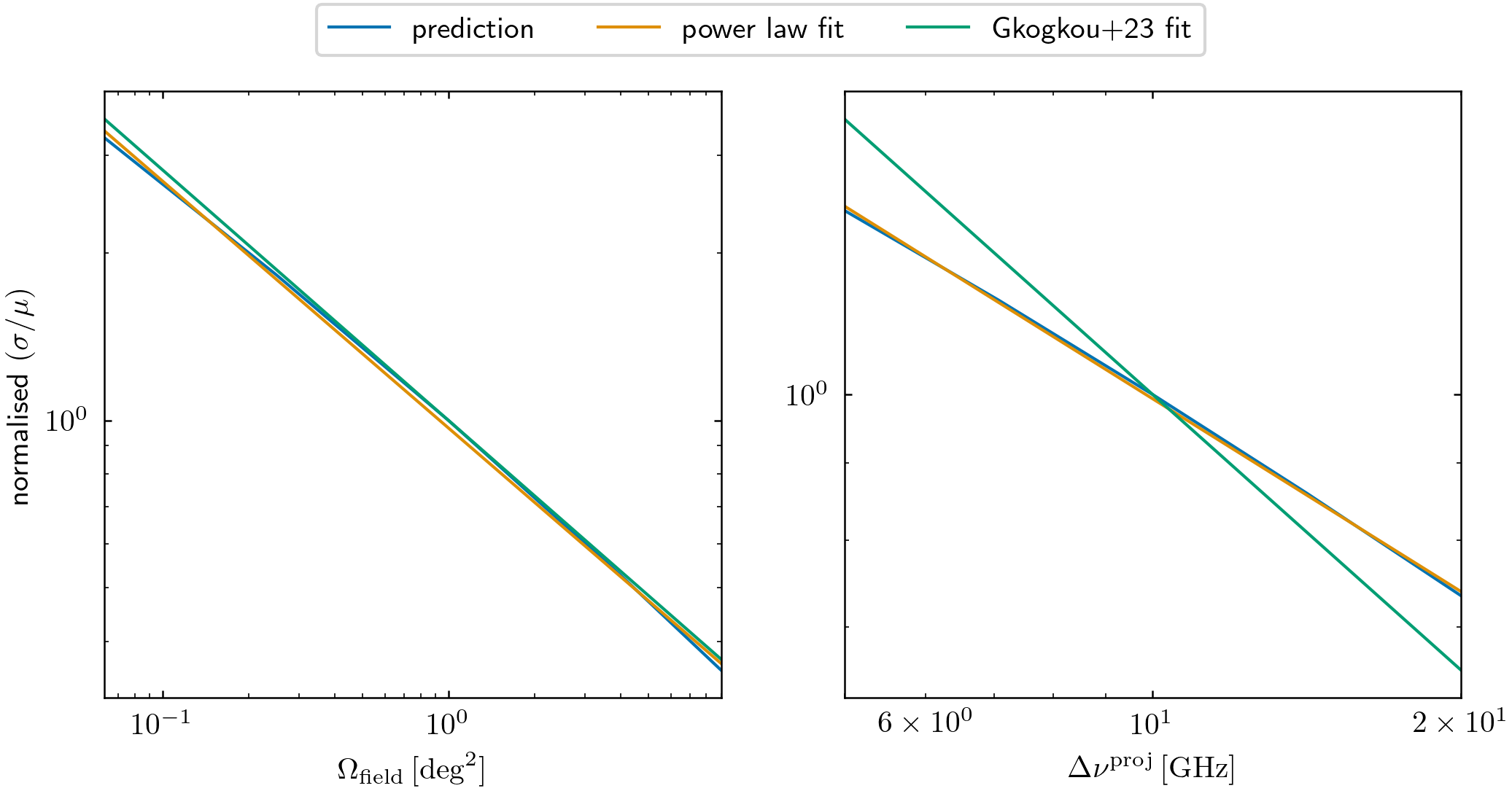}
    \caption{Comparison of the relative variance obtained by Ref.~\cite{G20} to the prediction for the SSC from our work. We also perform our own fit to compare to the previous result. All curves are normalised to an arbitrary reference survey with $\Delta\nu_0^\mathrm{proj}=10\,\mathrm{GHz}$ and $\Omega_0=1\,\mathrm{deg^2}$. \emph{Left}: Comparison of the relative uncertainty of the power spectrum as a function of the survey's angular size. We compare this to our own power law fit and the fit presented in Ref.~\cite{G20}. \emph{Right}: The same as \emph{left}, but as a function of the frequency bin width.} 
    \label{fig:concerto_powerlaw}
\end{figure}

Now let us compare the scaling relations of the relative uncertainty predicted by our model under these assumptions with those measured in Ref. \cite{G20}, using the fitting function from equation~\eqref{eq:fitscale} with $\Delta\nu^{\rm proj}_0 = 10\,$GHz, $\nu_0 = 300\,\mathrm{GHz}$, and $\Omega_0 = 1\,{\rm deg}^2$.\footnote{Note that in our case $\Delta\nu^{\rm proj}$ corresponds to the frequency bandwidth of the three-dimensional volume considered.} We illustrate this for the scaling of $\Delta \nu^{\rm proj}$ and $\Omega_\mathrm{field}$ in the shot-noise regime in figure~\ref{fig:concerto_powerlaw}, since these are the two dependences which are more directly transferable between the angular and three-dimensional power spectra. 

On the left-hand side of figure~\ref{fig:concerto_powerlaw}, we find excellent agreement for the scaling relations with the angular size of the survey $\Omega_\mathrm{field}$: a power law index of $\gamma=-0.44$ compared to the value of $\gamma=-0.45$ found in Ref.~\cite{G20}. On the right-hand side of figure~\ref{fig:concerto_powerlaw}, the comparison of the scaling for $\Delta \nu^{\rm proj}$ returns qualitative agreement: a power law index of $\beta=-0.43$ compared to $\beta=-0.62$. 

This was expected due to multiple possible reasons. Firstly, the differences between the three-dimensional power spectrum and the angular power spectrum. The reason the power law for the angular size works much better is that the equivalent $\sigma_V$ for projected density fields can be decomposed into distinct contributions. One of them is a perpendicular factor, analogous to the $W$ term in the power spectrum, while the other is related to radial selection and parallel modes~\cite{Takada_2013, euclid_ssc}. Since we can identify a rough correspondence to the first contribution as also present in the definition of $\sigma_V$ in equation \eqref{eq:cov_ssc}, we expected good agreement. On the other hand, the radial selection and the transformation between $k$ modes in $\mathrm{arcmin}^{-1}$ and $\mathrm{Mpc}^{-1}$ influence the scaling relations with $\Delta\nu$, which is why we did not expect an agreement as good as for $\Omega_\mathrm{field}$.
Secondly, the meaning of $\Delta\nu^\mathrm{proj}$ in the two analyses are different: in Ref.~\cite{G20} it denotes the number of frequency channels averaged over to obtain the final power spectrum of each subfield, rather than the line-of-sight extent of a three-dimensional volume.
Additionally, differences between our line-luminosity model and the simulation influence the comparison for $\Delta \nu^\mathrm{proj}$ through the integration over the redshift dependence. 
Finally, we note that the subfields of Ref.~\cite{G20} are not fully independent, being drawn from a single lightcone and, at the larger sizes, tiling it contiguously. Reference~\cite{G20} includes a note stating that this may lead to an underestimation of the variance in the luminosity function measurements due to the large-scale modes; the same caveat applies to the shot-noise power spectra.

\subsection{Relative contributions and dependence on survey volume}\label{subsec:relative}

As stated above, the dependence of the relative importance of the SSC is not trivial. Therefore, in this section we study the relative contributions to the diagonal of the covariance for different survey volumes. We consider a fixed-shape volume changing from $10^6\,\mathrm{Mpc}^3$ to $10^{10}\,\mathrm{Mpc}^3$, always centred at the same effective redshift. Since the logarithm of the response function is roughly constant and does not depend on the survey volume, we focus on changes in $\sigma_V^2$ (see Equation ~\ref{eq:sigmaomu}). We note that the Gaussian and non-Gaussian in-box contributions are inversely proportional to the volume, through the number of modes. Consequently, a convenient quantity to study is $V\sigma^2_V$, i.e., the product of the covariance contributions and the survey volume. 

\begin{figure}
    \centering
    \includegraphics[width=0.99\linewidth]{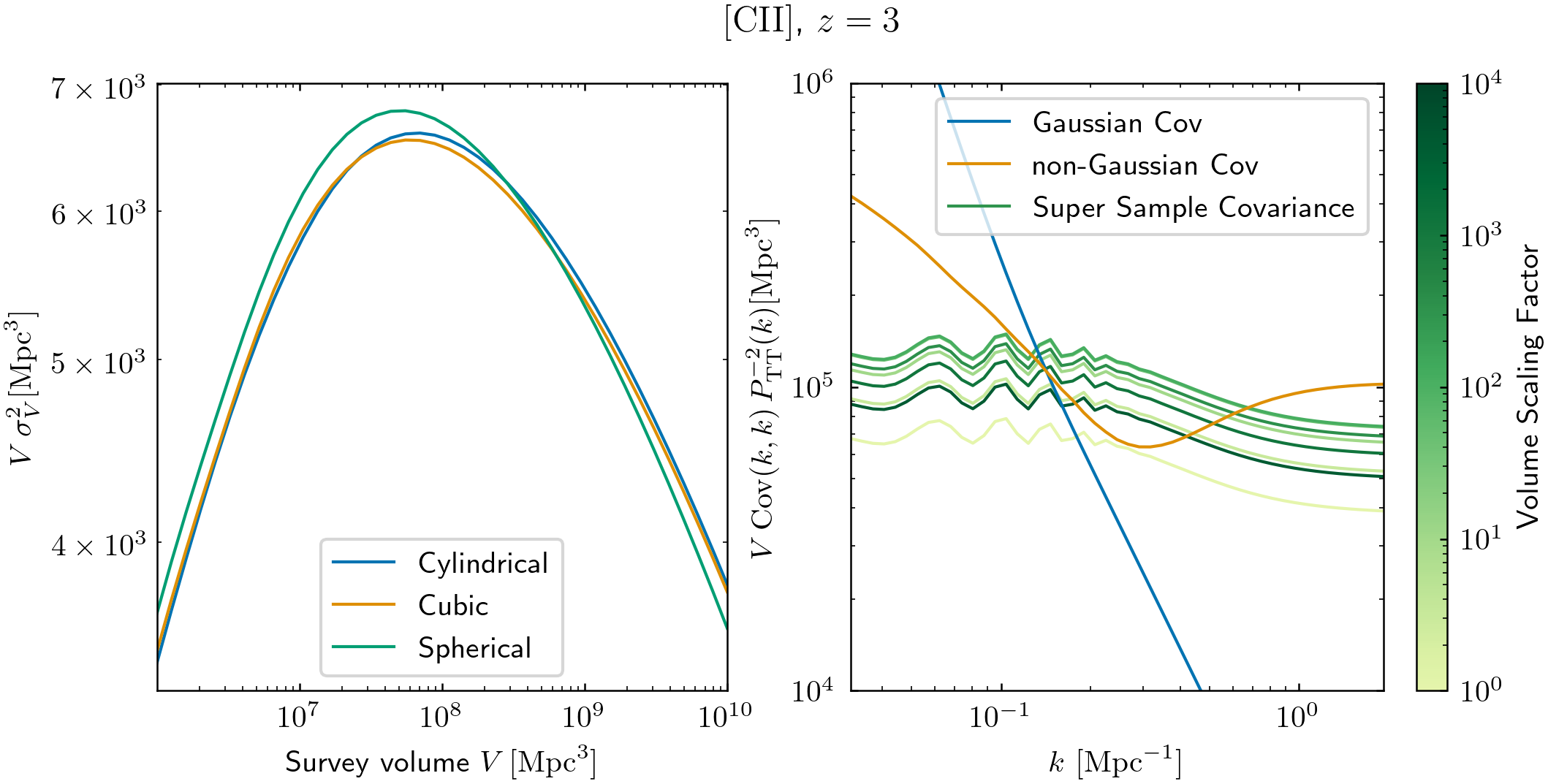}
    \caption{Scaling of the covariance contributions for a fictional [CII] experiment at redshift $z=3$. \emph{Left}:
    Dependence of the variance $\sigma_V^2$ times the volume as a function of the survey volume for different survey shapes. 
    \emph{Right}: The relative variance of the power spectrum normalised to the survey volume. An additional factor of the volume absorbs the scaling of the Gaussian and non-Gaussian Covariance, while the exact scaling of the SSC can be taken from the left plot. The different shades of green correspond to different volumes, with the minimum volume being $10^6$ Mpc$^3$.}
    %\AMcomment{Same comment as in figure 4, why did we switch the k unit?}}
    \label{fig:ssc_scaling}
\end{figure}

The left panel of figure~\ref{fig:ssc_scaling} shows $V\sigma^2_V$ as a function of the survey volume, for cubic, spherical and cylindrical\footnote{For the cylindrical setup, we choose a height-to-radius ratio of $\pi$, which yields the minimum variance for the cylindrical survey setup \cite{Takada_2013}.} geometries, which allows us to assess the impact of the survey shape. 
Within the four orders of magnitude in change of $V$, the product $V\,\sigma^2_V$ is quite constant: it changes by about a factor of two up to a maximum around $\sim 10^8\,\mathrm{Mpc}^3$, and starts falling again afterwards. The turnaround can be understood as stemming from whether there are modes at scales larger than the horizon at equality contributing to $\sigma_V^2$ or not, which manifests in the power spectrum as a change of slope. Indeed, $10^8\,\mathrm{Mpc}^3$ corresponds to a box of side $\simeq 460\,\mathrm{Mpc}$, i.e.\ $k\simeq 0.014\,\mathrm{Mpc}^{-1}$, comparable to $k_\mathrm{eq}$. The three curves for the three different survey geometries have roughly the same shape, with small differences in amplitude and slope.

This near-constancy underlies the claim, anticipated in section~\ref{sec:intro}, that the SSC does not lose relevance as the survey volume grows. The Gaussian and in-box contributions scale as $V^{-1}$ through the number of modes, while the SSC scales as $\sigma_V^2$; since $V\sigma_V^2$ varies by only a factor of two across four orders of magnitude in $V$, the \emph{ratio} between the SSC and the other contributions is nearly independent of the survey volume. Increasing the volume therefore reduces all contributions together, rather than suppressing the SSC preferentially.

We show the comparison of the diagonal elements of the covariance contributions times the survey volume over the power spectrum square in the right panel of figure~\ref{fig:ssc_scaling}, assuming the cylindrical survey case. The Gaussian covariance has the previously derived $N_i^{-1}\propto k^{-2}\,\Delta k^{-1}$ scaling, where $\Delta k$ is the width of a wavevector shell. On the other hand, the two normalised non-Gaussian contributions are roughly constant. While the volume scaling of the in-box covariance is exactly cancelled out, the SSC is scaled up and down for different volumes following the trend shown in the left panel, with the covariance $k$-dependent part given by the power spectrum response. As the product $V\,\sigma^2_V$ is only changing by a factor of two for our range of volumes, the transition wavenumber between Gaussian- and non-Gaussian-dominated covariance is quite stable. This model additionally predicts a large contribution from the non-Gaussian in-box covariance at the smallest scales, even overtaking that of the SSC. This large contribution is the one-halo term of the trispectrum.
For this model, we find that the transition from Gaussian-dominated to non-Gaussian-dominated covariance occurs around a wavenumber of $k\sim 0.1\,\mathrm{Mpc}^{-1}$. We stress that this depends strongly on the line under consideration and the survey setup. 
Finally, while the exact survey shape influences the scaling relations of the SSC amplitude, its effect on the comparison between Ref.~\cite{G20} and our work is expected to be much smaller than the difference of the summary statistics focused on in each work. 

\subsection{Forecasts for specific surveys}

\noindent 
To assess the relative importance of the three contributions to the covariance matrix in more realistic scenarios,  we perform a case study for two representative LIM experiments, MeerKLASS, measuring $21\,\mathrm{cm}$ line, and CCAT-PRIME, measuring [CII] line. The assumed survey specifications for these surveys are found in Table~\ref{tab:survey_specs}. For MeerKLASS, we differentiate between two different radio bands: the L-band and the UHF-band, which have slightly different system temperature and angular beam; we assume similar observation times for both. We consider a single patch for the MeerKLASS-UHF band survey. The full survey is planned to extend to $\Omega_\mathrm{field}\approx10^4\,\mathrm{deg}^2$, but first analyses will analyse each patch independently and combine them later. For CCAT-PRIME, we choose to study the end of reionisation (EoR) survey and a futuristic version with ten times better sensitivity. CCAT-PRIME results are given in terms of spectral flux density, for which we need to use

\begin{equation}
    C_\mathrm{LI} = \frac{c}{4\,\pi\,\nu\,H(z)}\:,
\end{equation}
instead of $C_{\rm LT}$, and replace the system temperature by the noise-equivalent intensity per pixel to compute the instrumental noise. 
The necessary extensions to the covariance expressions and definitions of survey properties are found in appendix \ref{sec:survey}, where we describe an effective treatment of the anisotropic resolution limits to keep the spherical symmetry assumed in the derivation of the covariance in sections~\ref{sec:theo_cov} and~\ref{sec:limhalos}.

\begin{table}
    \centering
    \begin{tabular}{|c|ccc|}
        \hline
        & {MeerKLASS}-L band &{MeerKLASS}-UHF band &{CCAT-PRIME} EOR Survey \\
        \hline
         Target line & $\mathrm{H_{I}}$, $994.7\,\mathrm{MHz}$& $\mathrm{H_{I}}$, $745\,\mathrm{MHz}$ & $\mathrm{[CII]}$, $280\,\mathrm{GHz}$\\
         Redshift $z$ & 0.42& 0.91& 5.79\\
         %\hline
         Survey area $\Omega_\mathrm{field}$ & $236\,\mathrm{deg}^2$ & $236\,\mathrm{deg}^2$& $4\,\mathrm{deg}^2$\\
         %\hline
         Bin width $\Delta\nu$& $52\,\mathrm{MHz}$ & $185\,\mathrm{MHz}$ & $40\,\mathrm{GHz}$\\
         %\hline
         Angular beam $\theta_\mathrm{FWHM}$& $1.38\,\mathrm{deg}$& $1.84\,\deg$ & $48\,\mathrm{arcsec}$\\
         %\hline
         Spectral resolution $\delta\nu$& $0.21\,\mathrm{MHz}$ & $0.21\,\mathrm{MHz}$ & $1.19\,\mathrm{GHz} $\\
         %\hline
         Voxel noise variance $\sigma_{T/I}^2$ & $5.6\times10^4\,\mu\mathrm{K}^2$ & $3.6\times10^4\,\mu\mathrm{K}^2$ & $4.2\times10^7\,\mathrm{Jy}^2\,\mathrm{sr}^{-2}$\\
         \hline
    \end{tabular}
    \caption{Assumed experimental specifications for the different surveys we are studying in this section.} 
    \label{tab:survey_specs}
\end{table}

\begin{figure}
    \centering
    \includegraphics[width=\linewidth]{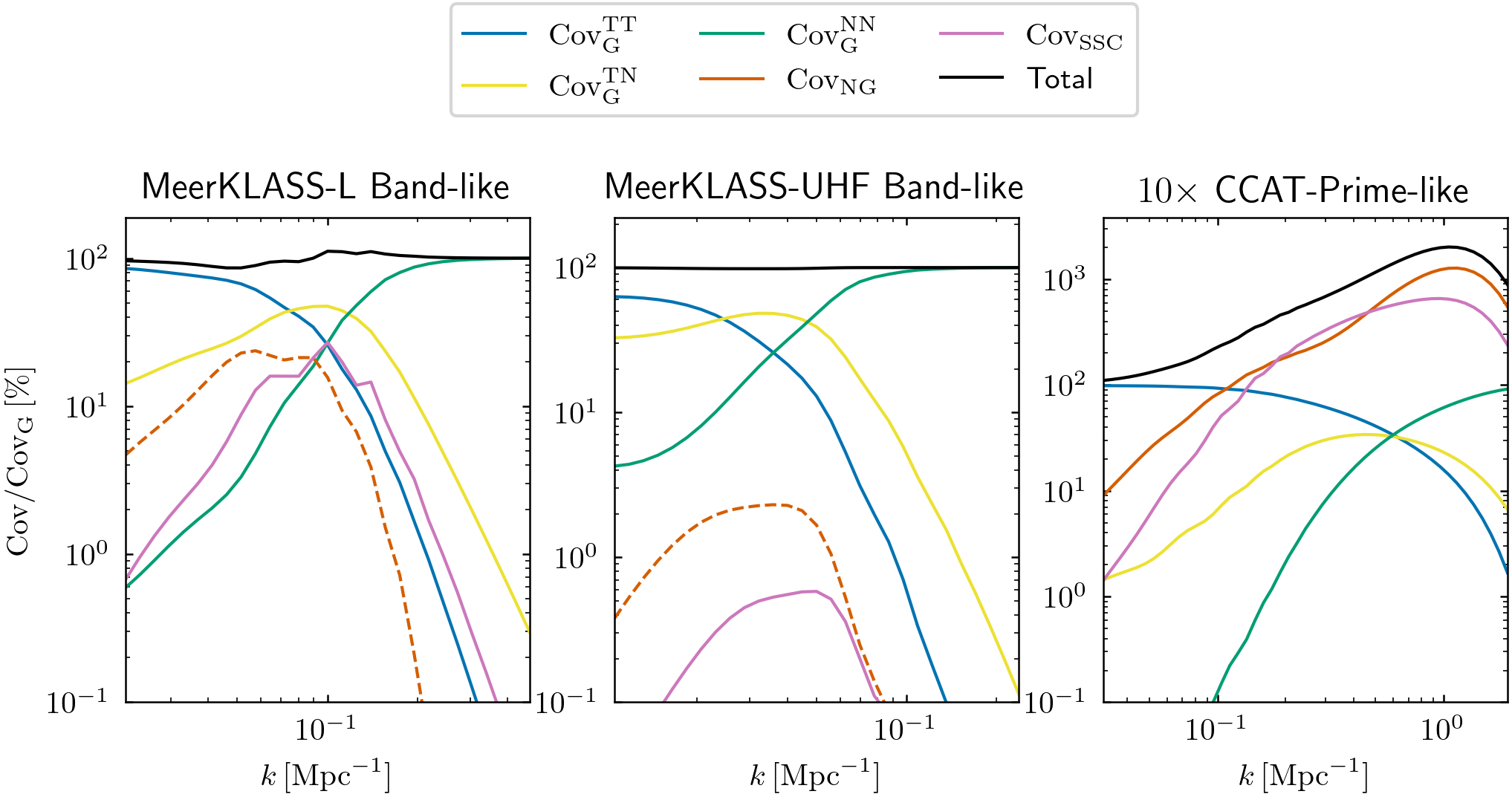}
    \caption{Contributions to the diagonal of the covariance normalised to the Gaussian covariance for different surveys. We show the individual Gaussian contributions as separate lines and compare them to the non-Gaussian covariances. Negative contributions to the covariance are denoted with dashed lines. On the \emph{left} plot, this is shown for our {MeerKLASS}-L band-like setup, on the \emph{center} for the UHF band, and on the \emph{right} for a {CCAT-prime}-like setup.}
    \label{fig:covarance_diagonal_CATT_MeerKlass}
\end{figure}
We focus first on the diagonal of the covariance, the contributions of which are shown in figure~\ref{fig:covarance_diagonal_CATT_MeerKlass}.
We have separated the Gaussian covariance into three distinct covariances as they all depend on $k$ differently. The signal-only, signal--cross--noise, and noise-only covariances are proportional to $P_{\rm TT}^2$, to $P_{\rm TT}$ and independent of the power spectrum, respectively. 

On large scales, we find that the signal-only covariance dominates for all setups, while both non-Gaussian covariances show a rising trend.

On the smallest scales, however, we see the general trend that all covariance contributions except for the noise-only contribution are heavily suppressed. This can be explained by the resolution limit, which suppresses the power spectrum and trispectrum while leaving the noise power spectrum untouched. For CCAT-Prime, the survey resolution is only starting to affect the power spectrum at the maximum wavenumber. However, the low noise plateau makes the noise-only covariance subdominant across all scales.

For the MeerKLASS L-band case, the noise covariance dominates the diagonal of the covariance at $k >0.2\,\mathrm{Mpc}^{-1}$. For scales smaller than that, the correlation of modes is reduced since the noise covariance is only diagonal. For  $k\gtrsim 0.08\,\mathrm{Mpc}^{-1}$, the non-Gaussian covariances contribute significantly to the diagonal. The in-box non-Gaussian covariance is negative for these scales since some of the theoretically predicted bias parameters are negative, and smaller than the SSC in amplitude. For these intermediate scales, the non-Gaussian covariance in sum is roughly 10\% of the total covariance, and we can expect mild correlation between the modes. 

On the other hand, for the {MeerKLASS}-UHF band-like setup, the non-Gaussian covariance is always in the order of 1\% of the Gaussian covariance. This is because of the larger instrumental beam and higher noise.

Note that when all patches are combined, we expect lower covariance.  
In the case of the futuristic CCAT-PRIME, the noise covariance is subdominant on all accessible scales. Thus, we expect there to be a large correlation between the modes for all $k\lesssim 0.1\,\mathrm{Mpc}^{-1}$. We can also see that the in-box non-Gaussian covariance and the SSC are similar in amplitude for all scales.

\begin{figure}
    \centering
    \includegraphics[width=0.99\linewidth]{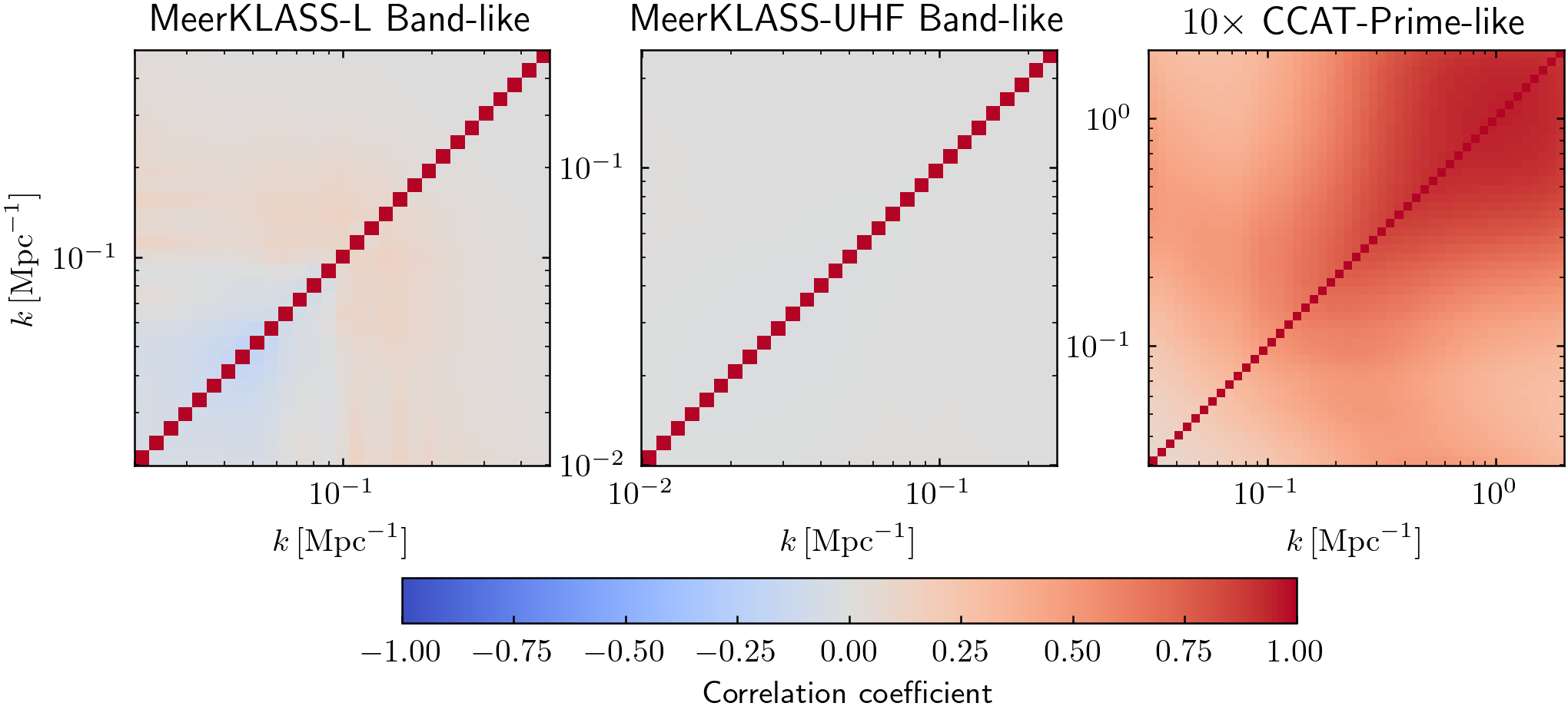}
    \caption{Correlation coefficients for the two MeerKLASS and CCAT-prime surveys.}
    \label{fig:meerklass_corr}
\end{figure}
We also show the correlation matrix for both cases in figure~\ref{fig:meerklass_corr} for the two {MeerKLASS} surveys on the left and centre plots, and the futuristic {CCAT-prime} on the right. For the L band, our previous intuition holds; starting at a scale of roughly $0.08\,\mathrm{Mpc}^{-1}$, there is considerable correlation between modes, while for much smaller scales around $0.2\,\mathrm{Mpc}^{-1}$ the modes decorrelate, due to the noise covariance taking over and the beam suppressing the signal. For the intermediate scales, the correlation between modes is quite large, with a correlation coefficient close to 40\%. On the other hand, the off-diagonal correlations are heavily suppressed for the UHF-band survey, and the covariance can be assumed to be mostly diagonal. For the futuristic CCAT-prime-like setup, the situation is flipped; there is a sizeable correlation for all scales, while scales smaller than $0.1\,\mathrm{Mpc}^{-1}$ are fully correlated. Contrary to the L-band forecast, we do not see a decorrelation at the smallest scales, as the suppression of power due to survey resolution is only starting to significantly affect the covariances.

In summary, whether the non-Gaussian contributions matter is governed by the noise level rather than by the survey volume. In the noise-dominated MeerKLASS-UHF configuration, they are negligible, and the covariance is effectively diagonal, whereas in the signal-dominated CCAT-prime configuration the in-box and super-sample terms are comparable to each other and correlate modes across the full range of accessible scales.

\section{Conclusion}
\label{sec:conc}
While LIM experiments provide a unique window into the LSS, they are, like any other cosmological probe, fundamentally limited by finite volumes to infer the cosmological mean. 
Accurate uncertainty estimation is crucial for robust detection and inference of cosmological and astrophysical parameters. This includes contributions from unconstrained modes at scales larger than the size of the survey to the covariance, i.e., the SSC. As LIM surveys enter the precision cosmology era, proper estimation of measurement uncertainties is key for extracting reliable cosmological constraints. Especially for low-noise scenarios and for small scales, we find that adding the non-Gaussian covariance is vital.

Current LIM experiments rely on mock simulations or analysis of the observations to estimate the covariance of the measurements, using e.g., jackknife methods. However, by definition, using the observed data cannot account for the influence of super survey modes, while their modelling in simulations significantly increases the computational requirements. This is why previous forecasts and analyses of LIM surveys, with the exception of the empirical study of Ref.~\cite{G20}, have neglected this additional uncertainty. Given the small volumes that some of these surveys involve, this might have large impacts.

In this paper, we provide the first derivation of the real-space LIM power spectrum SSC to account for this effect in LIM measurements. For this, we have extended the common LIM halo model with SPT to model higher-order correlators.

We have validated our theoretical prediction against painted LIM $N$-body simulations to high precision, reproducing the diagonal at the few-per-cent level over most of the $k$ range, degrading to $\lesssim 30$\% ($\lesssim 50$\%) in two intermediate $k$-bins when the bias parameters are fitted to the simulation (fixed to their predicted values). A byproduct of this work is the first derivation of the non-Gaussian in-box covariance for the LIM power spectrum, too, needed to quantify the relevance of the SSC. Thanks to our analytic modelling, and its implementation in the public code released with this work, the full covariance can be easily computed for any survey and line of interest. To further validate our derivations, we have compared our implementation with the empirical fits from Ref.~\cite{G20}, finding generally good agreement even if that work focuses on the angular power spectrum.

After validation, we have explored different experiments and lines to study the dependence of the relative contribution of the SSC. Our results suggest that extracting global cosmological information using LIM probes has the same limitations as traditional LSS probes, and on top of the in-box non-Gaussian covariance, a full assessment of the covariance must include the SSC unless the covariance is completely dominated by the noise contributions. For LIM surveys, this is especially important as they typically probe very small volumes and are thus limited to the smallest scales, where the non-Gaussian contributions to the covariance become significant. Note, however, that since all covariance contributions scale roughly similarly with volume, these scales will still be dominated by the SSC covariance as the survey volume grows.

As a first step, we have limited our derivations to the real-space power spectrum, neglecting redshift-space distortions and anisotropic resolution limits, which we account for using an effective treatment which keeps the spherical symmetry of the power spectrum. Given the particularities of LIM experiments, where map making, gridding, resolution limits and other observational effects may introduce strong anisotropies, future efforts must be invested to derive the full covariance for an axisymmetric three-dimensional Fourier power spectrum. We plan to do so in future work, as well as exploring the effect that signal loss due to foreground cleaning might have on the non-Gaussian covariance.

LIM surveys continue to improve in sensitivity and accuracy and a new generation of pathfinders and second-generation experiments are expected for the forthcoming years. Theoretical developments to enable optimal analyses must then keep up and be ready for the moment the observations are available and maximise the scientific return from this novel observational probe. 

\acknowledgments
SP and JLB acknowledge funding from the project UC-LIME (PID2022-140670NA-I00), financed by MCIN/AEI/ 10.13039/501100011033/FEDER, UE. SP acknowledges financial support from the 2023 Grants for predoctoral contracts 'Concepción Arenal' under the Universidad de Cantabria's programme for pre-doctoral research staff No. 084/2024. AMD acknowledges supported by the Agence Nationale de la Recherche (ANR) under grant No. ANR-23-CPJ1-0160-01. The authors acknowledge the use of the computer resources provided by the Spanish Supercomputing Network (RES) node at Universidad de Cantabria and the Institute of Physics of Cantabria (IFCA-CSIC).

\appendix
\section{Standard perturbation theory of large-scale structure}
\label{sec:appendixA}

To compute the SPT kernels for the halo model corrections, we insert equations \eqref{eq:matter_spt} and  \eqref{eq:velocity_spt} into \eqref{eq:simple_bias} and collect terms with the same power of $\delta_\mathrm{L}$. The resulting mode-coupling kernels are the same as presented in Ref.~\cite{Kobayashi:2023vpu}, but with the growth rate $f$ set to zero, since we are not including redshift-space distortions.\footnote{Following standard notation, we use $K$ instead of $Z$ for the real-space kernels.} Explicitly, we have
\begin{align}
    K_1(M) &= b_1(M)\:,\\
    K_2(\boldsymbol{q}_1,\boldsymbol{q}_2,M) &= b_1(M)\,F_2(\boldsymbol{q}_1, \boldsymbol{q}_2) + \frac{b_2(M)}{2} + b_{\mathcal{G}_2}(M)\,\mathcal{G}_2(\boldsymbol{q}_1,\boldsymbol{q}_2)\:,\\
    3\,K_3(\boldsymbol{q}_1,\boldsymbol{q}_2,\boldsymbol{q}_3,M) &= b_1(M)\,F_3(\boldsymbol{q}_1,\boldsymbol{q}_2,\boldsymbol{q}_3)+b_2(M)\,F_2(\boldsymbol{q}_2,\boldsymbol{q}_3)+\frac{b_3(M)}{6} \nonumber\\
    &+2\,b_{\mathcal{G}_2}(M)\,F_2(\boldsymbol{q}_2,\boldsymbol{q}_3)\,\mathcal{G}_2(\boldsymbol{q}_1,\boldsymbol{q}_2+\boldsymbol{q}_3) + \,b_{\delta\,\mathcal{G}_2}\,\mathcal{G}_2(\boldsymbol{q}_2,\boldsymbol{q}_3)+b_{\mathcal{G}_3}\,\mathcal{G}_3(\boldsymbol{q}_1,\boldsymbol{q}_2,\boldsymbol{q}_3)\nonumber\\
    &+2\,b_\Gamma(M)\left[F_2(\boldsymbol{q}_2,\boldsymbol{q}_3)-G_2(\boldsymbol{q}_2,\boldsymbol{q}_3)\right]\,\mathcal{G}_2(\boldsymbol{q}_1,\boldsymbol{q}_2+\boldsymbol{q}_3) +\:\text{perms.}\:,\label{eq:ap_z3}
\end{align}
where the permutations in Equations \eqref{eq:ap_z3} are the cyclic permutations of the wave vectors $\boldsymbol{q}_i$. The functions $\mathcal{G}_i$ are the Fourier-space operators of the Gallileons: 
\begin{align}
    \mathcal{G}_2(\boldsymbol{q}_1,\boldsymbol{q}_2) &= \cos^2\sphericalangle_{12} - 1\:,\\
    \mathcal{G
    }_3(\boldsymbol{q}_1, \boldsymbol{q}_2, \boldsymbol{q}_3) &= \cos\sphericalangle_{12}\,\cos\sphericalangle_{13}\,\cos\sphericalangle_{23} - \left[\cos^2\sphericalangle_{12}+\cos^2\sphericalangle_{13}+\cos^2\sphericalangle_{23}\right] + 1\:,
\end{align}
where $\sphericalangle_{ij}$ is the angle between $\boldsymbol{q}_i$ and $\boldsymbol{q}_j$. We take the expressions for $F_i$ and $G_i$ from Ref.~\cite{Bernardeau_2002}:
\begin{align}
    F_2(\boldsymbol{q}_1,\boldsymbol{q}_2) &= \frac{5}{7} + \frac{1}{2}\,\cos\sphericalangle_{12}\,\left(\frac{q_1}{q_2}+\frac{q_2}{q_1}\right) +\frac{2}{7}\,\cos^2\sphericalangle_{12}\:,\\
    G_2(\boldsymbol{q}_1,\boldsymbol{q}_2) &= \frac{3}{7} + \frac{1}{2}\,\cos\sphericalangle_{12}\,\left(\frac{q_1}{q_2}+\frac{q_2}{q_1}\right) +\frac{4}{7}\,\cos^2\sphericalangle_{12}\:,\\
    3\,F_3(\boldsymbol{q}_1,\boldsymbol{q}_2,\boldsymbol{q}_3)
    &= \frac{7}{18}\frac{\boldsymbol{q}_1\,\left(\boldsymbol{q}_1+\boldsymbol{q}_2+\boldsymbol{q}_3\right)}{q_1^2}\,F_2(\boldsymbol{q}_2,\boldsymbol{q}_3) + \frac{7}{18}\,\frac{\left(\boldsymbol{q}_2+\boldsymbol{q}_3\right)\,\left(\boldsymbol{q}_1+\boldsymbol{q}_2+\boldsymbol{q}_3\right)}{\|\boldsymbol{q}_2+\boldsymbol{q}_3\|^2}\,G_2(\boldsymbol{q}_2,\boldsymbol{q}_3) \nonumber\\
    &+\frac{2}{18}\,\frac{\|\boldsymbol{q}_1+\boldsymbol{q}_2+\boldsymbol{q}_3\|\,\boldsymbol{q}_1\left(\boldsymbol{q}_2+\boldsymbol{q}_3\right)}{q_1^2\|\boldsymbol{q}_2+\boldsymbol{q}_3\|^2}\,G_2(\boldsymbol{q}_2,\boldsymbol{q}_3) +\:\text{perms.}\:.
\end{align}
Like in the case of Equation \ref{eq:ap_z3}, the missing terms of $F_3$ are computed from the cyclic permutations of the wave vectors. 

\section{Computation of the halo trispectrum terms contributing to the covariance}
\label{sec:trispectrum_terms}
To compute the non-Gaussian in-box covariance, we need to evaluate the squeezed trispectrum. In particular, we are interested in the shell-averaged squeezed trispectrum:
\begin{equation}
    \mathrm{Cov}^\mathrm{NG}_{T_0}(k_i,k_j)\propto \frac{1}{N_i\,N_j}\sum_{\boldsymbol{k}_1\in\boldsymbol{V}_{\boldsymbol{k}_i}}\sum_{\boldsymbol{k}_2\in\boldsymbol{V}_{\boldsymbol{k}_j}}T_\mathrm{T}(\boldsymbol{k}_1,-\boldsymbol{k}_1,\boldsymbol{k}_2,-\boldsymbol{k}_2)\:.
\end{equation}

We approximate the shell average by integrating over the surfaces of two thin shells of radii $k_i$ and $k_j$ \cite{Wadekar:2019rdu}. Neglecting line-of-sight–dependent effects, the trispectrum depends only on the magnitudes of the vectors and the angle between them. Denoting this re-parametrised form by $T_0$ and exploiting axial symmetry, we align one vector with the $z$-axis and set the $y$-component of the other to zero. The covariance reduces to
\begin{equation}
    \mathrm{Cov}^\mathrm{NG}_{T_0}(k_i,k_j) \propto \frac{1}{2} \int_{-1}^{1}\:T_0(k_i,k_j, \cos\sphericalangle)\,\mathrm{d}\cos\sphericalangle\:.
\end{equation}

We can subdivide the integral into contributions from the same combinations of biases. 
The different contributions can be expanded into multiple terms, finding terms with their angular dependence residing entirely in the kernels discussed in the previous appendix, and terms involving the power spectrum evaluated at
\begin{equation}
    \frac{1}{2}\,\int_{-1}^1\:\mathcal{K}(k_i,k_j,\cos\sphericalangle)\,P\left(\sqrt{k_i^2+2\,\cos\sphericalangle\,k_i\,k_j+k_j^2}\right)\,\mathrm{d}\cos\sphericalangle\:,
\end{equation}
with $\mathcal{K}$ an arbitrary angle-dependent kernel. The former admit simple analytic solutions, whereas the latter generally do not.

A simple case is a term in the four-halo contribution to the covariance that connects three linear biases with one third-order local bias, denoted here as $\mathrm{Cov}_{T_0}^{3111}$,
\begin{equation}
   \mathrm{Cov}_{T_0}^{3111}(k_i,k_j) = \frac{2\,V_4}{V_2^2}\,\mathcal{I}_3^1(k_i)\mathcal{I}_1^1(k_i)\,\,P(k_i)\left[\mathcal{I}_1^1(k_j)\,P(k_j)\right]^2 +k_i\longleftrightarrow k_j\:,
\end{equation}
while a term containing an integral over the power spectrum appears in the four-halo contribution when connecting two linear biases with two quadratic biases, denoted here as $\mathrm{Cov}_{T_0}^{2211}$,
\begin{align}
    \mathrm{Cov}_{T_0}^{2211}(k_i,k_j) &= \frac{V_4}{V_2^2}\left\lbrace 2\,\left[\mathcal{I}_1^1(k_i)\,\mathcal{I}_2^1(k_j)\,P(k_i)\right]^2+4\,\mathcal{I}_1^1(k_i)\,\mathcal{I}_2^1(k_i)\,\mathcal{I}_1^1(k_j)\,\mathcal{I}_2^1(k_j)\,P(k_i)\,P(k_j)\right.\nonumber\\
    &+\left.2\,\left[\mathcal{I}_2^1(k_i)\,\mathcal{I}_1^1(k_j)\,P(k_j)\right]^2\right\}\,\frac{1}{2}\int_{-1}^{1}\:P\left(\sqrt{k_i^2+2\,k_i\,k_j\,\cos\sphericalangle+k_j^2}\right)\,\mathrm{d}\cos\sphericalangle\:. \label{eq:covariance_2211_example}
\end{align}

The kind of integrals similar to the one above require high numerical precision: the integration range spans scales from $|k_i-k_j|$ to $k_i+k_j$, and they exhibit a resonant behaviour around $k_i\sim k_j$. In a power-law Universe, however, the integrals can be evaluated analytically. We therefore apply the FFTLog method to decompose the power spectrum into a sum of power laws, following Ref.~\cite{Kobayashi:2023vpu}. This allows us to write the power spectrum as 
\begin{equation}
    P(k) = \sum_{n=0}^{N_\mathrm{FFT}} c_n k^{\alpha_n}\:, \label{eq:fftlog}
\end{equation}
where $N_\mathrm{FFT}$ is the order of the Fourier decomposition and $c_n$ and $\alpha_n$ are the complex Fourier coefficients and the FFTLog powers, respectively. We choose $N_\mathrm{FFT}$ to be $2^{10}$ and perform the FFTLog transformation from $10^{-4}$--$10^2\,\mathrm{Mpc}^{-1}$. We compute the power spectrum at the smallest and largest scales by extrapolating the linear power spectrum using a power law. This helps us to reduce ringing at the edges of our power spectrum. Using the decomposition \eqref{eq:fftlog}, we can compute the integrals over the power spectrum analytically. For example, the integral appearing in Equation \eqref{eq:covariance_2211_example} becomes:
\begin{equation}
    \frac{1}{2} \int_{-1}^{1}\:P\left(\sqrt{k_i^2+2\,k_i\,k_j\,\cos\sphericalangle+k_j^2}\right)\,\mathrm{d}\cos\sphericalangle = \sum_{n=0}^{N_\mathrm{FFT}} \,\frac{c_n}{\alpha_n+2}\,\frac{\left(k_i + k_j \right)^{\alpha_n+2}-\left|k_i -k_j\right|^{\alpha_n+2}}{k_i\,k_j}\:.
\end{equation}
For the other contributions, we obtained similar results using \texttt{Mathematica}.

\section{Simplified effective description of anisotropic resolutions}
\label{sec:survey}
To create a line-intensity map, the sky and frequencies are subdivided into pixels and channels, respectively. The resolution essentially limits the smallest scales from which information can be extracted, since they suppress the signal below its characteristic scales. The telescope beam is assumed to be Gaussian and constant across the area and frequencies of each experiment, described by its full-width-at-half-maximum $\theta_\mathrm{FWHM}$. 
The suppression of the power spectrum is then given by a multiplicative factor 
\begin{align}
    F_\perp(k,\mu) &= \exp\left[-\frac{1}{2}\,\left(1-\mu^2\right)\,k^2\,\sigma_\perp^2\right]\:,\\
    \sigma_\perp &= D_\mathrm{M}\,\frac{\theta_\mathrm{FWHM}}{\sqrt{8\,\log2}}\:, 
\end{align}
where $D_\mathrm{M}$ is the comoving angular diameter distance and $\mu$ is the cosine between the line of sight and the $k$-mode. Similarly, we model the spectral response of each channel as a Gaussian with standard deviation $\delta \nu$. The resulting suppression factor is given by
\begin{align}
    F_\parallel(k,\mu) &= \exp\left[-\frac{1}{2}\, \mu^2\,\,k^2\,\sigma_\parallel^2\right]\:,\\
    \sigma_\parallel &= \frac{1+z}{H / c}\frac{\delta\nu}{\nu_\mathrm{obs}}\:.
\end{align}
We assume cubic voxels in angular and frequency space, sized by $\theta_{\rm FWHM}$ and $\delta\nu$. The volume of a voxel is given by
\begin{equation}
    V_\mathrm{vox} = D_\mathrm{M}^2\,\theta_\mathrm{FWHM}^2\,\frac{1+z}{H/c}\,\frac{\delta\nu}{\nu_\mathrm{obs}}\,\sqrt{8\log2}\:.
\end{equation}

The two multiplicative factors $F_\parallel$ and $F_\perp$ are usually multiplied to the cylindrical power spectrum $P(k,\mu)$ before integrating over $\mu$ to obtain the Legendre multipoles (see e.g., Ref.~\cite{Bernal_2019}). This approach would, however, break the spherical symmetry assumed in section~\ref{sec:limhalos} for our derivation. We therefore treat the anisotropic resolution limits effectively and perform their spherical average separately, obtaining 
\begin{equation}
    F^2_0(k_i)=\frac{1}{N_i}\sum_{\boldsymbol{k}_1\in\boldsymbol{V}_i}\,F_\perp^2(\boldsymbol{k}_1)\,F_\parallel^2(\boldsymbol{k}_1)\:,
\end{equation}
to then apply it to the monopole of the real-space LIM intrinsic power spectrum.
Adding both of these observational effects to the description of the covariances modifies the expressions slightly. The theoretical power spectrum is given by
\begin{equation}
    P^\mathrm{obs}_\mathrm{TT}(k_i) =F^2_0(k_i)\,P_\mathrm{TT}(k_i)\:,
\end{equation}
and to compute the Gaussian covariance, we replace $P_\mathrm{TT}$ by $P^\mathrm{obs}_\mathrm{TT}$ in Equation \eqref{eq:Gaussian_Noise_cov}. On the other hand, for the SSC, we can use the fact that the suppression kernels have no true cosmological dependence. This is because the transformation from angles and frequencies to distances happens for a chosen fiducial cosmology. This cosmology is also the one that appears in the equations as $H$ and $D_\mathrm{M}$. Thus 
\begin{equation}
   \frac{\mathrm{d}P_\mathrm{TT}^\mathrm{obs}(k)}{\mathrm{d}\delta_\mathrm{b}} =  \frac{\mathrm{d}P_\mathrm{TT}(k)}{\mathrm{d}\delta_\mathrm{b}}\,F^2_0(k)\:.
\end{equation}
The modified SSC can be obtained by replacing this term in Equation \eqref{eq:cov_ssc}. Finally, for the in-box non-Gaussian covariance, 
the anisotropic nature of $F_\parallel$ and $F_\perp$ would lead to modifications of the angular average performed in Equation \eqref{eq:cov_ng}. As the angular isotropy is broken, the FFTlog method of computing the integrals would not suffice, and we would need to compute the integrals by direct integration. As this is beyond the scope of this first derivation of the real-space LIM power spectrum covariance, we use $F_0$ instead as an effective treatment. This way, we recover our integrals, while the squeezed trispectrum contribution becomes
\begin{equation}
    \mathrm{Cov}_{T_0}^\mathrm{NG}(k_i,k_j)\approx\frac{V_4}{V_2^2}\,F^2_0(k_i)\,F^2_0(k_j)\,\frac{1}{N_i\,N_j}\,\sum_{\boldsymbol{k}_1\in\boldsymbol{V}_{\boldsymbol{k}_i}}\sum_{\boldsymbol{k}_2\in\boldsymbol{V}_{\boldsymbol{k}_j}}T_\mathrm{T}(\boldsymbol{k}_1,-\boldsymbol{k}_1,\boldsymbol{k}_2,-\boldsymbol{k}_2)\:.
\end{equation}

\bibliographystyle{JHEP}
\bibliography{biblio.bib}

@article{Hadzhiyska:2026wts,
    author = "Hadzhiyska, Boryana and White, Martin",
    title = "{Fewer simulations, sharper covariances: Reducing mock covariance noise with Zeldovich approximation control variates}",
    eprint = "2605.28817",
    archivePrefix = "arXiv",
    primaryClass = "astro-ph.CO",
    month = "5",
    year = "2026"
}

@article{Farina:2026kji,
    author = "Farina, Antonio and Guidi, Massimo and Veropalumbo, Alfonso and Guida, Claudio",
    title = "{Denoising clustering covariance matrices with Rotational Invariant Estimators}",
    eprint = "2604.13851",
    archivePrefix = "arXiv",
    primaryClass = "astro-ph.CO",
    month = "4",
    year = "2026"
}

@article{Maus:2026wsb,
    author = "Maus, Mark and Baleato Lizancos, Ant{\'o}n and White, Martin and de Mattia, Arnaud and Chen, Shi-Fan",
    title = "{An analytic approximation to the covariance between pre- and post-reconstruction galaxy two-point statistics}",
    eprint = "2602.12343",
    archivePrefix = "arXiv",
    primaryClass = "astro-ph.CO",
    doi = "10.1088/1475-7516/2026/05/078",
    journal = "JCAP",
    volume = "05",
    pages = "078",
    year = "2026"
}

@article{eBOSS:2020wwo,
    author = "Zhao, Cheng and others",
    collaboration = "eBOSS",
    title = "{The completed SDSS-IV extended Baryon Oscillation Spectroscopic Survey: 1000 multi-tracer mock catalogues with redshift evolution and systematics for galaxies and quasars of the final data release}",
    eprint = "2007.08997",
    archivePrefix = "arXiv",
    primaryClass = "astro-ph.CO",
    doi = "10.1093/mnras/stab510",
    journal = "Mon. Not. Roy. Astron. Soc.",
    volume = "503",
    number = "1",
    pages = "1149--1173",
    year = "2021"
}

@article{Kitaura:2015uqa,
    author = "Kitaura, Francisco-Shu and others",
    title = "{The clustering of galaxies in the SDSS-III Baryon Oscillation Spectroscopic Survey: mock galaxy catalogues for the BOSS Final Data Release}",
    eprint = "1509.06400",
    archivePrefix = "arXiv",
    primaryClass = "astro-ph.CO",
    doi = "10.1093/mnras/stv2826",
    journal = "Mon. Not. Roy. Astron. Soc.",
    volume = "456",
    number = "4",
    pages = "4156--4173",
    year = "2016"
}

@article{Ereza:2023zmz,
    author = "Ereza, Julia and Prada, Francisco and Klypin, Anatoly and Ishiyama, Tomoaki and Smith, Alex and Baugh, Carlton M. and Li, Baojiu and Hern{\'a}ndez-Aguayo, C{\'e}sar and Ruedas, Jos{\'e}",
    title = "{The Uchuu-glam BOSS and eBOSS LRG lightcones: exploring clustering and covariance errors}",
    eprint = "2311.14456",
    archivePrefix = "arXiv",
    primaryClass = "astro-ph.CO",
    doi = "10.1093/mnras/stae1543",
    journal = "Mon. Not. Roy. Astron. Soc.",
    volume = "532",
    number = "2",
    pages = "1659--1682",
    year = "2024"
}

@article{Forero-Sanchez:2024bjh,
    author = "Forero-S{\'a}nchez, D. and others",
    title = "{Analytical and EZmock covariance validation for the DESI 2024 results}",
    eprint = "2411.12027",
    archivePrefix = "arXiv",
    primaryClass = "astro-ph.CO",
    reportNumber = "FERMILAB-PUB-24-0851-PPD",
    doi = "10.1088/1475-7516/2025/04/055",
    journal = "JCAP",
    volume = "04",
    pages = "055",
    year = "2025"
}

@article{Zhao:2024xit,
    author = "Zhao, Ruiyang and Koyama, Kazuya and Wang, Yuting and Zhao, Gong-Bo",
    title = "{Modeling the Covariance Matrix for the Power Spectra Before and After the BAO Reconstruction}",
    eprint = "2410.18524",
    archivePrefix = "arXiv",
    primaryClass = "astro-ph.CO",
    doi = "10.1088/1674-4527/ad8ba1",
    journal = "Res. Astron. Astrophys.",
    volume = "24",
    number = "12",
    pages = "125015",
    year = "2024"
}

@article{Taruya:2020qoy,
    author = "Taruya, Atsushi and Nishimichi, Takahiro and Jeong, Donghui",
    title = "{Covariance of the matter power spectrum including the survey window function effect: $N$ -body simulations versus fifth-order perturbation theory on grids}",
    eprint = "2007.05504",
    archivePrefix = "arXiv",
    primaryClass = "astro-ph.CO",
    reportNumber = "YITP-20-88",
    doi = "10.1103/PhysRevD.103.023501",
    journal = "Phys. Rev. D",
    volume = "103",
    number = "2",
    pages = "023501",
    year = "2021"
}

@article{Blot:2018oxk,
    author = "Blot, Linda and others",
    title = "{Comparing approximate methods for mock catalogues and covariance matrices II: Power spectrum multipoles}",
    eprint = "1806.09497",
    archivePrefix = "arXiv",
    primaryClass = "astro-ph.CO",
    doi = "10.1093/mnras/stz507",
    journal = "Mon. Not. Roy. Astron. Soc.",
    volume = "485",
    number = "2",
    pages = "2806--2824",
    year = "2019"
}

@article{Colavincenzo:2018cgf,
    author = "Colavincenzo, Manuel and others",
    title = "{Comparing approximate methods for mock catalogues and covariance matrices {\textendash} III: bispectrum}",
    eprint = "1806.09499",
    archivePrefix = "arXiv",
    primaryClass = "astro-ph.CO",
    doi = "10.1093/mnras/sty2964",
    journal = "Mon. Not. Roy. Astron. Soc.",
    volume = "482",
    number = "4",
    pages = "4883--4905",
    year = "2019"
}

@article{Schreiner:2024grf,
    author = "Schreiner, Greg and Krolewski, Alex and Joudaki, Shahab and Percival, Will J.",
    title = "{Super sample covariance and the volume scaling of galaxy survey covariance matrices}",
    eprint = "2411.16948",
    archivePrefix = "arXiv",
    primaryClass = "astro-ph.CO",
    doi = "10.1088/1475-7516/2025/02/022",
    journal = "JCAP",
    volume = "02",
    pages = "022",
    year = "2025"
}

@article{Li:2017qgh,
    author = "Li, Yin and Schmittfull, Marcel and Seljak, Uro{\v{s}}",
    title = "{Galaxy power-spectrum responses and redshift-space super-sample effect}",
    eprint = "1711.00018",
    archivePrefix = "arXiv",
    primaryClass = "astro-ph.CO",
    doi = "10.1088/1475-7516/2018/02/022",
    journal = "JCAP",
    volume = "02",
    pages = "022",
    year = "2018"
}

@article{Scoccimarro:1999kp,
    author = "Scoccimarro, Roman and Zaldarriaga, Matias and Hui, Lam",
    title = "{Power spectrum correlations induced by nonlinear clustering}",
    eprint = "astro-ph/9901099",
    archivePrefix = "arXiv",
    reportNumber = "CITA-98-62, FERMILAB-PUB-99-003-A",
    doi = "10.1086/308059",
    journal = "Astrophys. J.",
    volume = "527",
    pages = "1",
    year = "1999"
}

@article{Chan:2017fiv,
    author = "Chan, Kwan Chuen and Moradinezhad Dizgah, Azadeh and Nore{\~n}a, Jorge",
    title = "{Bispectrum Supersample Covariance}",
    eprint = "1709.02473",
    archivePrefix = "arXiv",
    primaryClass = "astro-ph.CO",
    doi = "10.1103/PhysRevD.97.043532",
    journal = "Phys. Rev. D",
    volume = "97",
    number = "4",
    pages = "043532",
    year = "2018"
}

@article{MoradinezhadDizgah:2021dei,
    author = "Moradinezhad Dizgah, Azadeh and Nikakhtar, Farnik and Keating, Garrett K. and Castorina, Emanuele",
    title = "{Precision tests of CO and  CII  power spectra models against simulated intensity maps}",
    eprint = "2111.03717",
    archivePrefix = "arXiv",
    primaryClass = "astro-ph.CO",
    doi = "10.1088/1475-7516/2022/02/026",
    journal = "JCAP",
    volume = "02",
    number = "02",
    pages = "026",
    year = "2022"
}

@ARTICLE{Lidz2008ApJ...680..962L,
       author = {{Lidz}, Adam and {Zahn}, Oliver and {McQuinn}, Matthew and {Zaldarriaga}, Matias and {Hernquist}, Lars},
        title = "{Detecting the Rise and Fall of 21 cm Fluctuations with the Murchison Widefield Array}",
      journal = {\apj},
         year = 2008,
        month = jun,
       volume = {680},
       number = {2},
        pages = {962-974},
          doi = {10.1086/587618},
archivePrefix = {arXiv},
       eprint = {0711.4373},
 primaryClass = {astro-ph},
       adsurl = {https://ui.adsabs.harvard.edu/abs/2008ApJ...680..962L}
}

@article{Pierre_Christian_2013,
doi = {10.1088/1475-7516/2013/09/014},
url = {https://doi.org/10.1088/1475-7516/2013/09/014},
year = {2013},
month = {sep},
publisher = {},
volume = {2013},
number = {09},
pages = {014},
author = {Pierre Christian and Abraham Loeb},
title = {Measuring the X-ray background in the reionization era with first generation 21 cm experiments},
journal = {Journal of Cosmology and Astroparticle Physics}
}

@article{Sinigaglia:2026pqe,
    author = "Sinigaglia, Francesco and Kitaura, Francisco-Shu",
    title = "{Cosmic variance or galaxy bias? Disentangling finite-volume and galaxy formation effects in cosmological analysis}",
    eprint = "2606.04830",
    archivePrefix = "arXiv",
    primaryClass = "astro-ph.CO",
    doi = "10.1103/1dqk-pq4q",
    month = "6",
    year = "2026"
}

@article{ concerto_overview,
	author = {{The CONCERTO Collaboration} and {Ade, P.} and {Aravena, M.} and {Barria, E.} and {Beelen, A.} and {Benoit, A.} and {Béthermin, M.} and {Bounmy, J.} and {Bourrion, O.} and {Bres, G.} and {De Breuck, C.} and {Calvo, M.} and {Cao, Y.} and {Catalano, A.} and {Désert, F.-X.} and {Durán, C.A.} and {Fasano, A.} and {Fenouillet, T.} and {Garcia, J.} and {Garde, G.} and {Goupy, J.} and {Groppi, C.} and {Hoarau, C.} and {Lagache, G.} and {Lambert, J.-C.} and {Leggeri, J.-P.} and {Levy-Bertrand, F.} and {Macías-Pérez, J.} and {Mani, H.} and {Marpaud, J.} and {Mauskopf, P.} and {Monfardini, A.} and {Pisano, G.} and {Ponthieu, N.} and {Prieur, L.} and {Roni, S.} and {Roudier, S.} and {Tourres, D.} and {Tucker, C.}},
	title = {A wide field-of-view low-resolution spectrometer at APEX: Instrument design and scientific forecast},
	DOI= "10.1051/0004-6361/202038456",
	url= "https://doi.org/10.1051/0004-6361/202038456",
	journal = {A\&A},
	year = 2020,
	volume = 642,
	pages = "A60",
}

@article{10.1117/1.JATIS.7.4.044004,
author = {Eric R. Switzer and Emily M. Barrentine and Giuseppe Cataldo and Thomas M. Essinger-Hileman and Peter A. R. Ade and Christopher J. Anderson and Alyssa Barlis and Jeffrey W. Beeman and Nicholas G. Bellis and Alberto D. Bolatto and Patrick C. Breysse and Berhanu T. Bulcha and Lee-Roger Chevres-Fernanadez and Chullhee Cho and Jake A. Connors and Negar Ehsan and Jason Glenn and Joseph E. Golec and James P. Hays-Wehle and Larry A. Hess and Amir E. Jahromi and Trevian Jenkins and Mark O. Kimball and Alan J. Kogut and Luke N. Lowe and Philip D. Mauskopf and Jeffrey J. McMahon and Mona Mirzaei and Harvey Moseley and Jonas W. Mugge-Durum and Omid Noroozian and Trevor M. Oxholm and Tatsat Parekh and Ue-Li Pen and Anthony R. Pullen and Maryam Rahmani and Mathias Ramirez and Florian Roselli and Konrad Shire and Gage L. Siebert and Adrian K. Sinclair and Rachel S. Somerville and Ryan C. Stephenson and Thomas R. Stevenson and Peter T. Timbie and Jared Termini and Justin Trenkamp and Carole E. Tucker and Elijah Visbal and Carolyn G. Volpert and Edward J. Wollack and Shengqi Yang and L. Y. Aaron Yung},
title = {{Experiment for cryogenic large-aperture intensity mapping: instrument design}},
volume = {7},
journal = {Journal of Astronomical Telescopes, Instruments, and Systems},
number = {4},
publisher = {SPIE},
pages = {044004},
year = {2021},
doi = {10.1117/1.JATIS.7.4.044004},
URL = {https://doi.org/10.1117/1.JATIS.7.4.044004}
}

@article{LujanNiemeyer:2024dyv,
    author = "Lujan Niemeyer, Maja",
    title = "{Effect of Ly{\ensuremath{\alpha}} Radiative Transfer on Intensity Mapping Power Spectra}",
    eprint = "2407.03060",
    archivePrefix = "arXiv",
    primaryClass = "astro-ph.CO",
    doi = "10.3847/1538-4357/ada8a3",
    journal = "Astrophys. J.",
    volume = "980",
    number = "2",
    pages = "250",
    year = "2025"
}

@article{LujanNiemeyer:2022cte,
    author = "Lujan Niemeyer, Maja and others",
    title = "{Ly{\ensuremath{\alpha}} Halos around [O iii]-selected Galaxies in HETDEX}",
    eprint = "2207.11098",
    archivePrefix = "arXiv",
    primaryClass = "astro-ph.GA",
    doi = "10.3847/2041-8213/ac82e5",
    journal = "Astrophys. J. Lett.",
    volume = "934",
    number = "2",
    pages = "L26",
    year = "2022"
}

@article{LujanNiemeyer:2022rby,
    author = "Lujan Niemeyer, Maja and others",
    title = "{Surface Brightness Profile of Lyman-{\ensuremath{\alpha}} Halos out to 320 kpc in HETDEX}",
    eprint = "2203.04826",
    archivePrefix = "arXiv",
    primaryClass = "astro-ph.GA",
    doi = "10.3847/1538-4357/ac5cb8",
    journal = "Astrophys. J.",
    volume = "929",
    number = "1",
    pages = "90",
    year = "2022"
}

@ARTICLE{silva2015ApJ...806..209S,
       author = {{Silva}, Marta and {Santos}, Mario G. and {Cooray}, Asantha and {Gong}, Yan},
        title = "{Prospects for Detecting C II Emission during the Epoch of Reionization}",
      journal = {\apj},
         year = 2015,
        month = jun,
       volume = {806},
       number = {2},
          eid = {209},
        pages = {209},
          doi = {10.1088/0004-637X/806/2/209},
archivePrefix = {arXiv},
       eprint = {1410.4808},
 primaryClass = {astro-ph.GA},
       adsurl = {https://ui.adsabs.harvard.edu/abs/2015ApJ...806..209S}
}

@ARTICLE{2001MNRAS.323....1S,
       author = {{Sheth}, Ravi K. and {Mo}, H.~J. and {Tormen}, Giuseppe},
        title = "{Ellipsoidal collapse and an improved model for the number and spatial distribution of dark matter haloes}",
      journal = {\mnras},
         year = 2001,
        month = may,
       volume = {323},
       number = {1},
        pages = {1-12},
          doi = {10.1046/j.1365-8711.2001.04006.x},
archivePrefix = {arXiv},
       eprint = {astro-ph/9907024},
 primaryClass = {astro-ph},
       adsurl = {https://ui.adsabs.harvard.edu/abs/2001MNRAS.323....1S}
}

@ARTICLE{nlbiases_2016JCAP...02..018L,
       author = {{Lazeyras}, Titouan and {Wagner}, Christian and {Baldauf}, Tobias and {Schmidt}, Fabian},
        title = "{Precision measurement of the local bias of dark matter halos}",
      journal = {\jcap},
         year = 2016,
        month = feb,
       volume = {2016},
       number = {2},
        pages = {018-018},
          doi = {10.1088/1475-7516/2016/02/018},
archivePrefix = {arXiv},
       eprint = {1511.01096},
 primaryClass = {astro-ph.CO},
       adsurl = {https://ui.adsabs.harvard.edu/abs/2016JCAP...02..018L}
}

@article{Bernal_2022,
   title={Line-intensity mapping: theory review with a focus on star-formation lines},
   volume={30},
   ISSN={1432-0754},
   url={http://dx.doi.org/10.1007/s00159-022-00143-0},
   DOI={10.1007/s00159-022-00143-0},
   number={1},
   journal={The Astronomy and Astrophysics Review},
   publisher={Springer Science and Business Media LLC},
   author={Bernal, José Luis and Kovetz, Ely D.},
   year={2022},
   month=sep }

@article{CHIME:2023til,
    author = "Amiri, Mandana and others",
    collaboration = "CHIME",
    title = "{A Detection of Cosmological 21 cm Emission from CHIME in Cross-correlation with eBOSS Measurements of the Ly{\ensuremath{\alpha}} Forest}",
    eprint = "2309.04404",
    archivePrefix = "arXiv",
    primaryClass = "astro-ph.CO",
    doi = "10.3847/1538-4357/ad0f1d",
    journal = "Astrophys. J.",
    volume = "963",
    number = "1",
    pages = "23",
    year = "2024"
}

@article{SKA:2018ckk,
    author = "Bacon, David J. and others",
    collaboration = "SKA",
    title = "{Cosmology with Phase 1 of the Square Kilometre Array: Red Book 2018: Technical specifications and performance forecasts}",
    eprint = "1811.02743",
    archivePrefix = "arXiv",
    primaryClass = "astro-ph.CO",
    doi = "10.1017/pasa.2019.51",
    journal = "Publ. Astron. Soc. Austral.",
    volume = "37",
    pages = "e007",
    year = "2020"
}

@article{euclid_ssc,
	author = {{Euclid Collaboration} and {Sciotti, D.} and {Gouyou Beauchamps, S.} and {Cardone, V. F.} and {Camera, S.} and {Tutusaus, I.} and {Lacasa, F.} and {Barreira, A.} and {Bonici, M.} and {Gorce, A.} and {Aubert, M.} and {Baratta, P.} and {Upham, R. E.} and {Carbone, C.} and {Casas, S.} and {Ilić, S.} and {Martinelli, M.} and {Sakr, Z.} and {Schneider, A.} and {Maoli, R.} and {Scaramella, R.} and {Escoffier, S.} and {Gillard, W.} and {Aghanim, N.} and {Amara, A.} and {Andreon, S.} and {Auricchio, N.} and {Baccigalupi, C.} and {Baldi, M.} and {Bardelli, S.} and {Bernardeau, F.} and {Bonino, D.} and {Branchini, E.} and {Brescia, M.} and {Brinchmann, J.} and {Capobianco, V.} and {Carretero, J.} and {Castander, F. J.} and {Castellano, M.} and {Castignani, G.} and {Cavuoti, S.} and {Cimatti, A.} and {Cledassou, R.} and {Colodro-Conde, C.} and {Congedo, G.} and {Conselice, C. J.} and {Conversi, L.} and {Copin, Y.} and {Corcione, L.} and {Courbin, F.} and {Courtois, H. M.} and {Cropper, M.} and {Da Silva, A.} and {Degaudenzi, H.} and {De Lucia, G.} and {Dinis, J.} and {Dubath, F.} and {Dupac, X.} and {Dusini, S.} and {Farina, M.} and {Farrens, S.} and {Fosalba, P.} and {Frailis, M.} and {Franceschi, E.} and {Fumana, M.} and {Galeotta, S.} and {Garilli, B.} and {Gillis, B.} and {Giocoli, C.} and {Grazian, A.} and {Grupp, F.} and {Guzzo, L.} and {Haugan, S. V. H.} and {Holmes, W.} and {Hook, I.} and {Hormuth, F.} and {Hornstrup, A.} and {Hudelot, P.} and {Jahnke, K.} and {Joachimi, B.} and {Keihänen, E.} and {Kermiche, S.} and {Kiessling, A.} and {Kunz, M.} and {Kurki-Suonio, H.} and {Lilje, P. B.} and {Lindholm, V.} and {Lloro, I.} and {Mainetti, G.} and {Maino, D.} and {Mansutti, O.} and {Marggraf, O.} and {Markovic, K.} and {Martinet, N.} and {Marulli, F.} and {Massey, R.} and {Maurogordato, S.} and {Medinaceli, E.} and {Mei, S.} and {Mellier, Y.} and {Meneghetti, M.} and {Meylan, G.} and {Moresco, M.} and {Moscardini, L.} and {Munari, E.} and {Neissner, C.} and {Niemi, S.-M.} and {Padilla, C.} and {Paltani, S.} and {Pasian, F.} and {Pedersen, K.} and {Pettorino, V.} and {Pires, S.} and {Polenta, G.} and {Poncet, M.} and {Popa, L. A.} and {Raison, F.} and {Rebolo, R.} and {Renzi, A.} and {Rhodes, J.} and {Riccio, G.} and {Romelli, E.} and {Roncarelli, M.} and {Saglia, R.} and {Sánchez, A. G.} and {Sapone, D.} and {Sartoris, B.} and {Schirmer, M.} and {Schneider, P.} and {Secroun, A.} and {Sefusatti, E.} and {Seidel, G.} and {Serrano, S.} and {Sirignano, C.} and {Sirri, G.} and {Stanco, L.} and {Starck, J.-L.} and {Steinwagner, J.} and {Tallada-Crespí, P.} and {Taylor, A. N.} and {Tereno, I.} and {Toledo-Moreo, R.} and {Torradeflot, F.} and {Valentijn, E. A.} and {Valenziano, L.} and {Vassallo, T.} and {Veropalumbo, A.} and {Wang, Y.} and {Weller, J.} and {Zacchei, A.} and {Zamorani, G.} and {Zoubian, J.} and {Zucca, E.} and {Biviano, A.} and {Boucaud, A.} and {Bozzo, E.} and {Di Ferdinando, D.} and {Farinelli, R.} and {Graciá-Carpio, J.} and {Mauri, N.} and {Scottez, V.} and {Tenti, M.} and {Akrami, Y.} and {Allevato, V.} and {Ballardini, M.} and {Blanchard, A.} and {Borgani, S.} and {Borlaff, A. S.} and {Burigana, C.} and {Cabanac, R.} and {Cappi, A.} and {Carvalho, C. S.} and {Castro, T.} and {Cañas-Herrera, G.} and {Chambers, K. C.} and {Cooray, A. R.} and {Coupon, J.} and {Davini, S.} and {Desprez, G.} and {Díaz-Sánchez, A.} and {Di Domizio, S.} and {Escartin Vigo, J. A.} and {Ferrero, I.} and {Finelli, F.} and {Gabarra, L.} and {Ganga, K.} and {Garcia-Bellido, J.} and {Gaztanaga, E.} and {Giacomini, F.} and {Gozaliasl, G.} and {Hildebrandt, H.} and {Jacobson, J.} and {Kajava, J. J. E.} and {Kansal, V.} and {Kirkpatrick, C. C.} and {Legrand, L.} and {Loureiro, A.} and {Macias-Perez, J.} and {Magliocchetti, M.} and {Martins, C. J. A. P.} and {Matthew, S.} and {Maurin, L.} and {Metcalf, R. B.} and {Migliaccio, M.} and {Monaco, P.} and {Morgante, G.} and {Nadathur, S.} and {Nucita, A. A.} and {Patrizii, L.} and {Pöntinen, M.} and {Popa, V.} and {Porciani, C.} and {Potter, D.} and {Pourtsidou, A.} and {Sereno, M.} and {Simon, P.} and {Spurio Mancini, A.} and {Stadel, J.} and {Teyssier, R.} and {Toft, S.} and {Tucci, M.} and {Valieri, C.} and {Valiviita, J.} and {Viel, M.}},
	title = {Euclid preparation - LII. Forecast impact of super-sample covariance on 3×2pt analysis with Euclid},
	DOI= "10.1051/0004-6361/202348389",
	url= "https://doi.org/10.1051/0004-6361/202348389",
	journal = {A\&A},
	year = 2024,
	volume = 691,
	pages = "A318",
}

@article{Wadekar:2019rdu,
    author = "Wadekar, Digvijay and Scoccimarro, Roman",
    title = "{Galaxy power spectrum multipoles covariance in perturbation theory}",
    eprint = "1910.02914",
    archivePrefix = "arXiv",
    primaryClass = "astro-ph.CO",
    doi = "10.1103/PhysRevD.102.123517",
    journal = "Phys. Rev. D",
    volume = "102",
    number = "12",
    pages = "123517",
    year = "2020"
}

@article{dePutter:2011ah,
    author = "de Putter, Roland and Wagner, Christian and Mena, Olga and Verde, Licia and Percival, Will",
    title = "{Thinking Outside the Box: Effects of Modes Larger than the Survey on Matter Power Spectrum Covariance}",
    eprint = "1111.6596",
    archivePrefix = "arXiv",
    primaryClass = "astro-ph.CO",
    doi = "10.1088/1475-7516/2012/04/019",
    journal = "JCAP",
    volume = "04",
    pages = "019",
    year = "2012"
}

@article{Kobayashi:2023vpu,
    author = "Kobayashi, Yosuke",
    title = "{Fast computation of the non-Gaussian covariance of redshift-space galaxy power spectrum multipoles}",
    eprint = "2308.08593",
    archivePrefix = "arXiv",
    primaryClass = "astro-ph.CO",
    doi = "10.1103/PhysRevD.108.103512",
    journal = "Phys. Rev. D",
    volume = "108",
    number = "10",
    pages = "103512",
    year = "2023"
}

@article{Takada_2013,
   title={Power spectrum super-sample covariance},
   volume={87},
   ISSN={1550-2368},
   url={http://dx.doi.org/10.1103/PhysRevD.87.123504},
   DOI={10.1103/physrevd.87.123504},
   number={12},
   journal={Physical Review D},
   publisher={American Physical Society (APS)},
   author={Takada, Masahiro and Hu, Wayne},
   year={2013},
   month=jun }

@article{COMAP:2021rny,
    author = "Chung, Dongwoo T. and others",
    collaboration = "COMAP",
    title = "{A Model of Spectral Line Broadening in Signal Forecasts for Line-intensity Mapping Experiments}",
    eprint = "2104.11171",
    archivePrefix = "arXiv",
    primaryClass = "astro-ph.CO",
    doi = "10.3847/1538-4357/ac2a35",
    journal = "Astrophys. J.",
    volume = "923",
    number = "2",
    pages = "188",
    year = "2021"
}

@article{Eggemeier:2018qae,
    author = "Eggemeier, Alexander and Scoccimarro, Roman and Smith, Robert E.",
    title = "{Bias Loop Corrections to the Galaxy Bispectrum}",
    eprint = "1812.03208",
    archivePrefix = "arXiv",
    primaryClass = "astro-ph.CO",
    doi = "10.1103/PhysRevD.99.123514",
    journal = "Phys. Rev. D",
    volume = "99",
    number = "12",
    pages = "123514",
    year = "2019"
}

@ARTICLE{G20,
       author = {{Gkogkou}, A. and {B{\'e}thermin}, M. and {Lagache}, G. and {Van Cuyck}, M. and {Jullo}, E. and {Aravena}, M. and {Beelen}, A. and {Benoit}, A. and {Bounmy}, J. and {Calvo}, M. and {Catalano}, A. and {Cora}, S. and {Croton}, D. and {de la Torre}, S. and {Fasano}, A. and {Ferrara}, A. and {Goupy}, J. and {Hoarau}, C. and {Hu}, W. and {Ishiyama}, T. and {Knudsen}, K.~K. and {Lambert}, J.-C. and {Mac{\'\i}as-P{\'e}rez}, J.~F. and {Marpaud}, J. and {Mellema}, G. and {Monfardini}, A. and {Pallottini}, A. and {Ponthieu}, N. and {Prada}, F. and {Roehlly}, Y. and {Vallini}, L. and {Walter}, F.},
        title = "{CONCERTO: Simulating the CO, [CII], and [CI] line emission of galaxies in a 117 deg$^{2}$ field and the impact of field-to-field variance}",
      journal = {Astronomy \& Astrophysics},
         year = 2023,
        month = feb,
       volume = {670},
          eid = {A16},
        pages = {A16},
          doi = {10.1051/0004-6361/202245151},
archivePrefix = {arXiv},
       eprint = {2212.02235},
 primaryClass = {astro-ph.CO},
       adsurl = {https://ui.adsabs.harvard.edu/abs/2023A&A...670A..16G}
}

@article{Chang:2026ake,
    author = "Chang, Tzu-Ching and Lidz, Adam",
    title = "{Line-Intensity Mapping}",
    eprint = "2602.03011",
    archivePrefix = "arXiv",
    primaryClass = "astro-ph.CO",
    month = "2",
    year = "2026"
}

@ARTICLE{2021MNRAS.502.1401M,
       author = {{Mead}, A.~J. and {Brieden}, S. and {Tr{\"o}ster}, T. and {Heymans}, C.},
        title = "{HMCODE-2020: improved modelling of non-linear cosmological power spectra with baryonic feedback}",
      journal = {\mnras},
         year = 2021,
        month = mar,
       volume = {502},
       number = {1},
        pages = {1401-1422},
          doi = {10.1093/mnras/stab082},
archivePrefix = {arXiv},
       eprint = {2009.01858},
 primaryClass = {astro-ph.CO},
       adsurl = {https://ui.adsabs.harvard.edu/abs/2021MNRAS.502.1401M}
}

@article{Asgari_2023,
   title={The halo model for cosmology: a pedagogical review},
   volume={6},
   ISSN={2565-6120},
   url={http://dx.doi.org/10.21105/astro.2303.08752},
   DOI={10.21105/astro.2303.08752},
   journal={The Open Journal of Astrophysics},
   publisher={Maynooth University},
   author={Asgari, Marika and Mead, Alexander J. and Heymans, Catherine},
   year={2023},
   month=nov }

@ARTICLE{1996ApJ...462..563N,
       author = {{Navarro}, Julio F. and {Frenk}, Carlos S. and {White}, Simon D.~M.},
        title = "{The Structure of Cold Dark Matter Halos}",
      journal = {\apj},
         year = 1996,
        month = may,
       volume = {462},
        pages = {563},
          doi = {10.1086/177173},
archivePrefix = {arXiv},
       eprint = {astro-ph/9508025},
 primaryClass = {astro-ph},
       adsurl = {https://ui.adsabs.harvard.edu/abs/1996ApJ...462..563N}
}

@ARTICLE{conc2019ApJ...871..168D,
       author = {{Diemer}, Benedikt and {Joyce}, Michael},
        title = "{An Accurate Physical Model for Halo Concentrations}",
      journal = {\apj},
         year = 2019,
        month = feb,
       volume = {871},
       number = {2},
          eid = {168},
        pages = {168},
          doi = {10.3847/1538-4357/aafad6},
archivePrefix = {arXiv},
       eprint = {1809.07326},
 primaryClass = {astro-ph.CO},
       adsurl = {https://ui.adsabs.harvard.edu/abs/2019ApJ...871..168D}
}

@article{COORAY_2002,
   title={Halo models of large scale structure},
   volume={372},
   ISSN={0370-1573},
   url={http://dx.doi.org/10.1016/S0370-1573(02)00276-4},
   DOI={10.1016/s0370-1573(02)00276-4},
   number={1},
   journal={Physics Reports},
   publisher={Elsevier BV},
   author={COORAY, A and SHETH, R},
   year={2002},
   month=dec, pages={1–129} }

@ARTICLE{coevohalo2018JCAP...07..029A,
       author = {{Abidi}, Muntazir Mehdi and {Baldauf}, Tobias},
        title = "{Cubic halo bias in Eulerian and Lagrangian space}",
      journal = {\jcap},
         year = 2018,
        month = jul,
       volume = {2018},
       number = {7},
          eid = {029},
        pages = {029},
          doi = {10.1088/1475-7516/2018/07/029},
archivePrefix = {arXiv},
       eprint = {1802.07622},
 primaryClass = {astro-ph.CO},
       adsurl = {https://ui.adsabs.harvard.edu/abs/2018JCAP...07..029A}
}

@misc{li2024modelingnonlinearpowerspectrum,
      title={Modeling the Nonlinear Power Spectrum in Low-redshift HI Intensity Mapping}, 
      author={Zhixing Li and Laura Wolz and Hong Guo and Steven Cunnington and Yi Mao},
      year={2024},
      eprint={2407.02131},
      archivePrefix={arXiv},
      primaryClass={astro-ph.CO},
      url={https://arxiv.org/abs/2407.02131}, 
}

@ARTICLE{sfr2013ApJ...762L..31B,
       author = {{Behroozi}, Peter S. and {Wechsler}, Risa H. and {Conroy}, Charlie},
        title = "{On the Lack of Evolution in Galaxy Star Formation Efficiency}",
      journal = {\apjl},
         year = 2013,
        month = jan,
       volume = {762},
       number = {2},
          eid = {L31},
        pages = {L31},
          doi = {10.1088/2041-8205/762/2/L31},
archivePrefix = {arXiv},
       eprint = {1209.3013},
 primaryClass = {astro-ph.CO},
       adsurl = {https://ui.adsabs.harvard.edu/abs/2013ApJ...762L..31B}
}

@ARTICLE{sfr2013ApJ...770...57B,
       author = {{Behroozi}, Peter S. and {Wechsler}, Risa H. and {Conroy}, Charlie},
        title = "{The Average Star Formation Histories of Galaxies in Dark Matter Halos from z = 0-8}",
      journal = {\apj},
         year = 2013,
        month = jun,
       volume = {770},
       number = {1},
          eid = {57},
        pages = {57},
          doi = {10.1088/0004-637X/770/1/57},
archivePrefix = {arXiv},
       eprint = {1207.6105},
 primaryClass = {astro-ph.CO},
       adsurl = {https://ui.adsabs.harvard.edu/abs/2013ApJ...770...57B}
}

@ARTICLE{spt_breakdown2014JCAP...01..010B,
       author = {{Blas}, Diego and {Garny}, Mathias and {Konstandin}, Thomas},
        title = "{Cosmological perturbation theory at three-loop order}",
      journal = {\jcap},
         year = 2014,
        month = jan,
       volume = {2014},
       number = {1},
          eid = {010},
        pages = {010},
          doi = {10.1088/1475-7516/2014/01/010},
archivePrefix = {arXiv},
       eprint = {1309.3308},
 primaryClass = {astro-ph.CO},
       adsurl = {https://ui.adsabs.harvard.edu/abs/2014JCAP...01..010B}
}

@article{Euclid:2025fby,
    author = "Euclid colaboration: Naidoo, K. and others",
    collaboration = "Euclid",
    title = "{Euclid preparation - LXXXIX. Accurate and precise data-driven angular power spectrum covariances}",
    eprint = "2506.09118",
    archivePrefix = "arXiv",
    primaryClass = "astro-ph.CO",
    doi = "10.1051/0004-6361/202555893",
    journal = "Astron. Astrophys.",
    volume = "708",
    pages = "A167",
    year = "2026"
}

@article{Mohammad:2021aqc,
    author = "Mohammad, Faizan G. and Percival, Will J.",
    title = "{Creating jackknife and bootstrap estimates of the covariance matrix for the two-point correlation function}",
    eprint = "2109.07071",
    archivePrefix = "arXiv",
    primaryClass = "astro-ph.CO",
    doi = "10.1093/mnras/stac1458",
    journal = "Mon. Not. Roy. Astron. Soc.",
    volume = "514",
    number = "1",
    pages = "1289--1301",
    year = "2022"
}

@article{Ibanez:2026kuk,
    author = "Ib{\'a}{\~n}ez, Marcos Pellejero and Alonso, David and Peacock, John A. and Zennaro, Matteo and Brieden, Samuel",
    title = "{CHEFT: A Hybrid Effective Field Theory halo model}",
    eprint = "2607.09571",
    archivePrefix = "arXiv",
    primaryClass = "astro-ph.CO",
    month = "7",
    year = "2026"
}

@article{Norberg:2008tg,
    author = "Norberg, Peder and Baugh, Carlton M. and Gaztanaga, Enrique and Croton, Darren J.",
    title = "{Statistical Analysis of Galaxy Surveys - I. Robust error estimation for 2-point clustering statistics}",
    eprint = "0810.1885",
    archivePrefix = "arXiv",
    primaryClass = "astro-ph",
    doi = "10.1111/j.1365-2966.2009.14389.x",
    journal = "Mon. Not. Roy. Astron. Soc.",
    volume = "396",
    pages = "19",
    year = "2009"
}

@article{Sarkar:2026rfv,
    author = "Sarkar, Debanjan and Foreman, Simon",
    title = "{Simulation-Based Priors for HI Bias from Halo Occupation Physics}",
    eprint = "2607.14206",
    archivePrefix = "arXiv",
    primaryClass = "astro-ph.CO",
    month = "7",
    year = "2026"
}

@article{Escoffier:2016qnf,
    author = "Escoffier, S. and Cousinou, M. -C. and Tilquin, A. and Pisani, A. and Aguichine, A. and de la Torre, S. and Ealet, A. and Gillard, W. and Jullo, E.",
    title = "{Jackknife resampling technique on mocks: an alternative method for covariance matrix estimation}",
    eprint = "1606.00233",
    archivePrefix = "arXiv",
    primaryClass = "astro-ph.CO",
    month = "6",
    year = "2016"
}

@article{Baumann_EFT_2012,
doi = {10.1088/1475-7516/2012/07/051},
url = {https://doi.org/10.1088/1475-7516/2012/07/051},
year = {2012},
month = {jul},
publisher = {},
volume = {2012},
number = {07},
pages = {051},
author = {Daniel Baumann and Alberto Nicolis and Leonardo Senatore and Matias Zaldarriaga},
title = {Cosmological non-linearities as an effective fluid},
journal = {Journal of Cosmology and Astroparticle Physics}
}

@ARTICLE{HI_powerlaw...05..004O,
       author = {{Obuljen}, Andrej and {Castorina}, Emanuele and {Villaescusa-Navarro}, Francisco and {Viel}, Matteo},
        title = "{High-redshift post-reionization cosmology with 21cm intensity mapping}",
      journal = {\jcap},
         year = 2018,
        month = may,
       volume = {2018},
       number = {5},
          eid = {004},
        pages = {004},
          doi = {10.1088/1475-7516/2018/05/004},
archivePrefix = {arXiv},
       eprint = {1709.07893},
 primaryClass = {astro-ph.CO},
       adsurl = {https://ui.adsabs.harvard.edu/abs/2018JCAP...05..004O}
}

@article{Obuljen:2018kdy,
    author = "Obuljen, Andrej and Alonso, David and Villaescusa-Navarro, Francisco and Yoon, Ilsang and Jones, Michael",
    title = "{The H I content of dark matter haloes at z {\ensuremath{\approx}} 0 from ALFALFA}",
    eprint = "1805.00934",
    archivePrefix = "arXiv",
    primaryClass = "astro-ph.CO",
    doi = "10.1093/mnras/stz1118",
    journal = "Mon. Not. Roy. Astron. Soc.",
    volume = "486",
    number = "4",
    pages = "5124--5138",
    year = "2019"
}

@ARTICLE{CCAT-p_specs,
       author = {{CCAT-Prime Collaboration} and {Aravena}, Manuel and {Austermann}, Jason E. and {Basu}, Kaustuv and {Battaglia}, Nicholas and {Beringue}, Benjamin and {Bertoldi}, Frank and {Bigiel}, Frank and {Bond}, J. Richard and {Breysse}, Patrick C. and {Broughton}, Colton and {Bustos}, Ricardo and {Chapman}, Scott C. and {Charmetant}, Maude and {Choi}, Steve K. and {Chung}, Dongwoo T. and {Clark}, Susan E. and {Cothard}, Nicholas F. and {Crites}, Abigail T. and {Dev}, Ankur and {Douglas}, Kaela and {Duell}, Cody J. and {D{\"u}nner}, Rolando and {Ebina}, Haruki and {Erler}, Jens and {Fich}, Michel and {Fissel}, Laura M. and {Foreman}, Simon and {Freundt}, R.~G. and {Gallardo}, Patricio A. and {Gao}, Jiansong and {Garc{\'\i}a}, Pablo and {Giovanelli}, Riccardo and {Golec}, Joseph E. and {Groppi}, Christopher E. and {Haynes}, Martha P. and {Henke}, Douglas and {Hensley}, Brandon and {Herter}, Terry and {Higgins}, Ronan and {Hlo{\v{z}}ek}, Ren{\'e}e and {Huber}, Anthony and {Huber}, Zachary and {Hubmayr}, Johannes and {Jackson}, Rebecca and {Johnstone}, Douglas and {Karoumpis}, Christos and {Keating}, Laura C. and {Komatsu}, Eiichiro and {Li}, Yaqiong and {Magnelli}, Benjamin and {Matthews}, Brenda C. and {Mauskopf}, Philip D. and {McMahon}, Jeffrey J. and {Meerburg}, P. Daniel and {Meyers}, Joel and {Muralidhara}, Vyoma and {Murray}, Norman W. and {Niemack}, Michael D. and {Nikola}, Thomas and {Okada}, Yoko and {Puddu}, Roberto and {Riechers}, Dominik A. and {Rosolowsky}, Erik and {Rossi}, Kayla and {Rotermund}, Kaja and {Roy}, Anirban and {Sadavoy}, Sarah I. and {Schaaf}, Reinhold and {Schilke}, Peter and {Scott}, Douglas and {Simon}, Robert and {Sinclair}, Adrian K. and {Sivakoff}, Gregory R. and {Stacey}, Gordon J. and {Stutz}, Amelia M. and {Stutzki}, Juergen and {Tahani}, Mehrnoosh and {Thanjavur}, Karun and {Timmermann}, Ralf A. and {Ullom}, Joel N. and {van Engelen}, Alexander and {Vavagiakis}, Eve M. and {Vissers}, Michael R. and {Wheeler}, Jordan D. and {White}, Simon D.~M. and {Zhu}, Yijie and {Zou}, Bugao},
        title = "{CCAT-prime Collaboration: Science Goals and Forecasts with Prime-Cam on the Fred Young Submillimeter Telescope}",
      journal = {The Astrophysical Journal Supplement Series},
         year = 2023,
        month = jan,
       volume = {264},
       number = {1},
          eid = {7},
        pages = {7},
          doi = {10.3847/1538-4365/ac9838},
archivePrefix = {arXiv},
       eprint = {2107.10364},
 primaryClass = {astro-ph.CO},
       adsurl = {https://ui.adsabs.harvard.edu/abs/2023ApJS..264....7C}
}

@ARTICLE{2018PhR...733....1D,
       author = {{Desjacques}, Vincent and {Jeong}, Donghui and {Schmidt}, Fabian},
        title = "{Large-scale galaxy bias}",
      journal = {\physrep},
         year = 2018,
        month = feb,
       volume = {733},
        pages = {1-193},
          doi = {10.1016/j.physrep.2017.12.002},
archivePrefix = {arXiv},
       eprint = {1611.09787},
 primaryClass = {astro-ph.CO},
       adsurl = {https://ui.adsabs.harvard.edu/abs/2018PhR...733....1D}
}

@article{Barreira:2017kxd,
    author = "Barreira, Alexandre and Schmidt, Fabian",
    title = "{Response Approach to the Matter Power Spectrum Covariance}",
    eprint = "1705.01092",
    archivePrefix = "arXiv",
    primaryClass = "astro-ph.CO",
    doi = "10.1088/1475-7516/2017/11/051",
    journal = "JCAP",
    volume = "11",
    pages = "051",
    year = "2017"
}

@article{Barreira:2017fjz,
    author = "Barreira, Alexandre and Krause, Elisabeth and Schmidt, Fabian",
    title = "{Complete super-sample lensing covariance in the response approach}",
    eprint = "1711.07467",
    archivePrefix = "arXiv",
    primaryClass = "astro-ph.CO",
    doi = "10.1088/1475-7516/2018/06/015",
    journal = "JCAP",
    volume = "06",
    pages = "015",
    year = "2018"
}

@article{Akitsu:2016leq,
    author = "Akitsu, Kazuyuki and Takada, Masahiro and Li, Yin",
    title = "{Large-scale tidal effect on redshift-space power spectrum in a finite-volume survey}",
    eprint = "1611.04723",
    archivePrefix = "arXiv",
    primaryClass = "astro-ph.CO",
    doi = "10.1103/PhysRevD.95.083522",
    journal = "Phys. Rev. D",
    volume = "95",
    number = "8",
    pages = "083522",
    year = "2017"
}

@article{Castorina:2020blr,
    author = "Castorina, Emanuele and Moradinezhad Dizgah, Azadeh",
    title = "{Local Primordial Non-Gaussianities and Super-Sample Variance}",
    eprint = "2005.14677",
    archivePrefix = "arXiv",
    primaryClass = "astro-ph.CO",
    reportNumber = "CERN-TH-2020-083, CERN-TH-2020-083 CERN-TH-2020-083",
    doi = "10.1088/1475-7516/2020/10/007",
    journal = "JCAP",
    volume = "10",
    pages = "007",
    year = "2020"
}

@ARTICLE{kids_cov,
       author = {{Reischke}, Robert and {Unruh}, Sandra and {Asgari}, Marika and {Dvornik}, Andrej and {Hildebrandt}, Hendrik and {Joachimi}, Benjamin and {Porth}, Lucas and {von Wietersheim-Kramsta}, Maximilian and {van den Busch}, Jan Luca and {St{\"o}lzner}, Benjamin and {Wright}, Angus H. and {Yan}, Ziang and {Bilicki}, Maciej and {Burger}, Pierre and {Chisari}, Nora Elisa and {Harnois-D{\'e}raps}, Joachim and {Georgiou}, Christos and {Heymans}, Catherine and {Jalan}, Priyanka and {Joudaki}, Shahab and {Kuijken}, Konrad and {Li}, Shun-Sheng and {Linke}, Laila and {Mahony}, Constance and {Sciotti}, Davide and {Tr{\"o}ster}, Tilman and {Yoon}, Mijin},
        title = "{KiDS-Legacy: Covariance validation and the unified ONECOVARIANCE framework for projected large-scale structure observables}",
      journal = {Astronomy \& Astrophysics,},
         year = 2025,
        month = jul,
       volume = {699},
          eid = {A124},
        pages = {A124},
          doi = {10.1051/0004-6361/202452592},
archivePrefix = {arXiv},
       eprint = {2410.06962},
 primaryClass = {astro-ph.CO},
       adsurl = {https://ui.adsabs.harvard.edu/abs/2025A&A...699A.124R}
}

@article{Matsubara_LPT_PhysRevD.77.063530,
  title = {Resumming cosmological perturbations via the Lagrangian picture: One-loop results in real space and in redshift space},
  author = {Matsubara, Takahiko},
  journal = {Phys. Rev. D},
  volume = {77},
  issue = {6},
  pages = {063530},
  numpages = {19},
  year = {2008},
  month = {Mar},
  publisher = {American Physical Society},
  doi = {10.1103/PhysRevD.77.063530},
  url = {https://link.aps.org/doi/10.1103/PhysRevD.77.063530}
}

@article{assb_10.1111/j.1365-2966.2006.11230.x,
    author = {Croton, Darren J. and Gao, Liang and White, Simon D. M.},
    title = {Halo assembly bias and its effects on galaxy clustering},
    journal = {Monthly Notices of the Royal Astronomical Society},
    volume = {374},
    number = {4},
    pages = {1303-1309},
    year = {2007},
    month = {01},
    issn = {0035-8711},
    doi = {10.1111/j.1365-2966.2006.11230.x},
    url = {https://doi.org/10.1111/j.1365-2966.2006.11230.x},
    eprint = {https://academic.oup.com/mnras/article-pdf/374/4/1303/2860822/mnras0374-1303.pdf},
}

@article{assb_PhysRevLett.116.041301,
  title = {Evidence of Halo Assembly Bias in Massive Clusters},
  author = {Miyatake, Hironao and More, Surhud and Takada, Masahiro and Spergel, David N. and Mandelbaum, Rachel and Rykoff, Eli S. and Rozo, Eduardo},
  journal = {Phys. Rev. Lett.},
  volume = {116},
  issue = {4},
  pages = {041301},
  numpages = {5},
  year = {2016},
  month = {Jan},
  publisher = {American Physical Society},
  doi = {10.1103/PhysRevLett.116.041301},
  url = {https://link.aps.org/doi/10.1103/PhysRevLett.116.041301}
}

@article{Cleary_2022,
   title={COMAP Early Science. I. Overview},
   volume={933},
   ISSN={1538-4357},
   url={http://dx.doi.org/10.3847/1538-4357/ac63cc},
   DOI={10.3847/1538-4357/ac63cc},
   number={2},
   journal={The Astrophysical Journal},
   publisher={American Astronomical Society},
   author={Cleary, Kieran A. and Borowska, Jowita and Breysse, Patrick C. and Catha, Morgan and Chung, Dongwoo T. and Church, Sarah E. and Dickinson, Clive and Eriksen, Hans Kristian and Foss, Marie Kristine and Gundersen, Joshua Ott and Harper, Stuart E. and Harris, Andrew I. and Hobbs, Richard and Ihle, Håvard T. and Kim, Junhan and Kocz, Jonathon and Lamb, James W. and Lunde, Jonas G. S. and Padmanabhan, Hamsa and Pearson, Timothy J. and Philip, Liju and Powell, Travis W. and Rasmussen, Maren and Readhead, Anthony C. S. and Rennie, Thomas J. and Silva, Marta B. and Stutzer, Nils-Ole and Uzgil, Bade D. and Watts, Duncan J. and Wehus, Ingunn Kathrine and Woody, David P. and Basoalto, Lilian and Bond, J. Richard and Dunne, Delaney A. and Gaier, Todd and Hensley, Brandon and Keating, Laura C. and Lawrence, Charles R. and Murray, Norman and Paladini, Roberta and Reeves, Rodrigo and Viero, Marco P. and Wechsler, Risa H.},
   year={2022},
   month=jul, pages={182} }

@article{COMAP:2018kem,
    author = "Ihle, H{\r{a}}vard Tveit and others",
    collaboration = "COMAP",
    title = "{Joint power spectrum and voxel intensity distribution forecast on the CO luminosity function with COMAP}",
    eprint = "1808.07487",
    archivePrefix = "arXiv",
    primaryClass = "astro-ph.CO",
    doi = "10.3847/1538-4357/aaf4bc",
    journal = "Astrophys. J.",
    volume = "871",
    number = "1",
    pages = "75",
    year = "2019"
}

@article{Breysse_2022,
   title={COMAP Early Science. VII. Prospects for CO Intensity Mapping at Reionization},
   volume={933},
   ISSN={1538-4357},
   url={http://dx.doi.org/10.3847/1538-4357/ac63c9},
   DOI={10.3847/1538-4357/ac63c9},
   number={2},
   journal={The Astrophysical Journal},
   publisher={American Astronomical Society},
   author={Breysse, Patrick C. and Chung, Dongwoo T. and Cleary, Kieran A. and Ihle, Håvard T. and Padmanabhan, Hamsa and Silva, Marta B. and Bond, J. Richard and Borowska, Jowita and Catha, Morgan and Church, Sarah E. and Dunne, Delaney A. and Eriksen, Hans Kristian and Foss, Marie Kristine and Gaier, Todd and Gundersen, Joshua Ott and Harris, Andrew I. and Hobbs, Richard and Keating, Laura and Lamb, James W. and Lawrence, Charles R. and Lunde, Jonas G. S. and Murray, Norman and Pearson, Timothy J. and Philip, Liju and Rasmussen, Maren and Readhead, Anthony C. S. and Rennie, Thomas J. and Stutzer, Nils-Ole and Viero, Marco P. and Watts, Duncan J. and Wehus, Ingunn Kathrine and Woody, David P.},
   year={2022},
   month=jul, pages={188} }

@article{Karkare_2022,
   title={SPT-SLIM: A Line Intensity Mapping Pathfinder for the South Pole Telescope},
   volume={209},
   ISSN={1573-7357},
   url={http://dx.doi.org/10.1007/s10909-022-02702-2},
   DOI={10.1007/s10909-022-02702-2},
   number={5–6},
   journal={Journal of Low Temperature Physics},
   publisher={Springer Science and Business Media LLC},
   author={Karkare, K. S. and Anderson, A. J. and Barry, P. S. and Benson, B. A. and Carlstrom, J. E. and Cecil, T. and Chang, C. L. and Dobbs, M. A. and Hollister, M. and Keating, G. K. and Marrone, D. P. and McMahon, J. and Montgomery, J. and Pan, Z. and Robson, G. and Rouble, M. and Shirokoff, E. and Smecher, G.},
   year={2022},
   month=mar, pages={758–765} }

@article{Modi:2019ewx,
    author = "Modi, Chirag and Castorina, Emanuele and Feng, Yu and White, Martin",
    title = "{Intensity mapping with neutral hydrogen and the Hidden Valley simulations}",
    eprint = "1904.11923",
    archivePrefix = "arXiv",
    primaryClass = "astro-ph.CO",
    doi = "10.1088/1475-7516/2019/09/024",
    journal = "JCAP",
    volume = "09",
    pages = "024",
    year = "2019"
}

@article{Feng:2016yqz,
    author = "Feng, Yu and Chu, Man-Yat and Seljak, Uros and McDonald, Patrick",
    title = "{FastPM: a new scheme for fast simulations of dark matter and haloes}",
    eprint = "1603.00476",
    archivePrefix = "arXiv",
    primaryClass = "astro-ph.CO",
    doi = "10.1093/mnras/stw2123",
    journal = "Mon. Not. Roy. Astron. Soc.",
    volume = "463",
    number = "3",
    pages = "2273--2286",
    year = "2016"
}

@article{Bethermin:2022lmd,
    author = "Bethermin, M. and others",
    title = "{CONCERTO: High-fidelity simulation of millimeter line emissions of galaxies and [CII] intensity mapping}",
    eprint = "2204.12827",
    archivePrefix = "arXiv",
    primaryClass = "astro-ph.GA",
    doi = "10.1051/0004-6361/202243888",
    journal = "Astron. Astrophys.",
    volume = "667",
    pages = "A156",
    year = "2022"
}

@article{Sato-Polito:2022wiq,
    author = "Sato-Polito, Gabriela and Kokron, Nickolas and Bernal, Jos{\'e} Luis",
    title = "{A multitracer empirically driven approach to line-intensity mapping light cones}",
    eprint = "2212.08056",
    archivePrefix = "arXiv",
    primaryClass = "astro-ph.CO",
    doi = "10.1093/mnras/stad2498",
    journal = "Mon. Not. Roy. Astron. Soc.",
    volume = "526",
    number = "4",
    pages = "5883--5899",
    year = "2023"
}

@article{Cunnington_2022,
   title={H<scp>i</scp> intensity mapping with MeerKAT: power spectrum detection in cross-correlation with WiggleZ galaxies},
   volume={518},
   ISSN={1365-2966},
   url={http://dx.doi.org/10.1093/mnras/stac3060},
   DOI={10.1093/mnras/stac3060},
   number={4},
   journal={Monthly Notices of the Royal Astronomical Society},
   publisher={Oxford University Press (OUP)},
   author={Cunnington, Steven and Li, Yichao and Santos, Mario G and Wang, Jingying and Carucci, Isabella P and Irfan, Melis O and Pourtsidou, Alkistis and Spinelli, Marta and Wolz, Laura and Soares, Paula S and Blake, Chris and Bull, Philip and Engelbrecht, Brandon and Fonseca, José and Grainge, Keith and Ma, Yin-Zhe},
   year={2022},
   month=oct, pages={6262–6272} }

@article{Cunnington:2025sdr,
    author = "Cunnington, Steven and others",
    title = "{Revealing cosmological fluctuations in 21cm intensity maps with MeerKLASS: from maps to power spectra}",
    eprint = "2510.27549",
    archivePrefix = "arXiv",
    primaryClass = "astro-ph.CO",
    month = "10",
    year = "2025"
}

@ARTICLE{1974ApJ...187..425P,
       author = {{Press}, William H. and {Schechter}, Paul},
        title = "{Formation of Galaxies and Clusters of Galaxies by Self-Similar Gravitational Condensation}",
      journal = {\apj},
         year = 1974,
        month = feb,
       volume = {187},
        pages = {425-438},
          doi = {10.1086/152650},
       adsurl = {https://ui.adsabs.harvard.edu/abs/1974ApJ...187..425P}
}

@ARTICLE{Jing_resampling_2005ApJ...620..559J,
       author = {{Jing}, Y.~P.},
        title = "{Correcting for the Alias Effect When Measuring the Power Spectrum Using a Fast Fourier Transform}",
      journal = {\apj},
         year = 2005,
        month = feb,
       volume = {620},
       number = {2},
        pages = {559-563},
          doi = {10.1086/427087},
archivePrefix = {arXiv},
       eprint = {astro-ph/0409240},
 primaryClass = {astro-ph},
       adsurl = {https://ui.adsabs.harvard.edu/abs/2005ApJ...620..559J}
}

@misc{niemeyer2025lyalphaintensitymappinghetdex,
      title={Ly{\ensuremath{\alpha}} Intensity Mapping in HETDEX: Galaxy-Ly{\ensuremath{\alpha}} Intensity Cross-Power Spectrum}, 
      author={Maja Lujan Niemeyer and Eiichiro Komatsu and José Luis Bernal and Chris Byrohl and Robin Ciardullo and Olivia Curtis and Daniel J. Farrow and Steven L. Finkelstein and Karl Gebhardt and Caryl Gronwall and Gary J. Hill and Matt J. Jarvis and Donghui Jeong and Erin Mentuch Cooper and Deeshani Mitra and Shiro Mukae and Julian B. Muñoz and Masami Ouchi and Shun Saito and Donald P. Schneider and Lutz Wisotzki},
      year={2025},
      eprint={2510.11427},
      archivePrefix={arXiv},
      primaryClass={astro-ph.CO},
      url={https://arxiv.org/abs/2510.11427}, 
}

@article{Paul:2023yrr,
    author = "Paul, Sourabh and Santos, Mario G. and Chen, Zhaoting and Wolz, Laura",
    title = "{A first detection of neutral hydrogen intensity mapping on Mpc scales at $z\approx 0.32$ and $z\approx 0.44$}",
    eprint = "2301.11943",
    archivePrefix = "arXiv",
    primaryClass = "astro-ph.CO",
    month = "1",
    year = "2023"
}

@article{CHIME:2025cee,
    author = "Amiri, Mandana and others",
    collaboration = "CHIME",
    title = "{Detection of the Cosmological 21 cm Signal in Auto-correlation at z {\textasciitilde} 1 with the Canadian Hydrogen Intensity Mapping Experiment}",
    eprint = "2511.19620",
    archivePrefix = "arXiv",
    primaryClass = "astro-ph.CO",
    month = "11",
    year = "2025"
}

@article{Carucci:2024qpm,
    author = "Carucci, Isabella P. and others",
    title = "{Hydrogen intensity mapping with MeerKAT: Preserving cosmological signal by optimising contaminant separation}",
    eprint = "2412.06750",
    archivePrefix = "arXiv",
    primaryClass = "astro-ph.CO",
    doi = "10.1051/0004-6361/202453461",
    journal = "Astron. Astrophys.",
    volume = "703",
    pages = "A222",
    year = "2025"
}

@article{10.1093/mnras/staf195,
    author = {MeerKLASS Collaboration  and Barberi-Squarotti, Matilde and Bernal, José L and Bull, Philip and Camera, Stefano and Carucci, Isabella P and Chen, Zhaoting and Cunnington, Steven and Engelbrecht, Brandon N and Fonseca, José and Grainge, Keith and Irfan, Melis O and Li, Yichao and Mazumder, Aishrila and Paul, Sourabh and Pourtsidou, Alkistis and Santos, Mario G and Spinelli, Marta and Wang, Jingying and Witzemann, Amadeus and Wolz, Laura},
    title = {MeerKLASS L-band deep-field intensity maps: entering the H<scp>i</scp> dominated regime},
    journal = {Monthly Notices of the Royal Astronomical Society},
    volume = {537},
    number = {4},
    pages = {3632-3661},
    year = {2025},
    month = {02},
    issn = {0035-8711},
    doi = {10.1093/mnras/staf195},
    url = {https://doi.org/10.1093/mnras/staf195},
    eprint = {https://academic.oup.com/mnras/article-pdf/537/4/3632/61743817/staf195.pdf},
}

@article{Bernal_2019,
   title={User’s guide to extracting cosmological information from line-intensity maps},
   volume={100},
   ISSN={2470-0029},
   url={http://dx.doi.org/10.1103/PhysRevD.100.123522},
   DOI={10.1103/physrevd.100.123522},
   number={12},
   journal={Physical Review D},
   publisher={American Physical Society (APS)},
   author={Bernal, José Luis and Breysse, Patrick C. and Gil-Marín, Héctor and Kovetz, Ely D.},
   year={2019},
   month=dec }

@article{Repp:2015jja,
    author = "Repp, Andrew and Szapudi, Istv{\'a}n and Carron, Julien and Wolk, Melody",
    title = "{The Impact of Non-Gaussianity upon Cosmological Forecasts}",
    eprint = "1506.00083",
    archivePrefix = "arXiv",
    primaryClass = "astro-ph.CO",
    doi = "10.1093/mnras/stv2212",
    journal = "Mon. Not. Roy. Astron. Soc.",
    volume = "454",
    number = "4",
    pages = "3533--3541",
    year = "2015"
}

@article{Harnois-Deraps:2012kbb,
    author = "Harnois-Deraps, Joachim and Pen, Ue-Li",
    title = "{Non-Gaussian Error Bars in Galaxy Surveys -- 2}",
    eprint = "1211.6213",
    archivePrefix = "arXiv",
    primaryClass = "astro-ph.CO",
    doi = "10.1093/mnras/stt413",
    journal = "Mon. Not. Roy. Astron. Soc.",
    volume = "431",
    pages = "3349",
    year = "2013"
}

@article{Mohammed:2016sre,
    author = "Mohammed, Irshad and Seljak, Uros and Vlah, Zvonimir",
    title = "{Perturbative approach to covariance matrix of the matter power spectrum}",
    eprint = "1607.00043",
    archivePrefix = "arXiv",
    primaryClass = "astro-ph.CO",
    reportNumber = "FERMILAB-PUB-16-247-A",
    doi = "10.1093/mnras/stw3196",
    journal = "Mon. Not. Roy. Astron. Soc.",
    volume = "466",
    number = "1",
    pages = "780--797",
    year = "2017"
}

@article{Li:2014jra,
    author = "Li, Yin and Hu, Wayne and Takada, Masahiro",
    title = "{Super-Sample Signal}",
    eprint = "1408.1081",
    archivePrefix = "arXiv",
    primaryClass = "astro-ph.CO",
    doi = "10.1103/PhysRevD.90.103530",
    journal = "Phys. Rev. D",
    volume = "90",
    number = "10",
    pages = "103530",
    year = "2014"
}

@article{Hamilton:2005dx,
    author = "Hamilton, Andrew J. S. and Rimes, Christopher D. and Scoccimarro, Roman",
    title = "{On measuring the covariance matrix of the nonlinear power spectrum from simulations}",
    eprint = "astro-ph/0511416",
    archivePrefix = "arXiv",
    doi = "10.1111/j.1365-2966.2006.10709.x",
    journal = "Mon. Not. Roy. Astron. Soc.",
    volume = "371",
    pages = "1188--1204",
    year = "2006"
}

@article{Li:2014sga,
    author = "Li, Yin and Hu, Wayne and Takada, Masahiro",
    title = "{Super-Sample Covariance in Simulations}",
    eprint = "1401.0385",
    archivePrefix = "arXiv",
    primaryClass = "astro-ph.CO",
    doi = "10.1103/PhysRevD.89.083519",
    journal = "Phys. Rev. D",
    volume = "89",
    number = "8",
    pages = "083519",
    year = "2014"
}

@article{Harnois-Deraps:2011ixh,
    author = "Harnois-Deraps, Joachim and Pen, Ue-Li",
    title = "{Non-Gaussian Error in Galaxy Survey (Part 1)}",
    eprint = "1109.5746",
    archivePrefix = "arXiv",
    primaryClass = "astro-ph.CO",
    doi = "10.1111/j.1365-2966.2012.21039.x",
    journal = "Mon. Not. Roy. Astron. Soc.",
    volume = "423",
    pages = "2288",
    year = "2012"
}

@article{Bernardeau_2002,
   title={Large-scale structure of the Universe and cosmological perturbation theory},
   volume={367},
   ISSN={0370-1573},
   url={http://dx.doi.org/10.1016/S0370-1573(02)00135-7},
   DOI={10.1016/s0370-1573(02)00135-7},
   number={1–3},
   journal={Physics Reports},
   publisher={Elsevier BV},
   author={Bernardeau, F. and Colombi, S. and Gaztañaga, E. and Scoccimarro, R.},
   year={2002},
   month=sep, pages={1–248} }

@article{Bertolini:2015fya,
    author = "Bertolini, Daniele and Schutz, Katelin and Solon, Mikhail P. and Walsh, Jonathan R. and Zurek, Kathryn M.",
    title = "{Non-Gaussian Covariance of the Matter Power Spectrum in the Effective Field Theory of Large Scale Structure}",
    eprint = "1512.07630",
    archivePrefix = "arXiv",
    primaryClass = "astro-ph.CO",
    doi = "10.1103/PhysRevD.93.123505",
    journal = "Phys. Rev. D",
    volume = "93",
    number = "12",
    pages = "123505",
    year = "2016"
}

\end{document}